%% file: example.tex
\documentclass[fleqn,usenatbib]{mnras}
\usepackage{newtxtext,newtxmath}
\usepackage{chngcntr}
\usepackage[T1]{fontenc}

\DeclareRobustCommand{\VAN}[3]{#2}
\let\VANthebibliography\thebibliography
\def\thebibliography{\DeclareRobustCommand{\VAN}[3]{##3}\VANthebibliography}

\usepackage{graphics,graphicx}	% Including figure files
\usepackage{amsmath}	% Advanced maths commands
\input{macros.tex}
\usepackage{tikz}
   \usetikzlibrary{positioning,spy}
\usepackage{graphicx}
\usepackage{subfig}
\usepackage{tikz,xcolor,hyperref}

\definecolor{lime}{HTML}{A6CE39}
\DeclareRobustCommand{\orcidicon}{%
	\begin{tikzpicture}
	\draw[lime, fill=lime] (0,0) 
	circle [radius=0.16] 
	node[white] {{\fontfamily{qag}\selectfont \tiny ID}};
	\draw[white, fill=white] (-0.0625,0.095) 
	circle [radius=0.007];
	\end{tikzpicture}
	\hspace{-2mm}
}

\foreach \x in {A, ..., Z}{%
	\expandafter\xdef\csname orcid\x\endcsname{\noexpand\href{https://orcid.org/\csname orcidauthor\x\endcsname}{\noexpand\orcidicon}}
}

\title[Enrichment in four galaxy groups]{Chemical abundances in the outskirts of nearby galaxy groups measured with joint $\suzaku$ and $\chandra$ observations}

\author[Sarkar et al.]{
Arnab Sarkar,$^{1,2}$\thanks{E-mail: \href{mailto:arnab.sarkar@uky.edu,arnab.sarkar@cfa.harvard.edu}{arnab.sarkar@uky.edu,arnab.sarkar@cfa.harvard.edu}}\orcidA{}
Yuanyuan Su,$^{1}$
Nhut Truong,$^3$
Scott Randall,$^{2}$\orcidC{}
Fran\c{c}ois Mernier,$^{4,5}$\orcidI{}
\newauthor
Fabio Gastaldello,$^{6}$\orcidD{}
Veronica Biffi,$^{7}$
and
Ralph Kraft.$^{2}$\orcidF{}\\
$^{1}$Physics and Astronomy, University of Kentucky, 505 Rose street, Lexington, KY 40506, USA\\
$^{2}$Center for Astrophysics $\vert$ Harvard \& Smithsonian , 60 Garden Street, Cambridge, MA 02138, USA\\
$^{3}$Max-Planck-Institut für Astronomie, Königstuhl 17, D-69117 Heidelberg, Germany\\
$^{4}$European Space Agency (ESA), European Space Research and Technology Centre (ESTEC), Keplerlaan 1, 2201 AZ Noordwijk, The Netherlands\\
$^{5}$SRON Netherlands Institute for Space Research, Sorbonnelaan 2, NL-3584 CA Utrecht, The Netherlands\\
$^{6}$Instituto di Astrofisica Spaziale e Fisica Cosmica (INAF-IASF), Milano, via A. Corti 12, I-20133 Milano, Italy\\
$^{7}$INAF — Osservatorio Astronomico di Trieste, via Tiepolo 11, I-34143 Trieste, Italy\\
}

\date{Accepted XXX. Received YYY; in original form ZZZ}

\begin{document}
\label{firstpage}
\pagerange{\pageref{firstpage}--\pageref{lastpage}}
\maketitle

\begin{abstract}
%\TODO{re-write}
We report results from deep {\sl Suzaku} and mostly snapshot {\sl Chandra} 
observations of four nearby galaxy groups: MKW4, Antlia, RXJ1159+5531, 
and ESO3060170. Their peak temperatures vary over 2--3\,keV, 
making them the smallest systems 
{with} gas properties constrained 
to their viral radii.
The average Fe abundance in the outskirts (R $>$ 0.25R$_{200}$) 
of their intragroup medium (IGrM) is $Z_{\rm Fe}=0.309\pm0.018$ $Z_\odot$
with $\chi^2$ = 14 for 12 degrees of freedom,
which is remarkably uniform and strikingly similar to that 
of massive galaxy clusters, 
and {is} fully consistent with the numerical predictions 
from the IllustrisTNG cosmological simulation. 
Our results support an early-enrichment scenario among galactic 
systems over an order of magnitude in mass, even before 
their formation.
When integrated out to R$_{200}$, we start to see a tension between 
the measured Fe content in ICM and what is expected from supernovae yields.
We further constrain their O, Mg, Si, S, and Ni abundances. 
{The abundance ratios of those elements
relative to Fe are consistent 
with the predictions (if available) from IllustrisTNG.}
Their Type Ia supernovae fraction varies between 14\%--21\%.
%broadly consistent with those of our Galaxy and  
%{ massive} cluster centers
A pure core collapsed supernovae enrichment at group outskirts can be ruled out.
%Group centers have a chemical composition similar to our Solar system and that of 
%{ massive} cluster centers,
%while the X/Fe abundance ratios of group outskirts appear to be super solar, in spite of the sizable uncertainties. 
%, implying different enrichment processes for group centers and outskirts.
%The origin of metals at group centers may be dominated by their brightest group galaxies.
Their cumulative iron-mass-to-light ratios within R$_{200}$ are half 
that of the Perseus cluster, 
which may imply that galaxy groups do not retain all of their
enriched gas due to their shallower gravitational potential wells, 
or that groups and clusters may have different star formation histories. 
\end{abstract}

%TC:ignore


%%%%%%%%%%%%%%%%%%%%%%%%%%%%%%%%%%%%%%%%%%%%%%%%%%

%%%%%%%%%%%%%%%%% BODY OF PAPER %%%%%%%%%%%%%%%%%%

\section{Introduction}\label{sec:intro}

Big Bang nucleosynthesis primarily produced
all the hydrogen and helium in the Universe, and 
{trace amounts of
a} few lighter elements like
Li and Be. Heavier elements are later forged in
stars.
%and likely be injected into the intracluster medium (ICM) by 
%the galactic winds and outflows powered by supernovae.
%\citep[e.g.,][]{2007A&A...465..345D}.
Clusters of galaxies, with their deep gravitational potential wells,
retain substantial X-ray emitting hot gas (T $\sim$ 10$^7$--10$^8$ K), the
so-called intracluster medium (ICM) \citep{1999MNRAS.305..834E}. 
%The ICM is enriched with metals ejected by galactic winds and
A dominant fraction of metals 
in the local Universe can be found in the ICM, 
making them a unique astrophysical
laboratory to probe 
nucleosynthesis and {the}
chemical enrichment
history of the Universe \citep{2018SSRv..214..123B,2018SSRv..214..129M}.

The enrichment processes in the ICM 
{remain} an open question. 
In the late enrichment 
scenarios (after the cluster is assembled), 
the ICM metal distributions are expected to be
non-uniform, with a significant amount of 
azimuthal
scatters \citep{2006A&A...452..795D}. 
For instance,
galactic materials can be stripped off the 
galaxies and deposited in the ICM as 
the
infalling galaxies interact with the dense gas
\citep[e.g.,][]{1972ApJ...176....1G,2010MNRAS.401.1670F}, which
enriches the ICM along the directions of 
infalling galaxies. 
Also, the metals dispersed into 
the ICM at later times should broadly follow the spatial 
distributions
of galaxies, leading to flat metal mass-to-light ratios
\citep{2013PASJ...65...10M}. 
Neither of these is consistent with the observations.
Nearby massive clusters, such as Perseus, show a remarkably homogeneous
Fe abundance of $\sim$ 0.3 $Z_{\odot}$ at R $>$ 
0.25R$_{200}$\footnote[2]{$R_{\Delta}$\: the radius from within which the average
matter density is $\Delta$ times the critical density of the Universe.} 
%\citep[see this review article][]{2021Univ....7..208G},
%such as Perseus 
%\citep{2013Natur.502..656W}, and
%other massive nearby clusters 
\citep{2017MNRAS.470.4583U, 2013Natur.502..656W}.
Previous studies, using $\chandra$, $\suzaku$, and $\xmm$ observations, 
have 
reported steep iron mass-to-light ratios (IMLRs) 
for several galaxy clusters out to 
0.2--0.5R$_{200}$ 
\citep[e.g.,][]{2007A&A...462..953M,2009PASJ...61S.365S,Simionescu1576}, 
which indicates that metal mass is more extended 
than the distribution of cluster galaxies.
Although the ICM can achieve a uniform { distribution} of metals 
in the late enrichment process via
{large-scale} sloshing motion, 
the steep entropy profiles of galaxy 
clusters may prevent the
efficient mixing up of metals across the large 
{radii} 
\citep{2006A&A...452..795D,2013Natur.502..656W,2014A&A...570A.117G}.

The observed uniform Fe abundances and steep IMLR profiles suggest
an early enrichment scenario, 
in which
most of the enrichment occurred in the proto-cluster
environment \citep[e.g.,][]{2010MNRAS.401.1670F,2017MNRAS.468..531B}.
This process is primarily 
driven by the galactic winds and 
active galactic nuclei (AGN) feedback
at the epoch of the peak star formation 
\citep[around $z$ $\sim$ 2-3;][]{2014ARA&A..52..415M}.
The strong galactic winds and AGN 
activity may have expelled much of the
metals from the interstellar medium (ISM) 
and uniformly mixed the intergalactic gas even
before the cluster was assembled.
%Later entropy
%profile became steeper, which prevents further efficient mixing of metals
%\citep{2006A&A...452..795D}. 

With {a} 
better understanding of 
stellar nucleosynthesis over the past
decades, it is now evident that most of the 
%heavy elements 
%(heavier than H, He,
%Li and Be) 
%are forged in 
%stars and dispersed into and beyond ISM by the 
%supernovae (SNe) explosions \citep{1957RvMP...29..547B}. 
%Two types of SNe are mostly responsible for 
%metal enrichment in ICM -- Type Ia
%supernovae (SNe Ia) and core-collapsed supernovae (SNcc).
%While the former one depends on the age of the elliptical
%galaxies, the latter one depends on the initial mass function (IMF)
%of progenitor galaxies. 
lighter elements,
like O, Mg, and Ne, are synthesized by massive stars and expelled into the ICM 
by core-collapsed supernovae (SNcc)
\citep[e.g.,][]{2006NuPhA.777..424N}, while
the heavier elements, like Ar, Ca, Fe, and Ni, are primarily produced
by Type Ia supernovae (SNe Ia) 
\citep[e.g.,][]{1999ApJS..125..439I,2021ApJ...907...12S}.
Elements {such as}
Si and S 
are produced 
relatively comparably
by SNe Ia and SNcc \citep[e.g.,][]{2007A&A...465..345D}. 
Therefore, the metal abundance pattern 
of heavier elements strongly depends on 
the SNe Ia 
explosion mechanism, and the abundances of lighter elements 
depend on the stellar initial mass function (IMF) 
of galaxies 
\citep{2015A&A...575A..37M,2016A&A...595A.126M}. 
The measurement of the abundance profiles of those metals 
is crucial to constrain supernovae models, and to probe the chemical enrichment
history of the ICM.

The $\alpha$-capture (such as O, Mg, Ne, Si, S) and 
Fe peak elements (such as Cr, Mn, Fe, Co, Ni)
can be observed from their
emission lines in X-ray. Most of these emission lines 
originate from the K and L -- shell transitions in highly ionized
plasma (\citealt{2020ApJ...901...68C,2020ApJ...901...69C}). 
Since the ICM is very close to collisional ionization equilibrium
and optically thin, the equivalent widths of those 
emission lines can easily be
converted to element abundances 
\citep{2010A&ARv..18..127B,2016A&A...592A.157M,2017MNRAS.470.4583U}. 
%It has been reported that
%Mg, Si, and S abundances are likely to follow the Fe abundances to increase
%\citep[e.g.,][]{2004A&A...420..135T,2009PASJ...61S.365S,2013ApJ...764..147M,2016A&A...595A.126M}.
%In contrast, O abundances approximately stays same. 
Using $\chandra$ observations, \citet{2007MNRAS.380.1554R} have 
found that 
{the}
Si/Fe ratio increases with radius, strongly
suggesting an excess of SNe Ia contribution at  
{the} cluster center,
while {the}
SNcc contribution dominates at the outskirts. 
Later on, however, it was observed in M87 that the Si abundance 
profile is even more centrally peaked 
than Fe \citep{2011MNRAS.418.2744M}.
Most recently, an extensive $\suzaku$ investigation
of the nearby Virgo cluster reveals 
{an} SNIa 
fraction of 12--37\% throughout its ICM, 
which challenges pure SNcc enrichment at the
cluster outskirts \citep{2015ApJ...811L..25S}.
The current view is 
{that
abundance profile is roughly the same for all elements.}
%It is desirable to expand the sample rather than drawing conclusions based on the ``cherry-picking" of targets. 
%for example \citet{2015ApJ...811L..25S} and 
%\citet{2017A&A...603A..80M}, have found a uniform SNe Ia-to-SNe ratios
%near the cluster 
%outskirts, which challenges the pure SNcc enrichment of the ICM
%at the outskirts.

%Several previous studies have measured metal abundances in various
%systems, including cool-core and non cool-core clusters.
%The $\asca$ observations of Centaurus cluster reported
%a iron peak at the cluster core, 
%which declines 
%to 0.2--0.3Z$_{\odot}$ at the outskirts \citep{1994PASJ...46L..55F}. 
%For instance,
%Fornax cluster 
%\citep{2002ApJ...574L.135B}, NGC 5044 \citep{2003ApJ...595..151B},
%Abell~4059 \citep{2015A&A...575A..37M}, and Abell~262 
%\citep{2008PASJ...60S.333S}
%have shown
%an excess Fe abundance of 0.6--1.5 Z$_{\odot}$ within the central
%region. While cool-core clusters retain a metallicity peak at the center, non 
%cool-core clusters do not show a significant
%metallicity peak at the centers \citep{2004A&A...419..837D}. The frequent
%merger events in non cool-core clusters disrupt its core and make 
%abundance profiles
%flatter at the center 
%\citep[e.g.,][]{2018ApJS..238...14M,2019MNRAS.483..540L}.

\begin{table}
\centering
\caption{Observational log} 
%\TODO{same for Antlia}
\label{tab:obs_log}
%\resizebox{0.8\columnwidth}{!}{%
\scalebox{0.9}{
\begin{tabular}{cccccc}
\hline
Name & Obs Id & Obs Date & Exp. (ks) \\
%%& RA ($^\circ$) & DEC ($^\circ$) & P.I.\\
%%\decimalcolnumbers
\hline
& \ \ \ \ \ \ \ \ \ \ \ \ \ \ \ \ \ \ \ $\suzaku$ & \\
\hline
MKW4 central & 808066010  & 2013 Dec 30 & 34.6 \\
%%& 181.1346 & 1.9097 & 
%%F. Gastaldello \\
MKW4 Offset 1 & 805081010  & 2010 Nov 30 & 77.23 \\
%%& 181.1270 & 2.2181 &
%%F. Gastaldello\\
MKW4 N2 & 808067010  & 
2010 Nov 30 & 97 \\
%%& 181.1504 & 2.5206 & Y. Su\\
MKW4 Offset 2 & 805082010  & 2010 Nov 30 & 80 \\
%%& 181.4270 & 1.8972 & Y.
%%Su\\
MKW4 E1 & 808065010  & 
2013 Dec 29 & 100 \\
%%& 181.7142 & 1.8506 & F. 
%%Gastaldello\\
MKW4 NE & 809062010  & 
2013 Dec 29 & 87.5 \\
%%& 181.4583 & 2.3361 & Y. Su\\
\hline
Antlia E0 & 802035010  & 2007 Nov 19 & 55 \\
%%& 181.1346 & 1.9097 & 
%%F. Gastaldello \\
Antlia E1 & 807066010  & 2012 Jun 13 & 20 \\
%%& 181.1270 & 2.2181 &
%%F. Gastaldello\\
Antlia E2 & 807067010  & 
2012 Jun 14 & 21 \\
%%& 181.1504 & 2.5206 & Y. Su\\
Antlia E3 & 807068010  & 2012 Jun 15 & 19 \\
%%& 181.4270 & 1.8972 & Y.
%%Su\\
Antlia E4 & 807069010  & 
2012 Jun 16 & 17 \\
%%& 181.7142 & 1.8506 & F. 
%%Gastaldello\\
Antlia E5 & 807070010  & 
2012 Jun 17 & 39 \\
Antlia EB & 807071010  & 
2012 Jun 18 & 38 \\
\hline
RXJ1159 N & 804051010 &
2009 May 02 & 84 \\
RXJ1159 S & 807064010 &
2012 May 27 & 81 \\
 & 807064020 &
2012 Dec 18 & 21 \\
RXJ1159 E & 809063010 &
2014 May 29 & 96 \\
RXJ1159 W & 809064010 &
2014 May 31 & 94 \\
\hline
ESO3060170 central & 805075010 &
2010 May & 27 \\
ESO3060170 offset & 805075060 &
2010 May & 70 \\
\hline
& \ \ \ \ \ \ \ \ \ \ \ \ \ \ \ \ \ \ \ $\chandra$ & \\
\hline
MKW4 central & 3234  & 2002 Nov 
24 & 30 \\
%%& 181.1283 & 1.9286 & Y. Fukazawa\\
MKW4 N2 & 20593  & 2019 Feb 25 & 
14 \\
%%& 181.1369 & 2.5209 & Y. Su\\
MKW4 E1 & 20592 & 2018 Nov 17 & 
15 \\
%%& 181.7146 & 1.8715 & Y. Su\\
MKW4 NE & 20591  & 2019 Mar 08 &
14 \\
\hline
Antlia E1 & 15090  & 2013 Nov 
20 & 7 \\
%%& 181.1283 & 1.9286 & Y. Fukazawa\\
Antlia E2 & 15089  & 2013 Nov 22 & 
7 \\
%%& 181.1369 & 2.5209 & Y. Su\\
Antlia E3 & 15088  & 2013 Jul 02 & 
7 \\
%%& 181.7146 & 1.8715 & Y. Su\\
Antlia E4 & 15086  & 2013 Nov 04 &
7 \\
Antlia E5 & 15085  & 2013 Apr 05 &
7 \\
Antlia EB & 15087  & 2013 Nov 04 &
7 \\
%%& 181.4548 & 2.3585 & Y. Su\\
\hline
RXJ1159 central & 4964 & 2004 Feb 11
& 76 \\
RXJ1159 NW & 14026 & 2012 Aug 09    
& 50 \\
RXJ1159 NE & 14473 & 2012 Aug 12
& 37 \\
& 14027 & 2012 Aug 09
& 13 \\
\hline
ESO3060170 central & 3188 &
2002 March 08 & 14.06\\
ESO3060170 offset & 17219 &
2015 Oct 09 & 9.84 \\
\hline
\end{tabular}
%}
}
%%\tablecomments{}
\end{table}

%Although, the ICM enrichment processes in massive cluster outskirts are 
%well established, 
The enrichment process in lower mass systems (groups) is even less 
{straightforward}
\citep{2021Univ....7..208G}.
With deep gravitational potential wells, galaxy clusters behave like
closed-box systems, holding all of the metals ever produced by the stars
inside their virial radius \citep[e.g.,][]{2007A&A...465..345D}.
%[{\color{red}are clusters still considered ``closed-box" if they are enriched before the gravitational collapse?}]. 
In contrast,
galaxy groups have shallow gravitational potential wells, which makes them
{more}
vulnerable to losing material via non-gravitational processes, such as galactic winds driven by 
supernovae (SNe) and AGN feedback 
\citep[e.g.,][]{2007MNRAS.380.1554R,2015A&A...573A.118L,2016A&A...592A..37T}. These processes may expel 
much
of 
{the}
enriched gas out of the system \citep{2021MNRAS.501.3767S}.
Using $\asca$ data, \citet{2001PASJ...53..401M} %showed that the IMLRs for 
%groups are systematically lower compared 
%to those of massive clusters.
%\citet{2009MNRAS.399..239R} 
found that the IMLRs in galaxy groups within 0.5\,R$_{200}$ 
{increase} with their masses, while those of galaxy 
clusters are not 
{mass-dependent},
and are systematically higher than groups.
%The metal abundances and 
%the stellar mass distributions are quite different
%between clusters and groups \citep{2012ARA&A..50..353K}. 
%The star formation rates in massive clusters are low compared to the 
%galaxy groups \citep[e.g.,][]{2011A&A...535A..78Z,2019MNRAS.483..540L}.
The metal budgets out to the virial radius
of groups {are} relatively unexplored, 
due to their 
low surface brightness. 
The emission from the outskirts of groups
is typically dominated by 
the X-ray background, which
makes the measurement of metal abundances
very 
challenging \citep{2009MNRAS.399..239R}. 
With
its stable particle background and higher spectral
sensitivity below 1 keV \citep{2007PASJ...59S..23K},
$\suzaku$ can measure
the metal abundances 
more precisely at
R$_{200}$ and beyond. 
The inclusion of snapshot $\chandra$ observations
resolves much of the Cosmic X-ray 
Background (CXB), which is crucial to 
pin down the systematic uncertainty.

%Observations with larger sample using $\xmm$,
%\citet{2004A&A...419....7D} revealed that the
%cool-core clusters are likely to show a central Fe peak,
%while non-cool core clusters have more flatter Fe profile at the
%center \citep[e.g.,][]{2007A&A...461...71P,2018SSRv..214..129M,2019MNRAS.483..540L}.

In this paper, we study 
the metal abundances of four galaxy groups: MKW4, 
Antlia, RXJ1159+5531 (hereafter RXJ1159), and 
ESO3060170, from their centers 
out to R$_{200}$, with $\suzaku$ and $\chandra$.
Their peak temperatures (just outside the group core at 
$\sim0.25$\,R$_{200}$) are 2.2\,keV, 2.3\,keV, 2.5\,keV, 
and 2.7\,keV, respectively.
To our knowledge, they are the lowest mass 
systems {with} their gas properties
constrained at R$_{200}$.
MKW4, RXJ1159, and ESO3060170 are cool core systems, 
while Antlia is a non-cool-core group. 
%We measure their abundance profiles of O, Mg, Si, S, Fe and Ni and their
%ratios with respect to Fe. We discuss in detail the implications of these
%measurements for the SNe fractions and chemical enrichment history of MKW4
%and Antlia out to R$_{200}$.
%This is the second paper in this series. 
The thermodynamic properties of these four galaxy groups
out to their virial radii are reported in 
\citet{2021MNRAS.501.3767S},  
\citet{2016ApJ...829...49W}, \citet{2015ApJ...805..104S},
and \citet{2013ApJ...775...89S}, respectively. 
Unlike massive galaxy clusters, these groups do not show obvious flattening of their entropy profiles at large radii. 
%which is in sharp contrast to the flattening entropy profiles 
%observed at the outskirts of massive galaxy clusters.
We use NASA/IPAC Extragalactic 
Database\footnote[3]{\url{http://ned.ipac.caltech.edu}}
to estimate the luminosity distance for MKW4,  
Antlia, 
RXJ1159,
and ESO3060170,
by adopting a 
cosmology of H$_0$ = 70 km s$^{-1}$ Mpc$^{-1}$, 
$\Omega_{\Lambda}$ = 0.7, 
and $\Omega_\textrm{m}$ = 0.3. 
Table \ref{tab:prop} outlines the 
physical properties of 
{our}
four groups.
%We adopt R$_{200}$ = 884 kpc for MKW4 
%\citep{2021MNRAS.501.3767S}, 887 kpc for
%Antlia \citep{2016ApJ...829...49W}, 871 kpc 
%for RXJ1159 \citep{2015ApJ...805..104S}, and
%1.15 Mpc for ESO3060170 \citep{2013ApJ...775...89S}.
{Throughout
this paper, we assume a solar abundance table 
of \citet{2009ARA&A..47..481A}.
All uncertainties are reported at 
{a}
68\% confidence level.}

\begin{table}
\centering
\caption{Properties of the groups} 
%\TODO{same for Antlia}
\label{tab:prop}
%\resizebox{0.8\columnwidth}{!}{%
\scalebox{0.9}{
\begin{tabular}{cccccc}
\hline
Name & peak temp & R$_{200}$ & z & scale (1$\arcsec$) & CC/NCC \\
& (keV) & (kpc) & & (kpc)&\\
%%\decimalcolnumbers
\hline
MKW4 & 2.2 & 884 & 0.02 & 0.443 & CC\\
Antlia & 2.3 & 887 & 0.009 & 0.213 & NCC\\
RXJ1159 & 2.5 & 871 & 0.081 & 1.583 & CC\\
ESO3060170 & 2.7 & 1150 & 0.0358 & 0.73 & CC\\
\hline
\end{tabular}
%}
}
%%\tablecomments{}
\end{table}

\section{Observations and Spectral analysis} \label{sec:back_sub}
MKW4 has been observed 
{out to the virial radius}
with $\suzaku$ and $\chandra$ in three 
directions
(north, east, and northeast).
Antlia and ESO3060170 have 
{each been mapped to the viral radius in one direction,
east and south, respectively.}
RXJ1159 has been observed 
{to the virial radius} with full azimuthal coverage. 
The detailed observations are listed in Table \ref{tab:obs_log}.
The exposure
corrected and background subtracted mosaic $\suzaku$ images of
four groups in the 0.5--2.0 keV energy band 
are shown in Figures \ref{fig:image_mkw4}, 
\ref{fig:image_antlia}, \ref{fig:image_rxj},
and \ref{fig:image_eso}, respectively.
We followed the
data reduction procedure
described in 
\citet{2021MNRAS.501.3767S} for MKW4, 
\citet{2016ApJ...829...49W}
for Antlia, \citet{2015ApJ...805..104S}
for RXJ1159, and
\citet{2013ApJ...775...89S} for ESO3060170. 
\begin{figure*}
\begin{tabular}{cc}
\includegraphics[width=0.47\textwidth]{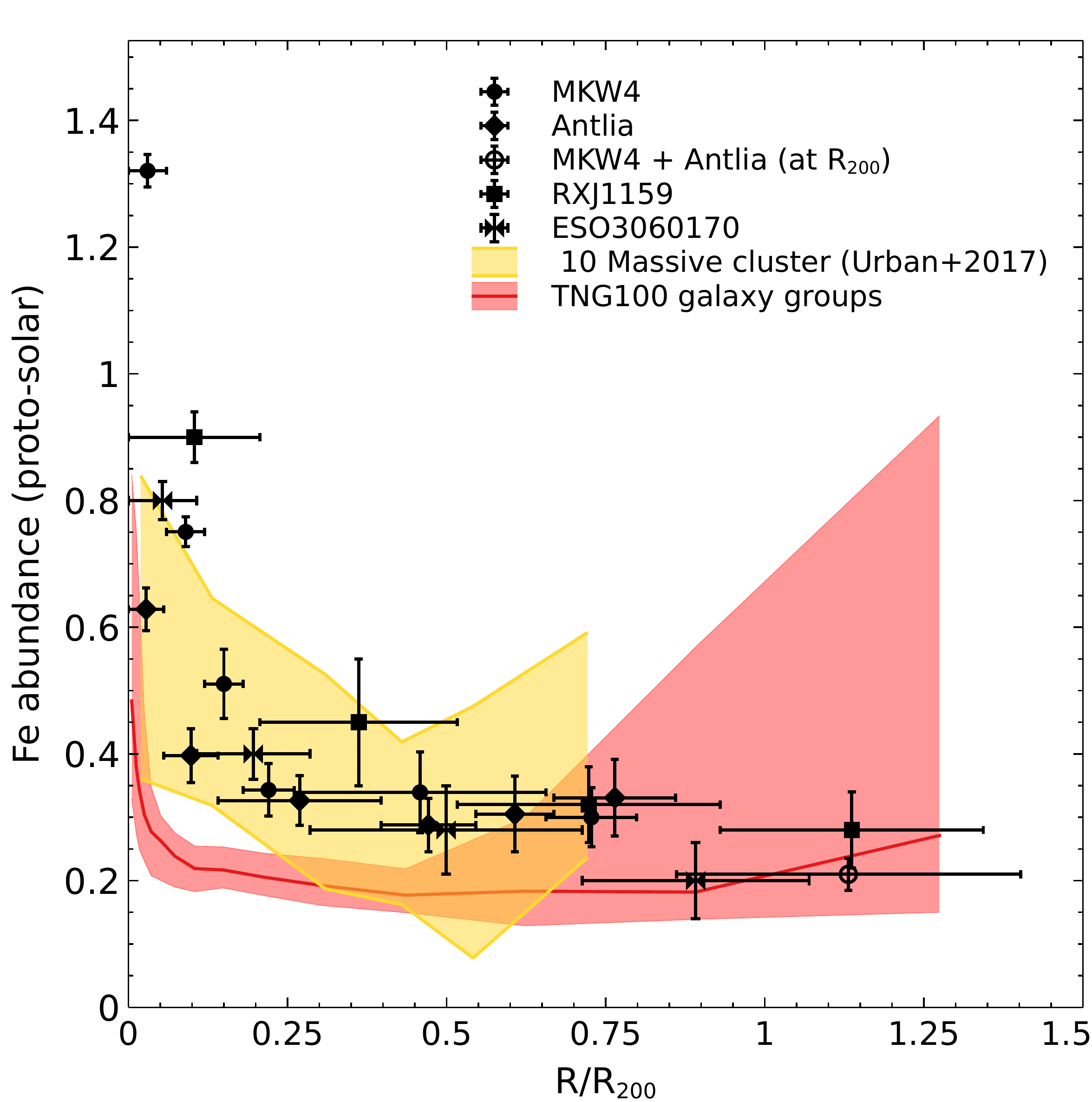} & \hspace{-10pt} \includegraphics[width=0.47\textwidth]{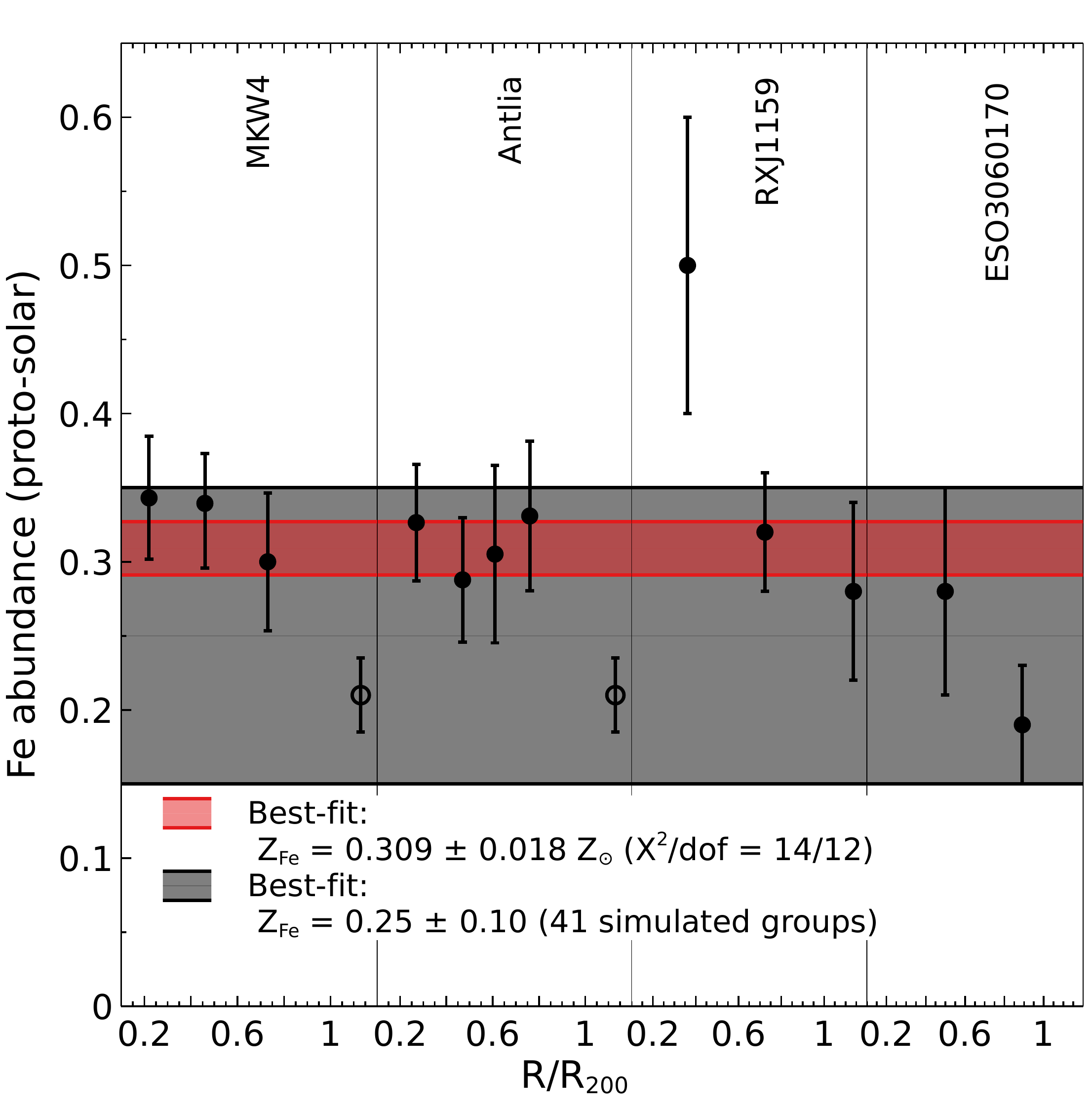} \\
%    &  \hspace{0pt} \\
\end{tabular}
\caption{Left: Fe profiles obtained from the 2T {\tt vapec} model. 
All abundances are in the solar unit. We assume
a solar abundance table of \citet{2009ARA&A..47..481A}. 
Yellow shaded region represents the measurement of 
Fe abundance in 10 massive clusters \citep{2017MNRAS.470.4583U}.
{Red shaded region shows the average Fe abundance
profile with 16--84 percentile error,
obtained from 41 galaxy groups in TNG100.}
Right: 
Iron abundances at R $>$ 0.25R$_{200}$ of individual groups. The red
shaded region indicates average iron abundance for 
0.25-1.0R$_{200}$ with 1$\sigma$ uncertainty.
The black shaded region represents average iron 
abundance at R $>$ 0.25R$_{200}$ for
41 galaxy groups in TNG100. Open circles represent
Fe abundance at R$_{200}$
of Antlia and MKW4, measured jointly.}
\label{fig:abun}
\end{figure*}
%We refer the interested reader to those papers for details.
In short, we extracted spectra from several annular regions,
from the center out to the virial radius of each group,
as marked in green annular sectors in Figures~\ref{fig:image_mkw4}, 
\ref{fig:image_antlia}, \ref{fig:image_rxj},
and \ref{fig:image_eso}.
We generated redistribution matrix files (RMF) 
and instrumental background files (NXB) for all regions and 
detectors. 
For each region, one ancillary response file (ARF) was generated 
using a $\beta$-profile image to model the ICM component; 
another ARF was produced for a uniform emission 
in a circular region of 20$\arcmin$ radius to model 
the sky X-ray background.

Spectral analysis was performed using {\tt\string 
XSPEC--12.10.1} and Cash statistics \citep{1979ApJ...228..939C}. 
We fit the spectra extracted from XIS0, XIS1, and XIS3 simultaneously.
For MKW4, the spectral fitting was 
restricted to the energy range of 0.4 - 7 keV for XIS1 and 
0.6 - 7 keV
for XIS0 and XIS3 \citep{2021MNRAS.501.3767S}, while the 
energy range of 0.5-7 keV was used for the other three groups.
We fit each spectrum with a
multi-component X-ray background model and a thermal emission model. 
For {the}
X-ray background model, we adopted --
{\tt phabs}({\tt pow$_{\rm resolved CXB}$} + {\tt pow$_{\rm unresolved CXB}$} + 
{\tt apec$_{\rm MW}$}) + {\tt apec$_{\rm LHB}$}. 
The {\tt pow$_{\rm resolved CXB}$} component represents the 
point sources resolved by $\chandra$\footnote[4]{
We produce synthetic $\suzaku$ observations for
each point source using {\tt xissim} based on 
their position and flux determined by $\chandra$. 
We extract spectra from these synthetic $\suzaku$ 
observations using the same extraction regions as the
radial profiles. 
Those spectra are then fit to a powerlaw model
with an index of 1.41 to obtain the normalization of 
{\tt pow$_{\rm resolved CXB}$}.}. 
A mosaic image of $\chandra$ observations
of each group in the 0.5--7.0 keV energy band 
is shown in Figures~\ref{fig:image_mkw4}, 
\ref{fig:image_antlia}, \ref{fig:image_rxj},
and \ref{fig:image_eso}, with
all resolved point sources marked with green elliptical regions. 
The second power-law component
({\tt pow$_{\rm unresolved CXB}$}) 
describes the unresolved point sources. 
The two thermal {\tt apec} components 
model the foreground emissions from Milky Way 
({\tt apec$_{\rm MW}$}) 
and Local Hot Bubble ({\tt apec$_{\rm LHB}$}).

{Two thermal components associated with 
a photoelectric absorption component - 
{\tt phabs}$\times$
({\tt vapec} + {\tt vapec}) were fitted to the spectra of each group
to model the ICM emission from each annulus}\footnote{
%Fe-bias is attributed to the fact that 1T plasma emission code either increasing temperature or decreasing Fe abundance or both while fitting
%the code generated excess Fe-L emission with the observation 
The Fe abundance measurement is biased low when fitting multi-phase gas with a single temperature model
\citep[e.g.,][]{1998MNRAS.296..977B,2000ApJ...539..172B,2021Univ....7..208G}. Given {\sl Suzaku}'s modest angular resolution, each radial bin covers a sizable physical region, likely associated with temperature variation even at cluster outskirts.
When fitting a single temperature spectrum generated with Xspec {\it fakeit} to our 2T model, no bias to the metal abundance measurement is found. Therefore, our model should not introduce any bias if even the gas is single phase.}. 
The galactic hydrogen column densities in the directions 
of MKW4, Antlia, RXJ1159, and ESO3060170
were obtained using 
{the} HEASARC N$_{\rm H}$ 
tool 
\footnote[5]{\url{http://heasarc.gsfc.nasa.gov/cgi-bin/Tools/w3nh/w3nh.pl}}.
We link the temperature of the cooler component  
of each region to half that of the hotter
component. The normalizations of both components are allowed to vary freely.
{The best-fit spectral models for the outermost regions
of the four groups are shown in Figure \ref{fig:spectra}.}

{For MKW4 and Antlia, 
we simultaneously fit spectra from regions 
{in} different 
azimuthal directions.} We let the temperatures 
and normalizations vary freely among different regions 
while linking the abundances of regions at the 
same distance from the group center. 
We allow the abundances of
O, Mg, Si, S, Fe, and Ni to vary freely, 
while {the} rest of 
the elements were varied collectively.
The abundances of Ar and Ca were linked to S.
{We are unable to
constrain the S abundance 
for E2 and E3 pointings of Antlia.
We, therefore, fixed 
the Si abundance to be equal to the S abundance.} 
For the outer three regions of MKW4 at R$_{200}$, 
and the outer two 
regions of Antlia at R$_{200}$ (E4 and E5), 
we did not find any 
good constraint on their metal abundances when fitted 
individually. Therefore, we performed a simultaneous fitting 
for all those 5 regions. Their temperatures and normalizations
were allowed to vary freely, but with their metallicities 
linked. The virial temperatures of MKW4 and Antlia are nearly 
the same. Both groups display entropy profiles rising 
to R$_{200}$. 
The distance of MKW4 is twice that of Antlia, 
which makes the outermost regions of MKW4 (R = 706-1200 kpc) cover 
almost the same physical radial range as the outermost two regions in 
Antilia (R = 828-1302 kpc). 
It is, therefore, reasonable to assume 
that they have similar metallicities at R$_{200}$.

For RXJ1159 and ESO3060170, we set their Ni abundance to 
1 Z$_{\odot}$ (the best fit for MKW4 and Antlia), as we 
could not find a good constraint when letting it vary freely.
%% Could you really not find a good spectral fit, or do you mean that you couldn't constraint the Ni abundance?
{The remaining elements were
allowed to vary as described 
for MKW4 and Antlia.}
{We also find it necessary to link the abundance of region 3 to
that of region 4, and the abundance of region 5 to that of region 6, for ESO3060170 
(the temperature and normalization of each region 
{are} free to vary)}.
%For the outer four regions of ESO3060170, we did not
%find good constraints on
%metallicity when fitted separately. We, therefore,
%simultaneously fitted region 4 with region 3 and region 6 with region 5.
%Their temperature and normalizations were allowed to vary
%independently, while their metallicities linked between two tied
%regions,
%assuming region 4 and region 6 have similar
%metal abundances to region 3 and region 5, respectively.
We did not perform any deprojection during spectral fitting 
since it may
introduce systematic noise 
\citep{10.1093/mnras/274.4.1093, 1999AJ....117.2398M}.
The best-fit Fe, O, Mg, Si, S, and Ni profiles of each group are shown in Figures 
\ref{fig:image_mkw4}, \ref{fig:image_antlia}, \ref{fig:image_rxj},
and \ref{fig:image_eso}. 

%%%%%%%%%%%%%%%%%%%%%%%%%%%%%%%%%%%%%%%%%%%%%%%%%%%%%%%%%%%%%%%%%%%%%%%%%%%%%%%

\section{Results} \label{sec:results}

%\TODO{read \citet{2016A&A...592A..37T}, \citet{2012ApJ...748...11H},\citet{2017A&A...603A..80M}}

\subsection{Fe abundance profiles}

{
We obtain abundance profiles of Fe
for four groups out to
 R$_{200}$, as shown in Figure~\ref{fig:abun} and Figures~\ref{fig:image_mkw4}, 
\ref{fig:image_antlia}, \ref{fig:image_rxj},
and \ref{fig:image_eso}.}
Within 0.25\,R$_{200}$, their Fe abundances 
increase towards their centers, with 
MKW4 showing the strongest gradient, reaching 1.3\,$Z_{\odot}$ in the center. 
This may be partially due to the smaller radial bin width that has been 
resolved for MKW4.
The Fe abundance profile of Antlia is much flatter 
at R $>$ 0.05R$_{200}$,
with the smallest central Fe abundance of 0.6\,$Z_{\odot}$ at R $<$ 0.05R$_{200}$.
Unlike the other three groups, Antlia is a non-cool-core system. 
The mechanism that has disrupted 
its core, such as {a}
major merger, 
may have redistributed its central metal content 
at R $>$ 0.05R$_{200}$
\citep[e.g.,][]{2018ApJS..238...14M,2019MNRAS.483..540L}. 

{Even though}
these systems show a variety of Fe abundances at their centers,
the Fe abundance profiles start to converge outside their cores. 
In the outskirts, the Fe abundances 
are mostly consistent within uncertainties. 
We measure an average Fe abundance of 0.309 $\pm$ 0.018 $Z_{\odot}$ ($\chi^2$ = 14 for 12 degrees of freedom) 
over the radii of 0.25--1.0\,R$_{200}$. 
This is strikingly similar to what was found for 
10 nearby massive clusters, with a 
$Z_{\rm Fe}$ of 0.316 $\pm$ 0.012 $Z_{\odot}$ ($\chi^2$ = 28.85 for 25 degrees of freedom) 
over 0.25--1.0\,R$_{200}$ \citep{2017MNRAS.470.4583U},
using the same Solar abundance table of \citet{2009ARA&A..47..481A}. 
%This result implies a similar enrichment process for groups and clusters. 

We compare the measured Fe abundance at R $>$ 0.25R$_{200}$
of our four groups with that of simulated groups,
as shown in Figure \ref{fig:abun}. 
We use IllustrisTNG100, a simulation run in the 
IllustrisTNG project 
\citep[e.g.,][]{2018MNRAS.477.1206N,2018MNRAS.475..676S,2018MNRAS.475..624N,2018MNRAS.480.5113M,2018MNRAS.475..648P}, 
to {predict} the Fe abundance in the outskirts of groups. 
The simulated volume is ~(110.7 Mpc)$^3$, providing 41 simulated groups, 
defined as Friend-of-Friend halos 
within the mass range M$_{200}$ $\sim$ 
5 $\times$ 10$^{13}$-- 4 $\times$ 10$^{14}$ M$_{\odot}$ 
at z=0. 
The abundance is measured within the range of [0.25-1]R$_{200}$, and 
is 
computed as an X-ray emission-weighted quantity, 
in which the X-ray emission of each simulated gas cell is obtained based 
on the gas thermodynamics, e.g., gas density, temperature, 
and metallicity, assuming an {\tt apec} plasma model. 
 %% One note on this is that the emission weighted quantity is often not a good match to the spectroscopic quantity that observers measure.  This came up at the Bertinoro conference.

{
%We compare the Fe radial profiles 
%of the four groups 
%with the average Fe profile 
%predicted by the TNG100 simulations.
%with the emission-weighted Fe profile of a sample of galaxy groups in TNG100,
%in the considered mass range,
%shown in red 
%shown in Figure \ref{fig:abun}.
%agrees well with our observations at R $>$ 0.5R$_{200}$.
At R $<$ 0.5R$_{200}$, simulations
predict lower Fe abundances by a factor of 
$\sim$ 2 compared to the observations. 
Similar discrepancies
have been found by \citet{2018MNRAS.478L.116M}
and \citet{2008A&A...487..461L}
while comparing the Fe abundance profiles 
with the hydrodynamic simulations.
\citet{2014MNRAS.438..195P} argue
that this discrepancy can be explained
by the outdated assumptions on the SNe yields,
the assumed initial mass function,
the fraction of binary systems, and/or
the SNe efficiency of releasing metals
into the ICM \citep{2018MNRAS.478L.116M}.
We test with two different SNIa yield models,
i.e., the
W7 model by \citet{1999ApJS..125..439I}, used in 
TNG simulations, and the yield model
by \citet{2006ApJ...645.1373B}, which best-fitted
with our observations, as discussed in
Section \ref{sec:yields}.
By assuming a SNIa fraction of
23\%, we find that the SNIa yield model of 
\citet{2006ApJ...645.1373B} produces 
$>$ 25\% more Fe than that of
the W7
model, which partially explains 
the discrepancy seen in Figure 
\ref{fig:abun} left.
In addition, our sample is dominated by cool 
core systems 
with concentrated metallicity distributions, while the
simulation sample has a more extensive
diversity of group cores.
Simulations show a large scatter in the Fe profile 
at R $>$ 0.5R$_{200}$,
as seen in Figure \ref{fig:abun} left. 
%While the former one reflects the diversity 
%of the group cores, 
%the latter could be 
This could be 
explained by the presence of satellites 
and high-density clumps at the outskirts.
The average Fe abundance of simulated groups 
in the 0.25R$_{200}$ $<$ R $<$ R$_{200}$ range
agrees with the observations, as shown in
Figure \ref{fig:abun} (right).}

%The profiles for Si and S decline
%from 1.61 Z$_{\odot}$ and 1.23 Z$_{\odot}$
%at group center to 0.49 Z$_{\odot}$ and 0.51 Z$_{\odot}$
%at $\sim$ 0.35R$_{500}$ and 0.2R$_{500}$, respectively.
%The Mg rather shows a flatter abundance profile 
%with a value of 0.63 Z$_{\odot}$ out to zR$_{500}$
%and then decreases to 0.25 Z$_{\odot}$ at R$_{500}$.
%The abundance profile for Ni does not show any significant
%changes from group center out to the R$_{500}$.
%\TODO{argue with the measurement of abundance profile
%of MKW4 \citep{2007MNRAS.380.1554R,2008PASJ...60S.333S}.}

\begin{figure*}
\begin{tabular}{ccc}
%\vspace{-40pt}\\
\includegraphics[width=0.42\textwidth]{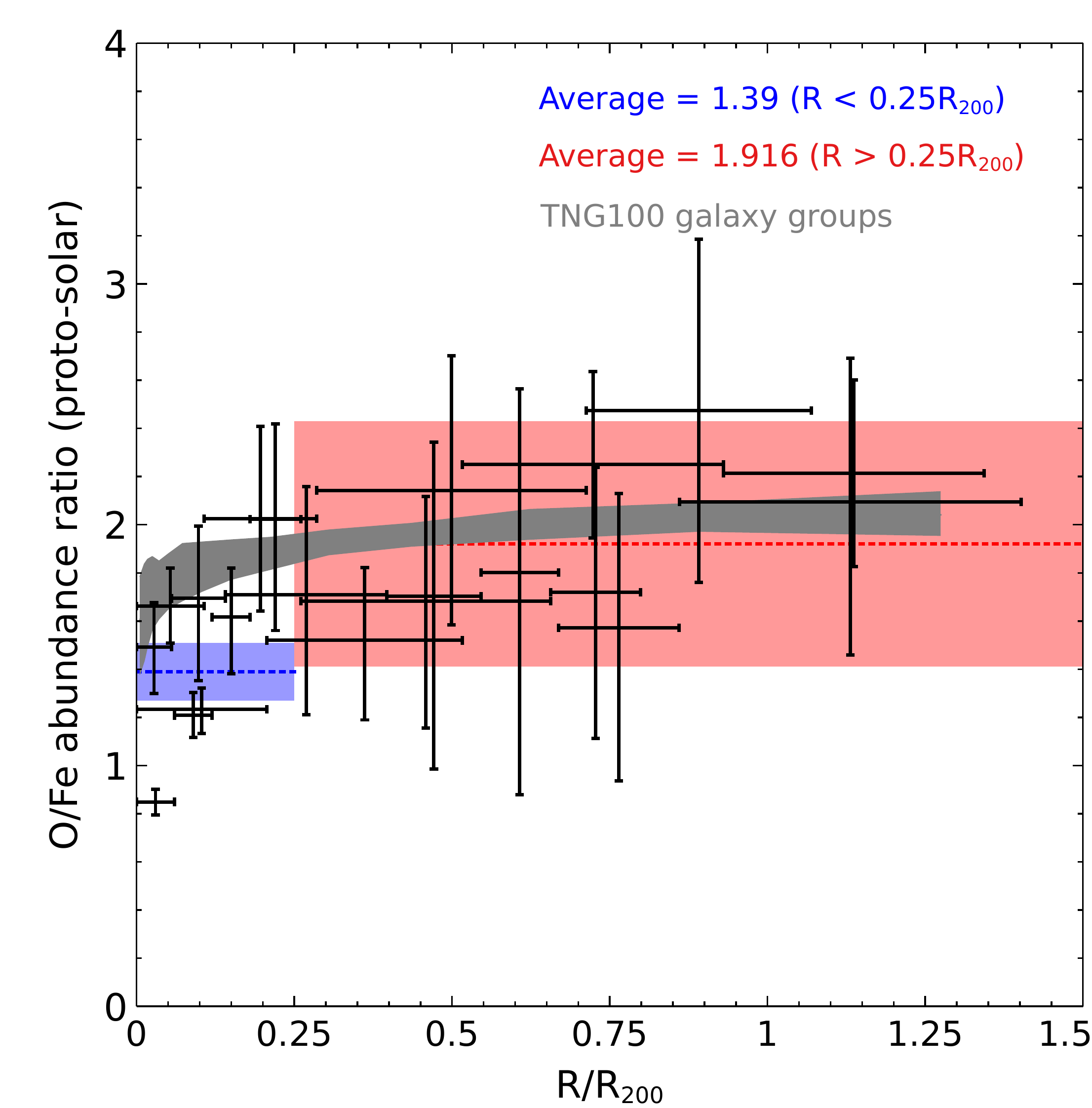}    &  \hspace{-10pt} \includegraphics[width=0.42\textwidth]{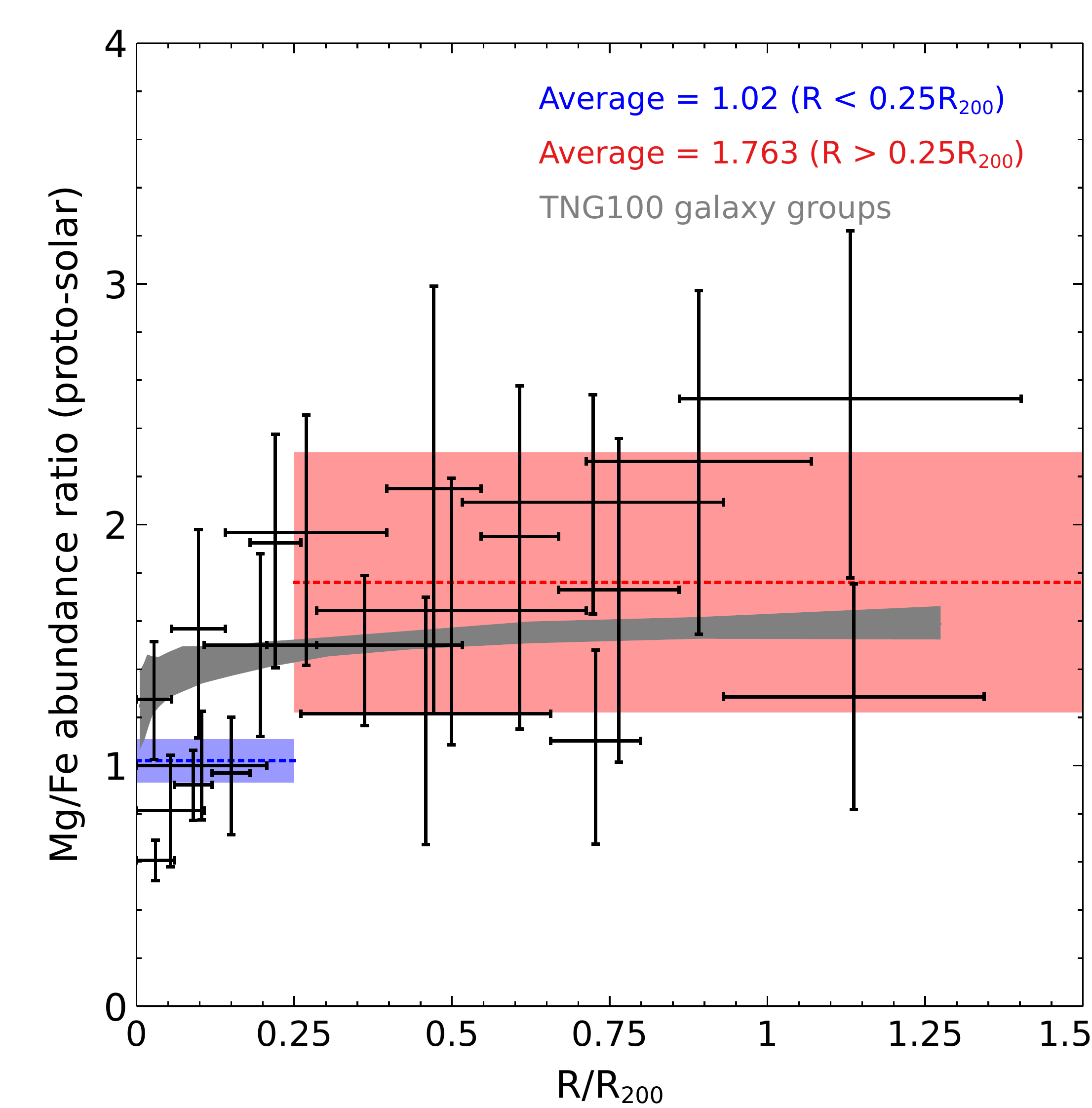} \\ \includegraphics[width=0.42\textwidth]{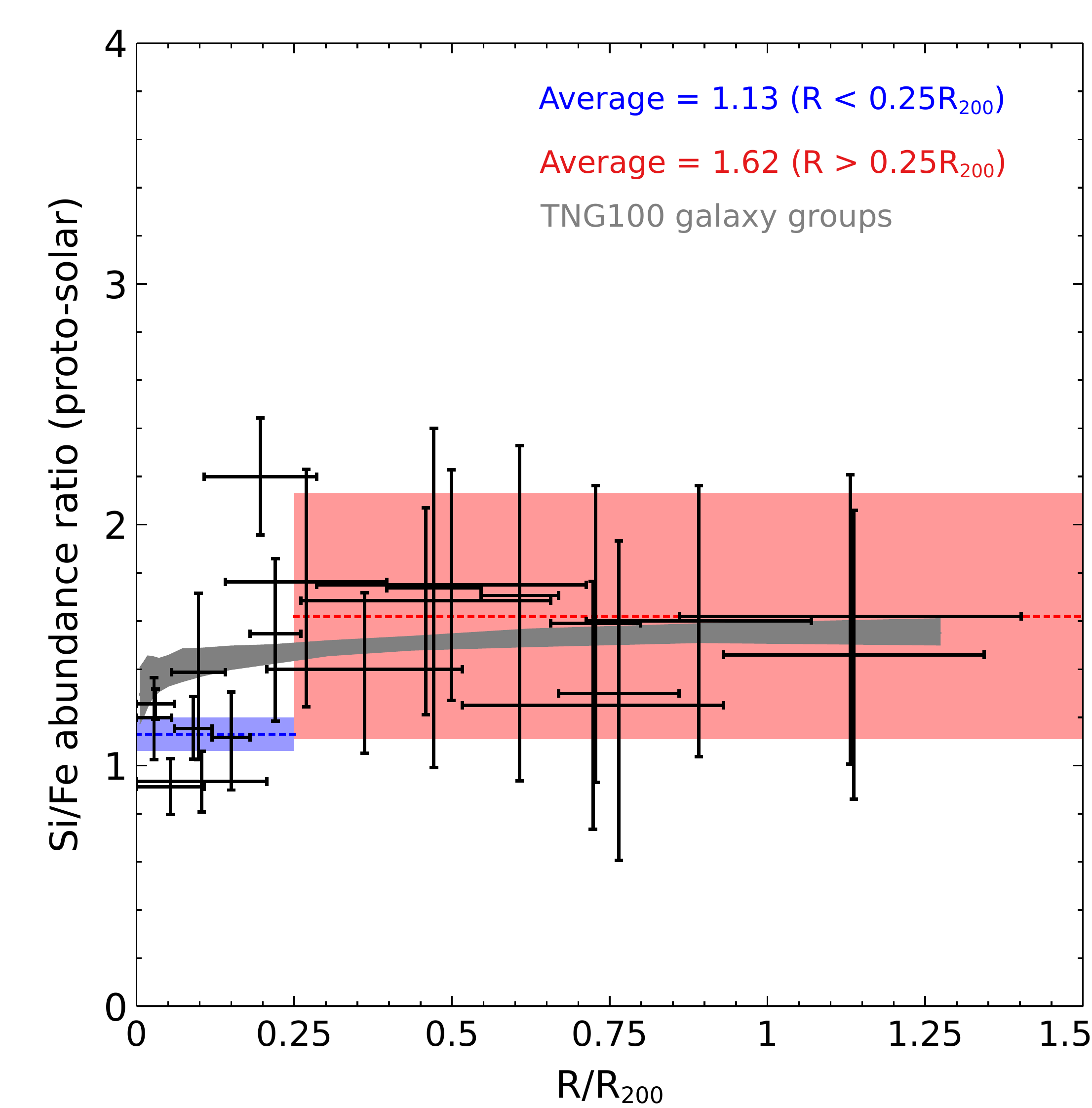}   &  \hspace{-10pt} \includegraphics[width=0.42\textwidth]{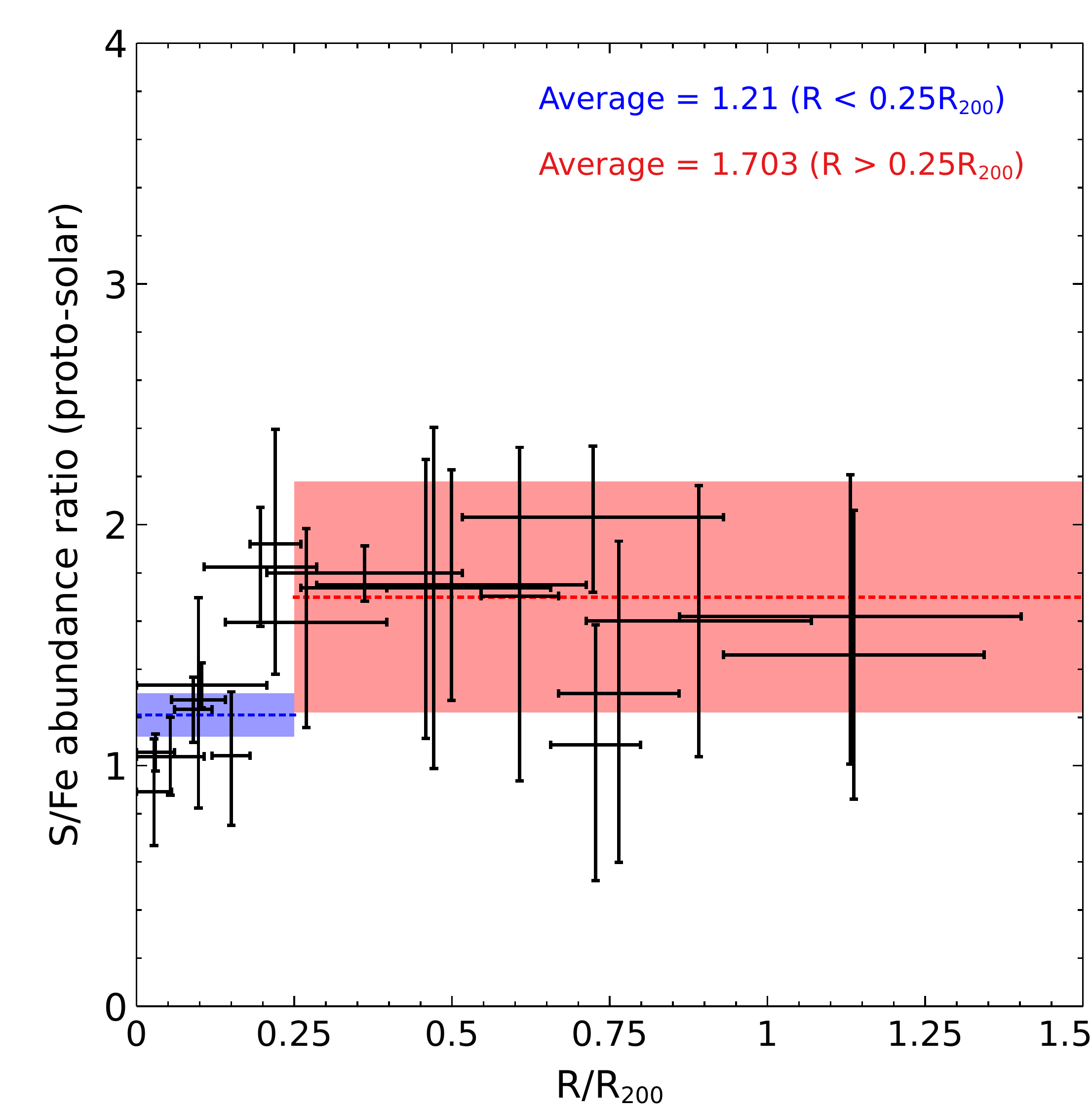}\\
\end{tabular}
\begin{tabular}{cc}
%\includegraphics[width=0.32\textwidth]{ni_fe_ratio_total.pdf} \\
%    &  \hspace{0pt} \\
\end{tabular}
\caption{X/Fe ratios as a function of radius for individual metals.
{
The red shaded region represents the average X/Fe 
excluding central R $<$ 0.25R$_{200}$ region. 
The blue shaded region 
represents the average X/Fe 
for the central R $<$ 0.25R$_{200}$ region.
%The vertical dotted line denotes 0.25R$_{200}$.
Black shaded region shows the 
average X/Fe profile 
of 41 galaxy groups in TNG100.}
}
\label{fig:x/w}
\end{figure*}

\subsection{Chemical composition}
In addition to Fe, we have further constrained the 
abundance profiles of O, Mg, S, Si, and Ni,
as shown in Figures \ref{fig:image_mkw4},
\ref{fig:image_antlia}, \ref{fig:image_rxj}, and \ref{fig:image_eso}.
%Unlike Fe, which shows a sudden change in the profile slopes between R $<$ 0.25R$_{200}$ and 0.25R$_{200}$ $<$ R $<$ R$_{200}$, the gradient change of other elements is relatively mild. 
We derive their abundance ratios relative to
Fe as a function of radius from the centers out to
the virial radii, as shown in Figure 
\ref{fig:x/w}. 
The uncertainties are calculated 
using Monte Carlo simulations with 2000 - 3000
realizations. 
We observe that the X/Fe ratios are consistent with the solar 
abundance within R $<$ 0.25R$_{200}$, 
with O/Fe = 1.39 $\pm$ 0.12 ($\chi^2$/degrees of freedom = 16.7/6),
Mg/Fe = 1.02 $\pm$ 0.09 ($\chi^2$/degrees of freedom = 17.6/5), 
Si/Fe = 1.13 $\pm$ 0.07 ($\chi^2$/degrees of freedom = 10.5/6), 
S/Fe = 1.21 $\pm$ 0.09 ($\chi^2$/degrees of freedom = 11.6/7), 
and Ni/Fe = 1.82 $\pm$ 0.29 ($\chi^2$/degrees of freedom = 16.5/4), 
which is similar to what have 
been measured for group centers such as the CHEERS project 
\citep{2018MNRAS.478L.116M}. 
The X/Fe ratios over 0.25R$_{200}$ $<$ R $<$ R$_{200}$ 
are somewhat super solar with O/Fe = 1.92 $\pm$ 0.51 
($\chi^2$/degrees of freedom = 5.2/12),
Mg/Fe = 1.76 $\pm$ 0.54 ($\chi^2$/degrees of freedom = 10/12), 
Si/Fe = 1.62 $\pm$ 0.51 ($\chi^2$/degrees of freedom = 8/12), 
S/Fe = 1.70 $\pm$ 0.48 ($\chi^2$/degrees of freedom = 6/12), 
and Ni/Fe = 3.60 $\pm$ 1.50 ($\chi^2$/degrees of freedom = 3.4/6). 

{
We compare the O/Fe, Mg/Fe, and Si/Fe ratios of
our four groups with that of simulated groups, as shown
in Figure \ref{fig:x/w}. 
Since TNG simulations do not
include S abundance, 
we could not compare the measured S/Fe ratio
with the simulation.
Our measured abundance ratios
for O/Fe, Mg/Fe, and Si/Fe
are consistent with the simulated groups
within their 1$\sigma$ error bars. 
}
%The difference in the chemical composition 
%between group centers and group outskirts may reflect their 
%different metal origins and enrichment processes. 

\begin{figure*}
\begin{tabular}{cc}
\includegraphics[width=0.45\textwidth]{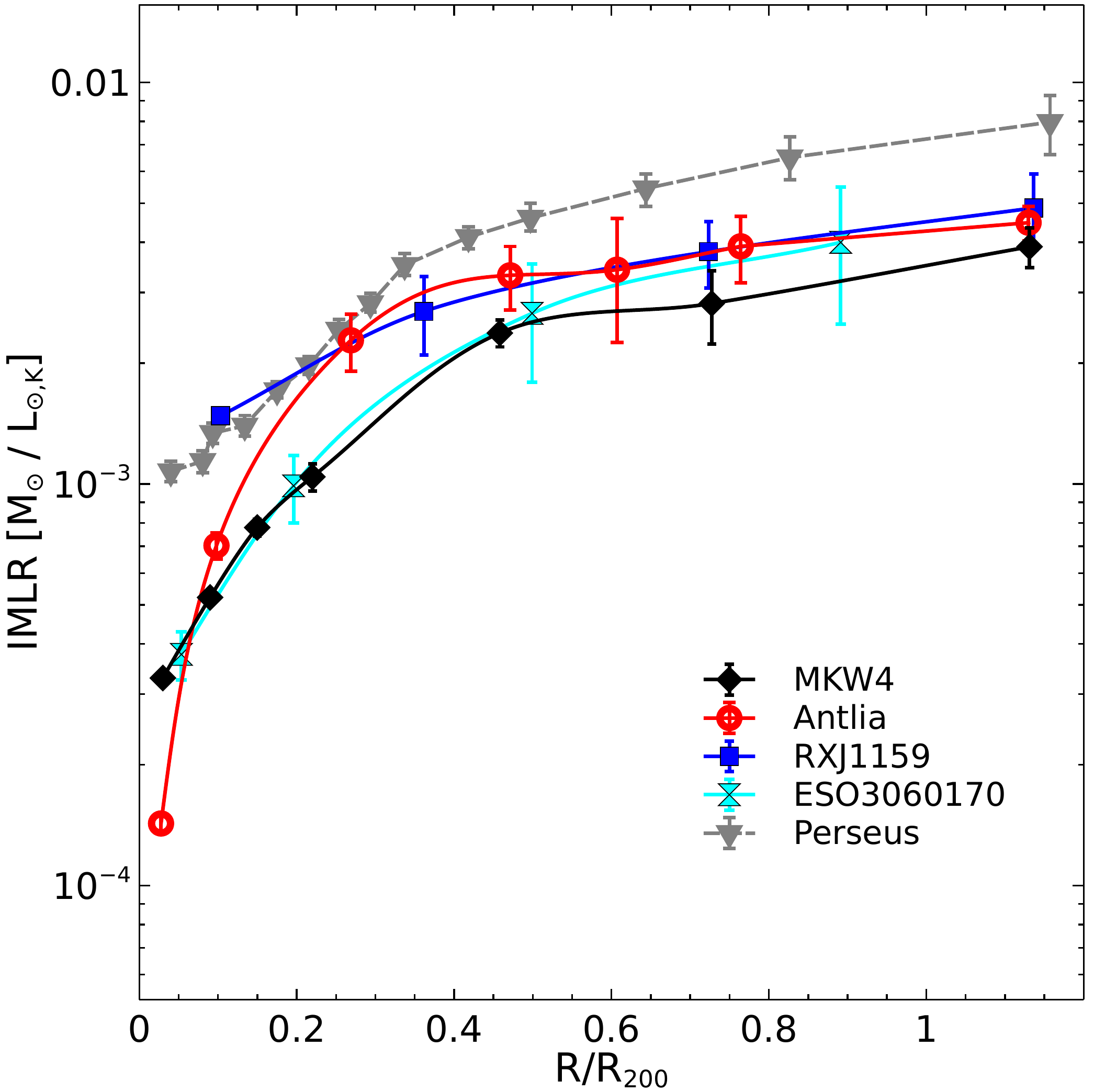}     &   \includegraphics[width=0.45\textwidth]{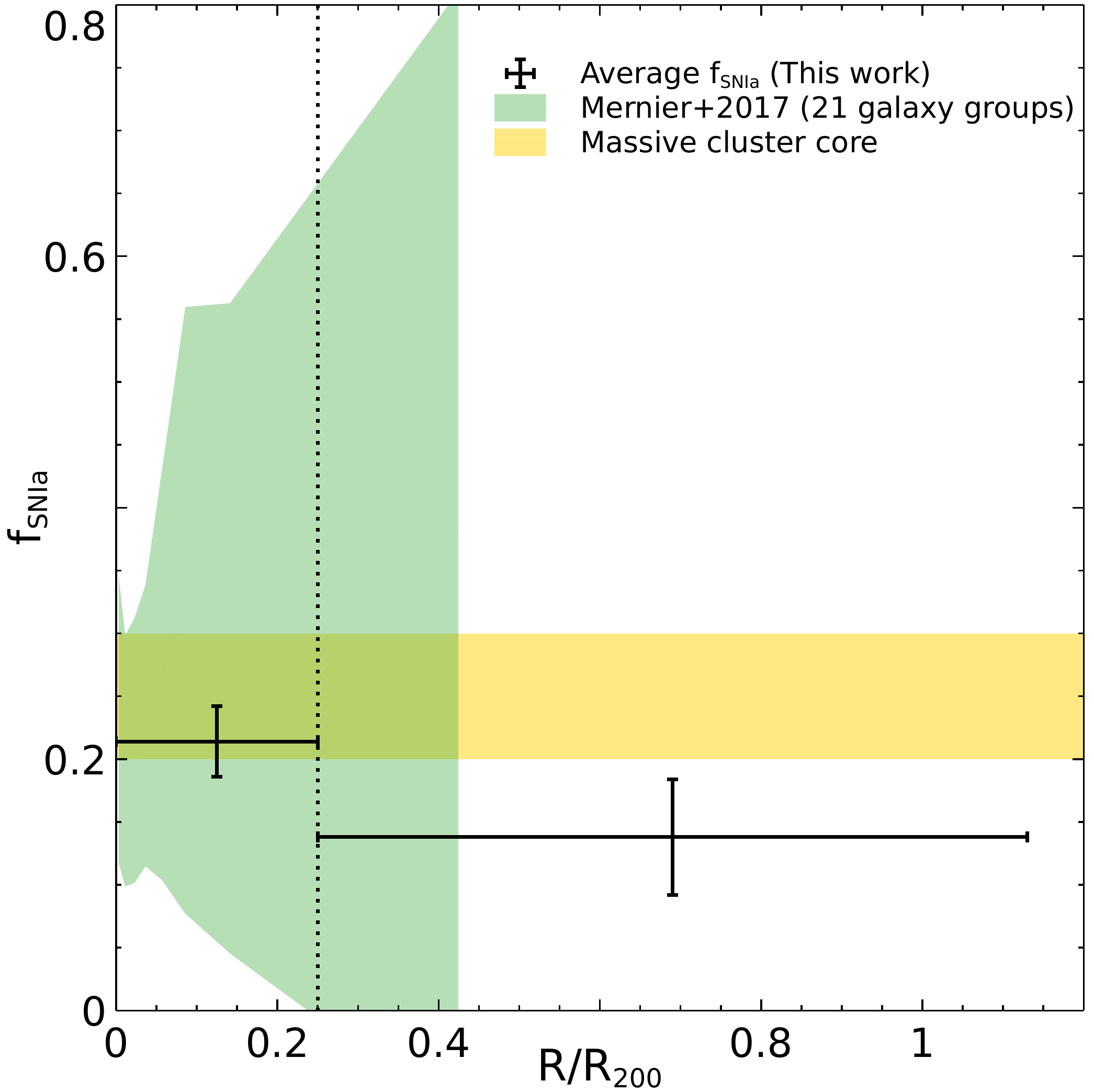}\\
\end{tabular}
\caption{{\it Left:} Radial profiles of IMLR with K$_{\rm s}$-band luminosity for groups in our sample and the Perseus cluster. 
We obtain IMLR of 
%Hydra A cluster from \citet{2012PASJ...64...95S},
%Abell~262 from \citet{2009PASJ...61S.365S}, 
%AWM 7 from \citet{2008PASJ...60S.333S}, Coma from
%\citet{2013ApJ...764..147M}, 
Perseus from \citet{2013PASJ...65...10M}.
%NGC 1550 and
%HCG 62 from \citet{2014ApJ...781...36S}. 
%The conversion R$_{180}$ $\simeq$ 1.7R$_{500}$
%is adopted from \citep{2017A&A...603A..80M}.
{\it Right:} 
The averaged f$_\textrm{SNIa}$ 
for the four groups at R $<$ 0.25R$_{200}$ and 0.25R$_{200}$ $<$ R $<$ R$_{200}$. 
Green shaded region
represents the measurement scatter obtained from a large number of groups 
\citep{2017A&A...603A..80M}. Yellow shaded region indicates the limit of
f$_\textrm{SNIa}$ at the massive cluster core \citep{2007A&A...465..345D}.
Vertical dotted line indicates 0.25R$_{200}$. 
}
\label{fig:imlr_vs_temp}
\end{figure*}

\begin{figure*}
\begin{tabular}{cc}
\includegraphics[width=0.45\textwidth]{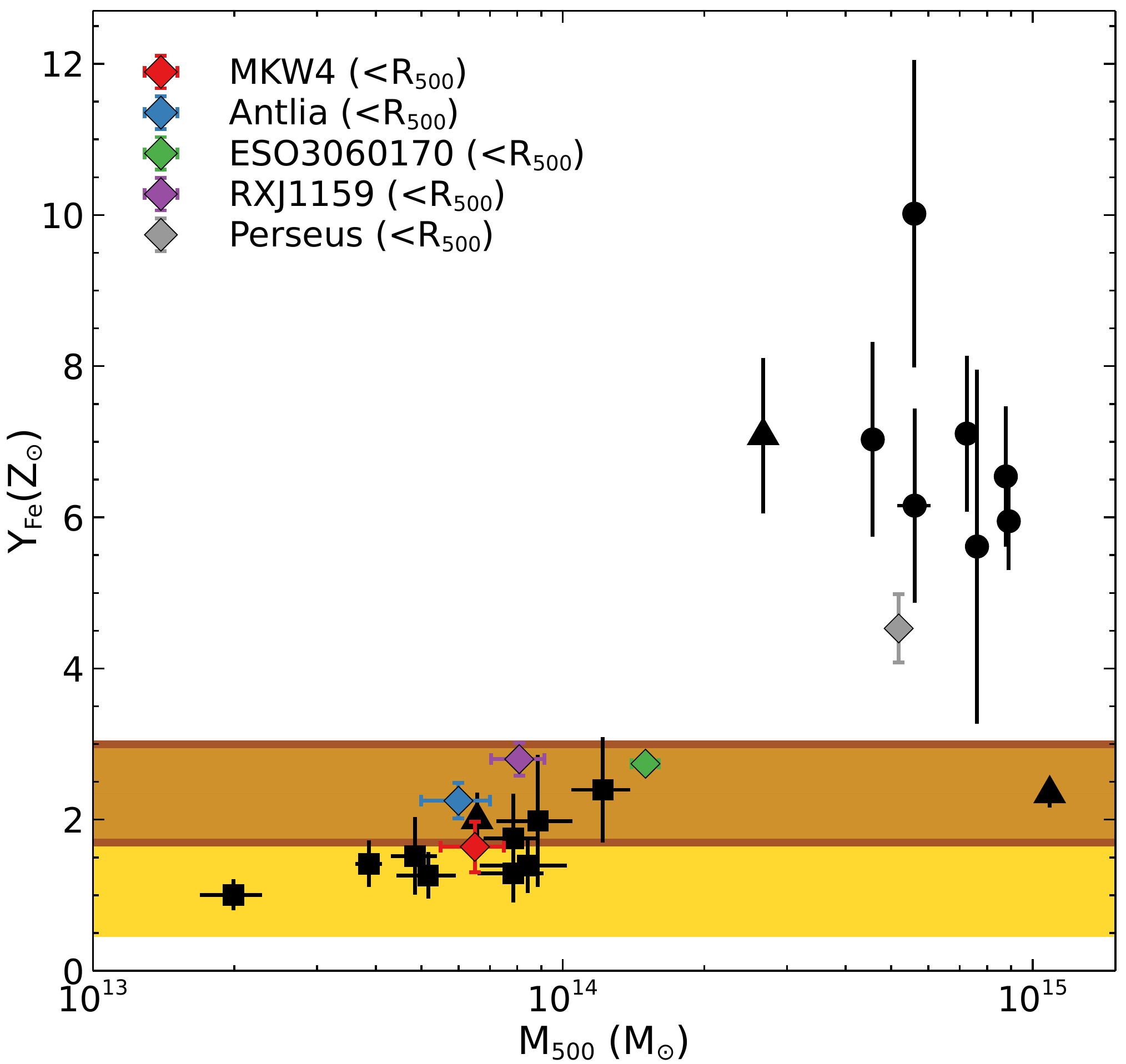} & \includegraphics[width=0.45\textwidth]{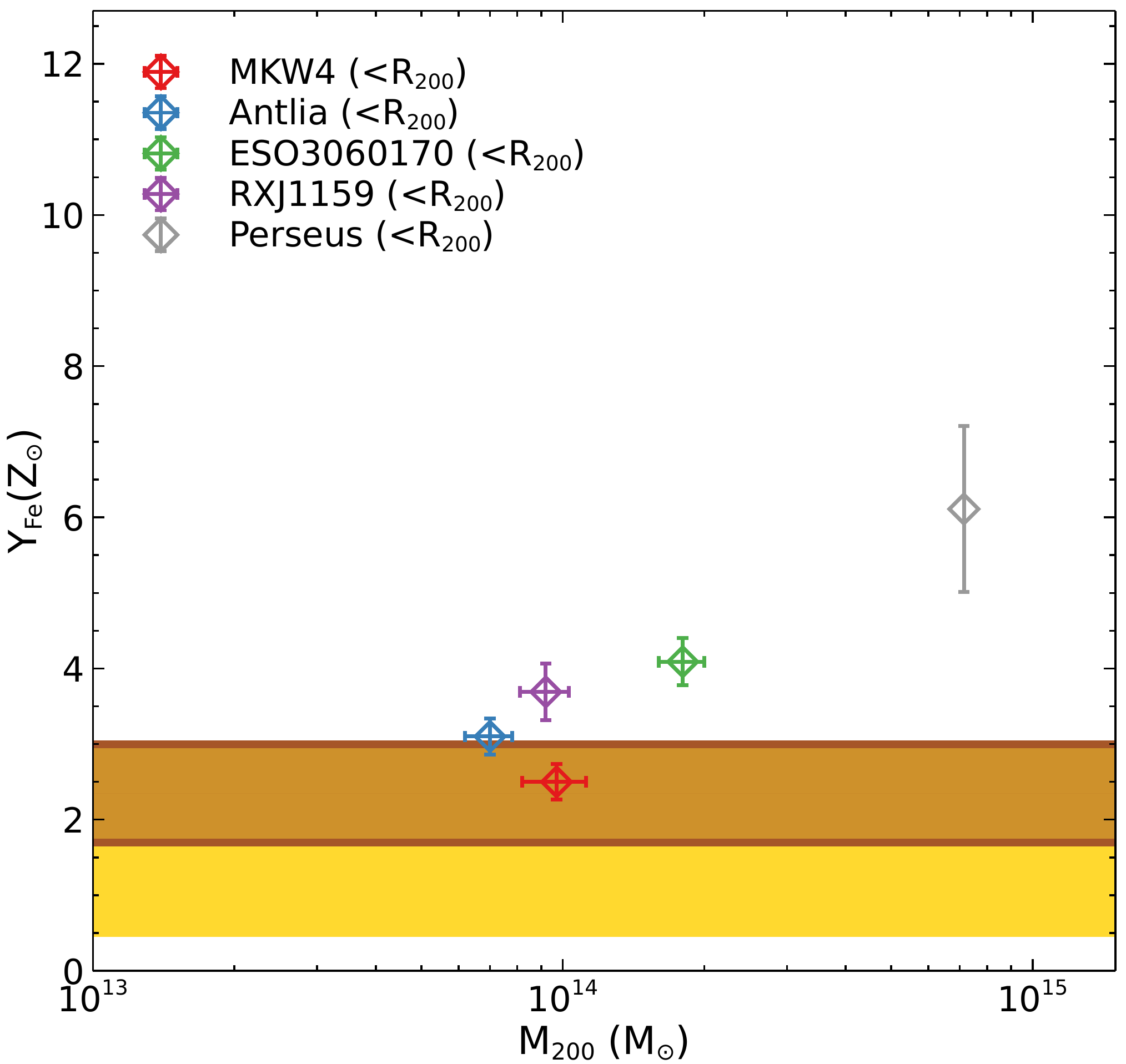}
\end{tabular}
\caption{{\it Left:}
Effective Fe yield as a function
of cluster/group mass within R$_{500}$. 
%Filled
%data points are Fe yield at R$_{500}$ and 
%open data points are Fe yield at R$_{200}$. 
Circles represent Fe yield for the cluster
in the sample of \citet{2021A&A...646A..92G},
squares represent the groups in the
sample of \citet{2014MNRAS.444.3581R},
triangles represent NGC 1550, 
Hydra A, and Coma 
in the sample of \citet{2014ApJ...781...36S}.
Yellow shaded region shows the expected value
estimated from the SN yields derived from 
\citet{2017ApJ...848...25M} and \citet{2014MNRAS.444.3581R}.
The brown shaded region shows the expected value
calculated assuming higher SNIa rate from 
\citet{2021MNRAS.502.5882F}.
{\it Right:} Effective Fe yield as a function
of cluster/group mass within R$_{200}$.
}
\label{fig:fe_yield}
\end{figure*}

\subsection{Iron mass-to-light ratios}
%\TODO{rewrite the method of calculating metal mass.}
We investigate the IMLR from
the centers out to the virial radii of these groups to compare
the Fe distribution in the IGrM with their stellar mass profiles. 
We use the metal abundance profiles obtained in 
this work, and the deprojected density profiles of MKW4, Antlia, RXJ1159,
and ESO3060170 derived in \citet{2021MNRAS.501.3767S}, 
\citet{2016ApJ...829...49W}, \citet{2015ApJ...805..104S},
and \citet{2013ApJ...775...89S},
respectively,
to calculate the accumulated iron mass 
within a specific radius for each group.
%\citep[e.g.,][]{2004A&A...419....7D}.
%\begin{figure}
%\includegraphics[width=0.45\textwidth]{metal_mass.pdf}
%\caption{Metal mass as a function of R/R$_{500}$.}
%\label{fig:metal_mass}
%\end{figure}
%\begin{figure*}
%\begin{tabular}{ccc}
%\vspace{-40pt}\\
%\includegraphics[width=0.36\textwidth]{MKW4_east_rad_average_model1.pdf}    &  \hspace{-22pt} 
%\includegraphics[width=0.36\textwidth]{MKW4_east_rad_average_model2.pdf} & \hspace{-22pt} 
%\includegraphics[width=0.36\textwidth]{MKW4_east_rad_average_model5.pdf}\\
%\end{tabular}
%\caption{Best fit SN models to the X / Fe abundances.}
%\label{fig:average_sn_fit}
%\end{figure*}
%Our estimated metal mass for Fe, Mg, S, and Si are
%consistent with each other within R$_{500}$, where 
%the O mass is $\sim$ 1 dex higher relative to those
%of other elements. Similar large O mass is also observed
%in other galaxy clusters and groups
%\citep{2007PASJ...59S.327M,2007A&A...462..953M,2008PASJ...60S.333S,2009PASJ...61S.365S}.
We estimate the K$_{\rm s}$-band
luminosity of each group using 
{the} Two Micron All-Sky Survey 
(2MASS\footnote[3]{\url{https://irsa.ipac.caltech.edu/applications/2MASS/IM/interactive.html}}).
We obtain a 2$^\circ$ $\times$ 2$^\circ$ mosaic image centered at 
each group from the 2MASS data catalog.
We adopt the Galactic extinction (A$_{\rm K}$) values for
both groups from NASA/IPAC Extragalactic 
Database\footnote[2]{\url{http://ned.ipac.caltech.edu}}.
We deproject the K$_{\rm s}$-band
luminosity with radius, and the resultant luminosity
profiles are used to calculate the cumulative IMLR.
%We obtain a stellar mass of $\sim$ 1.6 $\times$ 10$^{12}$ M$_{\odot}$
%within R$_{200}$ of MKW4 and $\sim$ 1.42 $\times$ 10$^{12}$ M$_{\odot}$
%within R$_{200}$ of Antlia. Similar stellar mass $<$ R$_{200}$ of Antlia
%is also obtained by \citet{2016ApJ...829...49W}.
Figure~\ref{fig:imlr_vs_temp} shows the IMLR
of MKW4, Antlia, RXJ1159, and ESO3060170 from their
centers to their virial radii. 
Radial profiles of the 
IMLR increase with 
{the} radius for all groups,
suggesting that their metal distribution is more
extended than the galaxy light out to R$_{200}$.
Their IMLRs within R$_{500}$ are within the scatter 
of what has been found for other groups 
\citep[e.g.,][]{2009PASJ...61S.365S}, 
although that of RXJ1159 at 0.1\,R$_{200}$ is exceptionally high. 
%\citep[e.g.,][]{2012PASJ...64...95S,2013ApJ...775...89S}. 
%We obtain an IMLR of $\sim$ 
%2.5 $\times$ 10$^{-3}$ M$_{\odot}$/L$_{\rm K \odot}$
%within R$_{500}$ of MKW4, which 
%is similar to those of other cool-core
%groups
%\citep[e.g.,][]{2009PASJ...61S.365S,2013ApJ...775...89S}.
%We find the IMLR of Antlia to be $\sim$ 
%3.5 $\times$ 10$^{-3}$ M$_{\odot}$/L$_{\rm K \odot}$ 
%within R$_{500}$, similar to the non-cool core cluster, Coma 
%\citep[e.g.,][]{2013ApJ...764..147M}. 
The IMLR profiles of these groups converge to (4-5)$\times$ 
10$^{-3}$ M$_{\odot}$/L$_{\rm K \odot}$ at R$_{200}$, 
which is half that of the Perseus cluster. 
%\begin{figure}
%\includegraphics[scale=0.35]{IMLR.pdf}
%\vspace{-10pt}
%\caption{Iron mass-to-light ratios of MKW4 (red)
%and Antlia (blue) as function a
%%of radius with respect to the K-band luminosity.}
%\label{fig:metal_lk}
%\end{figure}

We estimate the Fe yield (Y$_{\rm Fe}$), which is
the ratio of total Fe mass released by stars to the
total stellar mass formed for a given stellar population,
as follows \citep{2014MNRAS.444.3581R,2021A&A...646A..92G,2021Univ....7..208G}

\begin{equation}
    \ \ \ \ \ \ \ \ \ \ \ \rm Y_{\rm Fe,500} = \frac{M^{\rm star}_{\rm Fe,500} + M_{\rm Fe,500}}{M_{\rm star,500}(0)}\\
%    \rm Y_{\rm Fe,200} = \frac{M^{\rm star}_{\rm Fe,200} + M_{\rm Fe,200}}{M_{\rm star,200}(0)},
\end{equation}
where M$_{\rm Fe,500}$ is the Fe mass within R$_{500}$,
M$^{\rm star}_{\rm Fe,500}$ is the Fe mass
locked into the stars within R$_{500}$, M$_{\rm star,500}(0)$
is the mass of gas within R$_{500}$ 
that went into stars and whose present mass is reduced 
by a factor r$_{0}^{-1}$, i.e., 
M$_{\rm star,500}(0)$ = r$_{0}$M$_{\rm star,500}$ \citep{2021Univ....7..208G},
where r$_{0}$ is the return factor. We adopt r$_{0}$ = 1/0.58 following 
\citet{2014MNRAS.444.3581R} and \citet{2021Univ....7..208G}.
Figure \ref{fig:fe_yield} shows the effective Fe yield
of four galaxy groups within R$_{500}$ and R$_{200}$, respectively.
We compare {the} 
Fe yield of four groups within R$_{500}$ with the
sample of clusters studied in \citet{2021A&A...646A..92G}, 
the sample of groups studied in \citet{2014MNRAS.444.3581R},
and the systems studied in \citet{2014ApJ...781...36S}.
We adopt the expected values 
{estimated by}
\citet{2021Univ....7..208G}.
The Fe yields within R$_{200}$
for these four groups are significantly 
higher than that within R$_{500}$,
which
{suggests} that at the group scale
feedback redistributes metals, and 
may push (or prevent from collapse) a consistent fraction of 
them out to R$_{200}$ and beyond.

\subsection{Supernovae yields}{\label{sec:yields}}

The yields of SNe Ia and SNcc are
{significant} contributors
to the metal enrichment of the ICM. We fit our
measured metallicity distributions to 
the metallicity patterns 
predicted
by several theoretical SNe Ia and SNcc nucleosynthesis models, following \citet{2016A&A...595A.126M}.
% We follow 
%\citet{2016A&A...595A.126M} to reproduce 
%the abundance
%patterns from 
%several theoretical SNe Ia and SNcc %yields. 
We fit the average metal abundance ratios of 
O/Fe, Mg/Fe, Si/Fe, and S/Fe for these groups
to a combination of 
a single SNe Ia and a single SNcc model, and allow the relative 
number of SNe Ia over the total number of SNe (denoted as 
f$_{\rm SNIa}$) free to vary
%responsible for the enrichment
\citep{2006A&A...449..475W}.
\citet{2016A&A...595A.126M} 
and \citet{2019MNRAS.483.1701S} pointed out that existing SNe yield models
cannot successfully reproduce the observed Ni/Fe ratios, which in turn
introduces {significant}
errors in fitting parameters. We, therefore, 
exclude the Ni/Fe ratio from the fitting process. 
{The $\alpha$-capture elements--O and Mg are 
primarily produced by the SNcc, while
both SNcc and SNe Ia produces Si and S. Thus, their relative
abundances compared to Fe, a major SNe Ia product,
are sensitive to the fitting parameters.}
%\begin{table}
%\caption{Results of SN fits to the average element abundance showed in Table %\ref{tab:abun}}
%\resizebox{1\columnwidth}{!}{%
%\begin{tabular}{ccccccc}
%\hline
%SNe Ia model$^{\ddagger}$ \  & SNcc model \  & a \  & b \  & $\frac{\textrm{SNe Ia}}{\textrm{SNe Ia + SNcc}}$ \  & $\frac{\textrm{SNe Ia}}{\textrm{SNcc}}$\  & $\chi^2$/d.o.f\\
%\decimalcolnumbers
%\hline
%Iw99\_W7 & No06\_0.004 & 5.02 $\times\ 10^7$ & 2.09 $\times\ 10^8$ & 0.19 $\pm$ 0.20 & 0.24 & 0.09 / 4 \\\\ 
%\hline
%Ma10\_C-DEF & No13\_SNcc\_0.02 & 8.53 $\times\ 10^7$ & 2.30 $\times\ 10^8$ & 0.30 $\pm$ 0.20 & 0.40 & 0.10 / 4 \\\\
%\hline
%Ma10\_C-DEF & No13\_SNcc\_0.008 & 9.5 $\times\ 10^7$ & 2.09 $\times\ 10^8$ & 0.31 $\pm$ 0.18 & 0.45 & 0.11 / 4 \\\\
%\hline
%Iw99_W7 & No06_0.004 & 5.02 \times $10^7$ & 2.09 \times $10^8$ & 0.19 \pm\ 0.20 & 0.24 & 0.1 / 4 \\\\
%& Le18\_300-3-c3 & Ch04\_0 & 0.03 & 0.03 & 15.0 / 4\\\\
%\hline
%Iw99_W7 & No06_0.004 & 5.02 \times $10^7$ & 2.09 \times $10^8$ & 0.19 \pm\ 0.20 & 0.24 & 0.1 / 4 \\\\
%\hline
%\end{tabular}
%}
%\\
%\flushleft
%$^{\ddagger}$ Three best-fit SN models with increasing $\chi^2$ / d.o.f value.
%\label{tab:average_sn}
%\end{table}
Our measured average abundance patterns for MKW4 and
Antlia fit best with the
delayed detonation (DDT) SNe Ia model of 
\citet{2006ApJ...645.1373B}, and with the ``classical" mass-dependent
SNcc model of \citet{2006NuPhA.777..424N} with a progenitor metallicity of Z$_{\rm init}$ = 0.02 Z$_{\odot}$. 
For RXJ1159 and ESO3060170, we obtain a best-fit
by using {the} 
SNe Ia model of \citet{2006ApJ...645.1373B} and
{the} SNcc model of \citet{2013ARA&A..51..457N} and 
\citet{2016ApJ...821...38S}, respectively.
{Throughout this paper, 
we assume the SNcc yields are produced by the population
of massive stars having a Salpeter IMF, 
\citep{1955ApJ...121..161S} with a common Z$_{\rm init}$ 
\citep[e.g.,][]{2016A&A...595A.126M}.}
We next fit
the metal abundance patterns at each radial bin of four
groups to obtain their f$_{\rm SNIa}$
from the centers out to the virial radii. 
%to those 
%best-fit SNe Ia and SNcc models. Thus,
%we constrain the relative 
%number of SNe Ia over the total number of SNe (denoted as 
%f$_{\rm SNIa}$)
%responsible 
%for the metal enrichment in the ICM of these groups.
Figure~\ref{fig:imlr_vs_temp} shows the f$_{\rm SNIa}$ averaged 
over two radial bins, from 0 - 0.25R$_{200}$ and  0.25 - 1.0R$_{200}$.
The derived f$_{\rm SNIa}$ decreases from 0.214 $\pm$ 0.028 at 
their centers to 0.138 $\pm$ 0.046 at the outskirts.
%which suggests a uniform 
%contribution of SNe Ia (SNcc) products in the ICM within the uncertainties.
Those values are broadly consistent with those of the core of massive
galaxy clusters of 20\%--30\%
\citep[][]{2007A&A...465..345D,2009A&A...493..409S} 
and our Milky Way of 15\%
\citep{1995MNRAS.277..945T}. 
{We note that the SNe yields 
constitute
a crucial source of bias in the above analysis. 
The SNe yield models suffer from
uncertainties up to a 
factor of 2 
\citep[e.g.,][]{2009MNRAS.399..574W},
and as a consequence, the
derived estimation of f$_{\rm SNIa}$ should be
interpreted with caution 
\citep[e.g.,][]{2009A&A...508..565D,2018MNRAS.478L.116M}.}

\section{Systematic uncertainties}
We test our results against possible systematic
uncertainties introduced during spectral analysis, 
following \citet{2021MNRAS.501.3767S}.
One of the most {critical}
and uncertain components of the
X-ray background 
is the CXB. To examine the systematic uncertainty associated with the CXB, we
vary the
normalization of the unresolved CXB component by 10\%, with a fixed power-law slope of 
$\Gamma$ = 1.41. We obtain that the associated impacts on all the metal abundances
are within the statistical error limit. 
Next, we estimate the systematic uncertainties introduced by the two
foreground components-- MW and LHB. We vary their best-fit
normalizations by 10\%, which does not 
significantly change the measured
metal abundances. Finally, we experiment with a 20\% variation in 
Galactic column density (N$_{\rm H}$), which also has
no significant impact on the measured metal abundances.

\section{Discussion}
Using joint $\suzaku$ and $\chandra$ observations, we derived the
abundance profiles of Fe, O, Mg, S, Si, and Ni for four nearby galaxy groups,
from their centers out to their virial radii. 
To our knowledge, this is the first time that such 
measurements have been made for such low-mass systems. 
In the previous section, we show that our results are stable 
against various sources of systematic uncertainties. 
Below we compare the results with what we know from massive
galaxy clusters, and discuss the implications of our findings.

\subsection{Early enrichment scenario}
Previous studies have reported 
a universal Fe abundance of $\sim$ 0.3 Z$_{\odot}$
at {the} 
outskirts of nearby massive clusters using $\suzaku$ observations 
\citep[e.g.,][]{2013Natur.502..656W,2017MNRAS.470.4583U}.
It has also been shown that no significant redshift evolution of 
the global metallicity (R $<$ R$_{500}$) is found for high-redshift 
samples 
\citep[e.g.,][]{2016ApJ...826..124M,2017MNRAS.472.2877M,2020A&A...637A..58L}.
{These findings point to an 
early-enrichment scenario,
where the ICM is 
enriched during the star formation peak at $z=2-3$,
before the 
formation of galaxy clusters.}
%Recent state-of-the-art cosmological simulations indicate 
%that the uniform and homogeneous enrichment of the 
%ICM gas at the larger radius is related to the early-enrichment scenario
%\citep[e.g.,][]{2010MNRAS.401.1670F,2017MNRAS.468..531B,2018MNRAS.474.2073V}. 
The state-of-the-art hydrodynamical simulations have verified
that the ICM can be substantially enriched during the peak of
star formation at $2<z<3$ 
\citep[e.g.,][]{2017MNRAS.468..531B,2018SSRv..214..123B}. 
Early AGN feedback may have played an 
{essential} role in displacing
the metal-enriched gas produced within 
galaxies throughout the ICM \citep{2019MNRAS.484.2896T}. 
%Although the simulation results are in contrast 
%with the observational 
%results by \citet{2007MNRAS.380.1554R}, 
%they are aligned with the recent observations by 
%\citet{2017A&A...603A..80M} and \citet{2019MNRAS.483..540L}.
%None of these observations reach up to R$_{200}$ or beyond. 

We are able to constrain the metallicity profiles out to R$_{200}$ for 
four galaxy groups, with temperatures of $\sim2.5$\,keV and with 
masses of an 
order of magnitude smaller than those of massive clusters. We found a 
nearly flat Fe abundance profile over the radii of 
0.25-1.0 R$_{200}$, as shown in Figure~\ref{fig:abun}, 
with an average Z$_{\rm Fe}$= 0.309 $\pm$ 0.018 Z$_{\odot}$, 
which is remarkably 
similar to 
{what is found in} massive clusters. 
This Fe $\sim$ 0.3 Z$_{\odot}$ uniform abundance distribution 
has also been found at the outskirts of two poor clusters,  
UGC 03957 \citep{2016A&A...592A..37T} 
and Virgo \citep{2015ApJ...811L..25S}, 
whose peak temperatures are $\sim3.5$\,keV, 
which is between groups and clusters. 

In addition to Fe, other elements (O, Mg, Si, S, and Ni) 
have nearly constant abundance throughout the outskirts of the 
groups, despite the larger uncertainties. 
Our findings are consistent with the same enrichment 
mechanism working among systems at various 
mass scales. 
Low mass systems 
form earlier than more massive systems in a
hierarchical Universe.
For metals to be deposited and well mixed before the 
gravitational collapse of galaxy groups, 
the timeline of the early-enrichment scenario, including the 
formation of SNIa (and their progenitor stars),
needs to be pushed back further
\citep[e.g.,][]{2014ARA&A..52..107M,10.1093/mnras/sty363}.
%[{\color{red}more specific time scale from theory?}].    

\subsection{Early enrichment population}
  
A conundrum in the enrichment study is that the total metals 
observed in rich clusters cannot be explained by the visible 
stellar component, as shown in Figure~\ref{fig:fe_yield}
\citep[also see][]{2021arXiv210504638E,2021A&A...646A..92G}.          
This discrepancy invokes an external and universal source of
metals even {\it before} the star formation episode -- 
an Early Enrichment Population (EEP), which is not visible today 
\citep{2017MNRAS.472.2877M}. 
The time scale of this enrichment epoch is not well constrained,
but possibly at $3<z<10$, and such an early stellar
population may be directly observable with JWST.    

Our findings demonstrate that once the Fe abundance
is measured out to $R_{200}$, the current stellar 
content of galaxy groups is insufficient to produce 
their total metals. Nevertheless, within
R$_{500}$ and R$_{200}$, the effective Fe yields of 
groups are significantly smaller than those of rich clusters.   
%By combining stellar with the ICM measurement,
%we are able to take an inventory of Fe in four 
%groups and compare them with the clusters
%and theoretical expectations.
%We find that the effective Fe yields within 
%R$_{500}$ of four groups are consistent 
%with the theoretical expectations. 
%However, they are
%significantly lower than that of massive
%cluster, which signifies that the efficiency for
%Fe production in clusters may be higher 
%with a significant contribution by pop III stars or
%pair-instability supernovae
%\citep[e.g.,][]{2021A&A...646A..92G,2021Univ....7..208G}.
%Within R$_{200}$, the higher measured
%Fe yield of RXJ1159, ESO3060170, and
%Antlia compared to theoretical prediction suggests
%that the total Fe mass in the IGrM do not depend
%only on the total amount of Fe produced, but also
%on the different feedback processes, 
%pushing a substantial fraction of metal out
%to R$_{200}$ and beyond 
%\citep[][]{2010MNRAS.401.1670F,2018ApJS..238...14M,2021Univ....7..208G}.
In addition to Fe yields, this difference between groups and 
clusters is also reflected in their iron-mass-to-light ratios. 
As shown in Figure~\ref{fig:imlr_vs_temp}, we have derived 
the  IMLR profiles of these 
groups out to R$_{200}$. 
Compared to the Perseus cluster, galaxy groups 
contain twice 
as much stellar light relative to their metal content. 
%The formation of the EEP may be mass dependent as more massive protostellar clouds, presumably associated with more massive clusters rather than groups, collapse to form stars first. Alternatively,
The small IMLRs of groups relative to clusters may be attributed
to the relatively shallower potential wells of groups,
making them unable to retain all the enriched gas
against non-thermal processes such as AGN feedback. 
As shown in \citet{2021MNRAS.501.3767S}, the accumulated gas
mass to total mass ratios within R$_{200}$ are consistent with 
the cosmic baryon fraction for galaxy clusters, 
which are systematically higher than those of galaxy groups. 
Furthermore, 
groups and clusters may have different star formation 
{histories}.
The IGrM has a lower 
temperature and density than the ICM, providing a 
less hostile environment for the member galaxies. 
The higher star formation rate in galaxy groups 
may partially explain the small IMLR \citep{2008MNRAS.391..110D}. 
%\citet{2019MNRAS.484.2896T} 
%show that the substantial and un-suppressed 
%star formation activity in the group scale can 
%efficiently utilize much of its metal-enriched gas. 
%This process reduces the gas Fe content, 
%resulting in a lower IMLR. 
%Numerical simulations by \citet{2008MNRAS.391..110D} 
%also observe a 
%higher star formation rate in their 
%simulated groups. 
%Together with these studies, our results 
%may indicate that the higher star formation 
%rate in MKW4 and Antlia than Perseus could 
%reduce the IMLR at the larger radii, 
%as seen in Figure \ref{fig:imlr_vs_temp}.

\subsection{Chemical composition of galaxy groups}

The additional constraints on O, Mg, Si, and S allow us 
to derive the chemical components and SNIa fraction of galaxy groups. 
The abundance ratios of O/Fe, S/Fe, Si/Fe, and Mg/Fe
of the four groups in our sample are shown in Figure~\ref{fig:x/w}. 
Primarily caused by the abrupt change in the slope of 
the Fe abundance profile, we observe different 
chemical compositions between group centers and group outskirts. 
The X/Fe ratios within 0.25R$_{200}$ are consistent with
the Solar chemical composition, for which we derive a f$_{\rm SNIa}$ of 
$0.214\pm0.028$. This is in agreement with the average gas abundance
ratios of cluster cores \citep{2007A&A...465..345D} and
44 nearby systems within R $<$ R$_{500}$
(dominated by galaxy clusters) in the CHEERS sample
\citep{2017A&A...603A..80M}. 
Similar chemical composition for gas has been
found at the centers of the Perseus cluster 
using the microcalorimeter on 
board {\sl Hitomi} \citep{2019MNRAS.483.1701S},
as well as in NGC~1404, 
an early-type galaxy, through a joint {\sl XMM-Newton} EPIC, RGS, 
and $\chandra$-ACIS study \citep{2022arXiv220105161M}. 
It is intriguing that ICM, IGrM, and even ISM metallicities 
trace the same metal abundance pattern as our Solar System 
rather than their dominant galaxies. 
This further implies
a common origin of the metals among these systems.  
Alternatively, this also means that the stellar population we see today 
in clusters and groups (i.e., mainly red-and-dead population) has 
very little to do with the metal content of their hot halos.

Over the radii of 0.25-1.0R$_{200}$,
the X/Fe ratios are generally super solar, ranging 
from 1.6 to 1.9, for which we derive that the
SNIa fraction is $0.138\pm0.046$, allowing us to
rule out a pure SNcc enrichment by 3$\sigma$. 
The Fe abundance at the outskirts is unlikely to 
be significantly underestimated since we have 
employed a 2T model to mitigate the ``Fe-bias", and 
we have obtained a measured $Z_{\rm Fe}$ similar to 
that of the rich clusters. 
%Fe-bias is attributed to the fact that 1T plasma emission code either increasing temperature or decreasing Fe abundance or both while fitting
%the code generated excess Fe-L emission with the observation \citep[e.g.,][]{1998MNRAS.296..977B,2000ApJ...539..172B,2021Univ....7..208G}.
It is worth noting that 
the significant uncertainties and scatter of 
the measurements
do not allow us to claim {an apparent}
discrepancy from the 
solar value for S/Fe, Si/Fe, and Mg/Fe, while the O 
line can not be well resolved by CCD-like detectors such as XIS. 
{An} extensive $\suzaku$ study of the
outskirts of the Virgo
cluster reveals a super solar value for Mg/Fe of 1.2--1.6,
while their Si/Fe and S/Fe ratios are 0.6-0.8 and 0.8-1.2,
respectively. 
The X/Fe ratios may scatter among different systems. 
The results can be further complicated by the various
sources of systematic effects associated with the observations
of the background-dominated regimes. 
%The discrepancy may be caused by unknown systematic effects. 
%If the $f_{\rm SNIa}$ and X/Fe discrepancy is genuine and physical, it may imply a different enrichment history for group centers and outskirts. 
%, which may invoke the contributions from the BGG, infalling galaxies, and the intracluster light as well as a certain combination of the two early-enrichment scenarios discussed above.

\section{summary}
We analyzed joint $\suzaku$ and $\chandra$ observations of 
four nearby groups -- MKW4, Antlia, RXJ1159+5531, and ESO3060170. 
$\suzaku$ observations were used to constrain their gas properties, 
and $\chandra$ data were used to refine the results by
mitigating the uncertainties introduced by the CXB. 
We have derived abundance profiles of O, Mg, Si, S, Fe, 
and Ni out to the R$_{200}$ for the first time for 
such low-mass systems. 
Our results are summarized below. 
\begin{itemize}
%   \item Using the $\chandra$ observations, we are able to resolve 
%    much of the point sources, particularly at the group outskirts.
%       We model the contamination due to the point sources in the Suzaku data,  which significantly reduces the uncertainties in measuring metal abundances.

    \item The metal abundances in the central region of Antlia are 
    significantly lower than those of MKW4, RXJ1159+5531, and ESO3060170. 
    Unlike the other three systems,
    Antlia is a non-cool-core group. 
    {As} observed in massive system equivalents 
    (i.e., non-cool-core clusters), more frequent merging
   events may have disrupted its core and erased much of the
   centrally peaked metallicity.

    \item At group outskirts (0.25-1.0R$_{200}$), 
    we measure flat metal abundance profiles for all
    four groups, with an average Fe abundance of 
    Z$_{\rm Fe}=0.309\pm0.018$ Z$_{\odot}$ 
    ($\chi^2$ = 14 for 12 degrees of freedom),
    which is remarkably consistent with what was 
    found in rich clusters, and predictions from 
    IllustrisTNG for galaxy groups. 
    The uniform metal distribution suggests that the 
    same early enrichment process may have been at work 
    before the gravitational collapse of groups and clusters. 
    Early AGN feedback is likely to have played an essential 
    role in transporting metals from galaxies to the hot 
    gas as well as efficiently mixing the enriched gas 
    throughout the proto cluster/group gas at $z=2-3$.
    It is also plausible that an early-enrichment
    population is responsible for contaminating the 
    proto cluster/group gas before the formation
    of the current stellar components at $3<z<10$.

    \item We derive the accumulated iron mass-to-light ratios 
    from the center out to the R$_{200}$ of these groups. 
    Their IMLR porfiles increase with radius, 
    indicating that the enriched gas is more extended than 
    the stellar distribution. 
    Near R$_{200}$, the accumulated IMLR profiles of these groups are 
    consistent with each other at 4-5$\times$ 10$^{-3}$ M$_{\odot}$/L$_{\rm K \odot}$. 
%    suggesting similar early-enrichment processes.
    However, these values are
    %The IMLRs of these groups are 
    significantly lower 
    than the accumulated IMLR of Perseus at R$_{200}$, which may imply that 
    groups could not retain all of their enriched gas 
    due to their shallower potential wells, and/or a
    halo-mass dependent star formation history. 
    
    \item We compare the effective Fe yields of groups 
    within $R_{500}$ and $R_{200}$, and the theoretical expectation. 
    The iron yields measured out to R$_{200}$ of these four groups
    show tension with respect to the expected values from empirical 
    supernovae yields, and in contrast with what can be observed within R$_{500}$. 
    A significant amount of metals are found at the outskirts of groups. 
    When the Fe abundance is measured out to $R_{200}$, 
    the visible stellar component in the groups is 
    insufficient to produce the observed metals,
    which may pose a challenge for the chemical enrichment models 
    in the same fashion as the one already known for clusters.

    \item The O/Fe, Mg/Fe, Si/Fe, and S/Fe ratios at the
    group centers are nearly solar and consistent with 
    the chemical composition measured at cluster centers. 
    The hot gas in the systems that are orders of magnitude
    different in mass  
    traces the same metal abundance pattern as
    our Solar System 
    rather than their dominant galaxies. 
    The derived SNIa fraction is 14\% at their outskirts, 
    allowing {us} to rule out a pure core collapsed 
    supernovae enrichment before the gravitational collapse of galaxy groups.   
    %The group outskirts are consistent with super solar X/Fe ratios in spite of their sizable uncertainties. 
    %It is desirable to expand the sample and better understand various systematic uncertainties, particularly with future X-ray instruments such as {\it XRISM}, {\it Athena}, and {\it Lynx}. 
    
\end{itemize}

\section*{Acknowledgements}
{We are grateful to Dan Maoz
for the insightful comments that greatly
helped to improve the paper.}
A.\ S. and Y.\ S. were supported by Chandra X-ray Observatory 
grant GO1-22126X, NASA grant 80NSSC21K0714, and NSF grant 2107711.
SRON Netherlands Institute for Space Research 
is supported financially by NWO.

\section*{Data Availability}
{ The data underlying this article will be shared
on reasonable request to the corresponding author.}

\bibliographystyle{mnras}
\bibliography{example} % if your bibtex file is called example.bib

%%%%%%%%%%%%%%%%%%%%%%% individual images %%%%%%%%%%%%%%%%%%%%%%%%%%%%%%%%%%%%%%%%%%%
\appendix
\renewcommand{\thefigure}{A1}

\begin{figure*}
\begin{tabular}{ccc}
%\hspace{-113pt}
\includegraphics[width=0.4\textwidth]{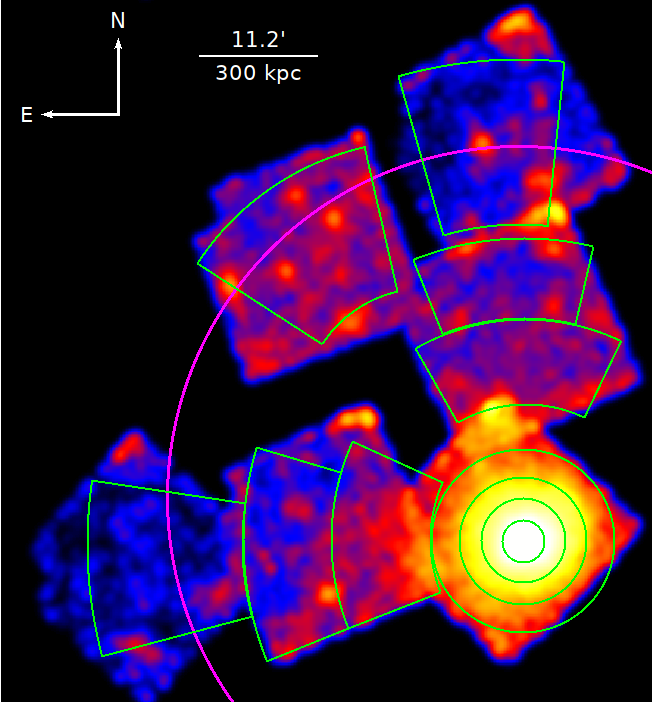} & \hspace{30pt} \includegraphics[width=0.4\textwidth]{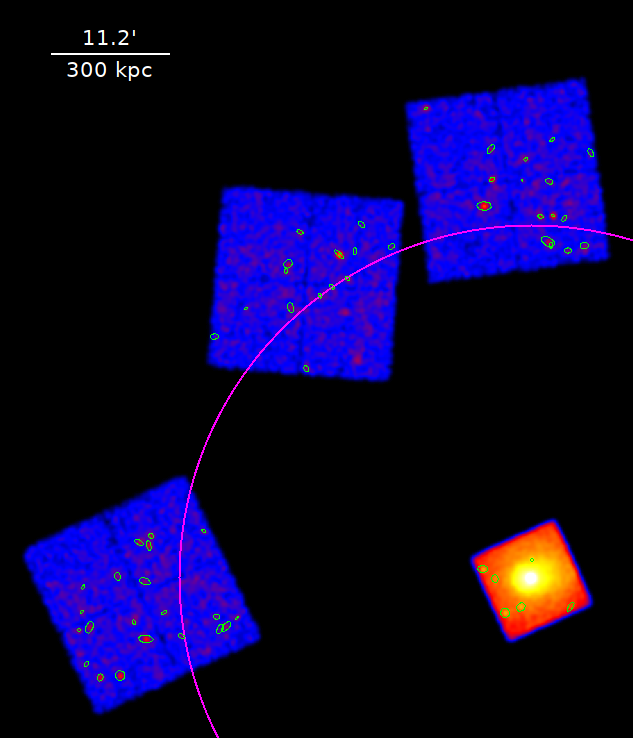}\\
 \hspace{-50pt}\includegraphics[width=0.35\textwidth]{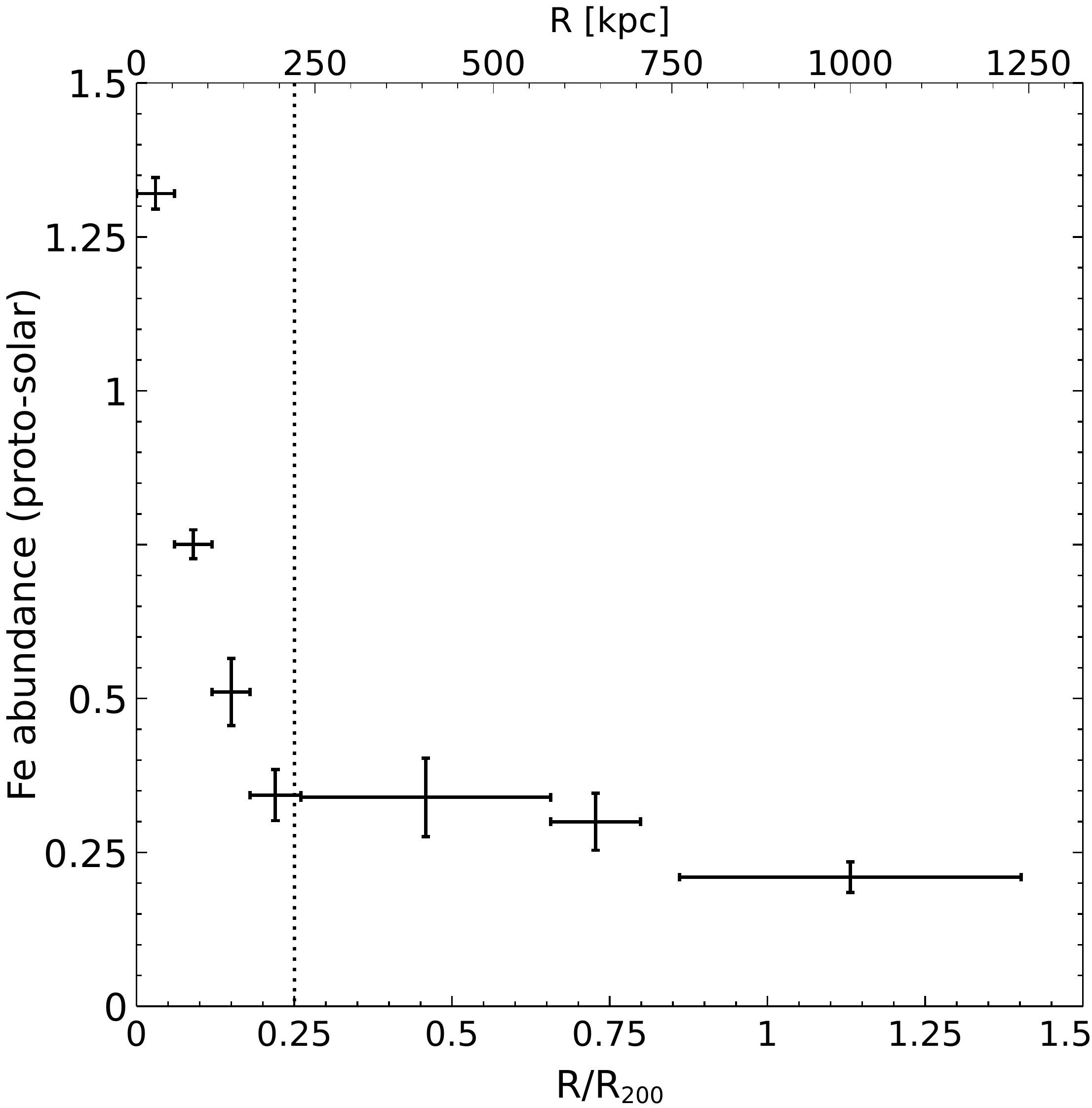} & \hspace{-130pt} \includegraphics[width=0.35\textwidth]{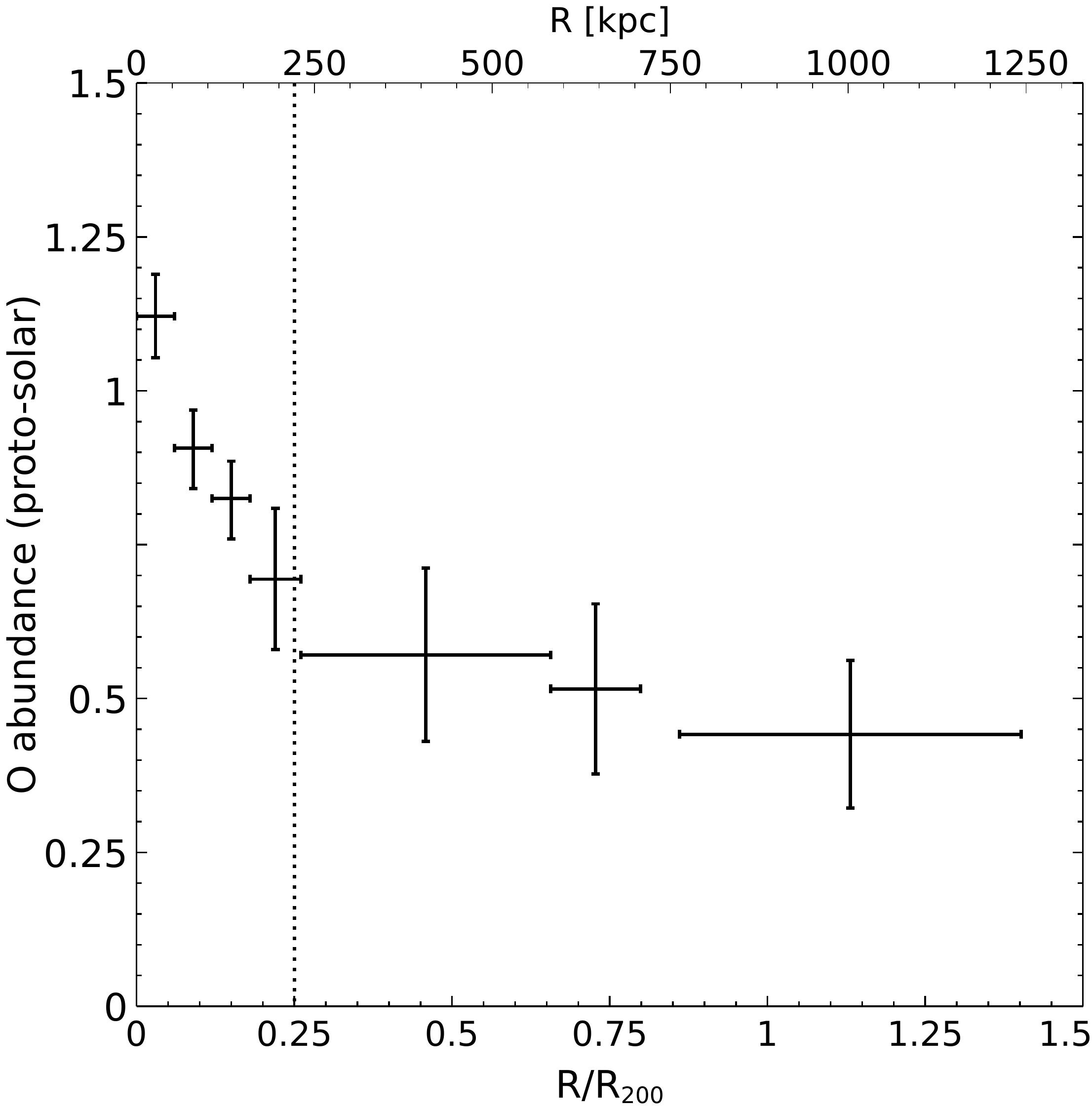} & \hspace{-100pt} \includegraphics[width=0.35\textwidth]{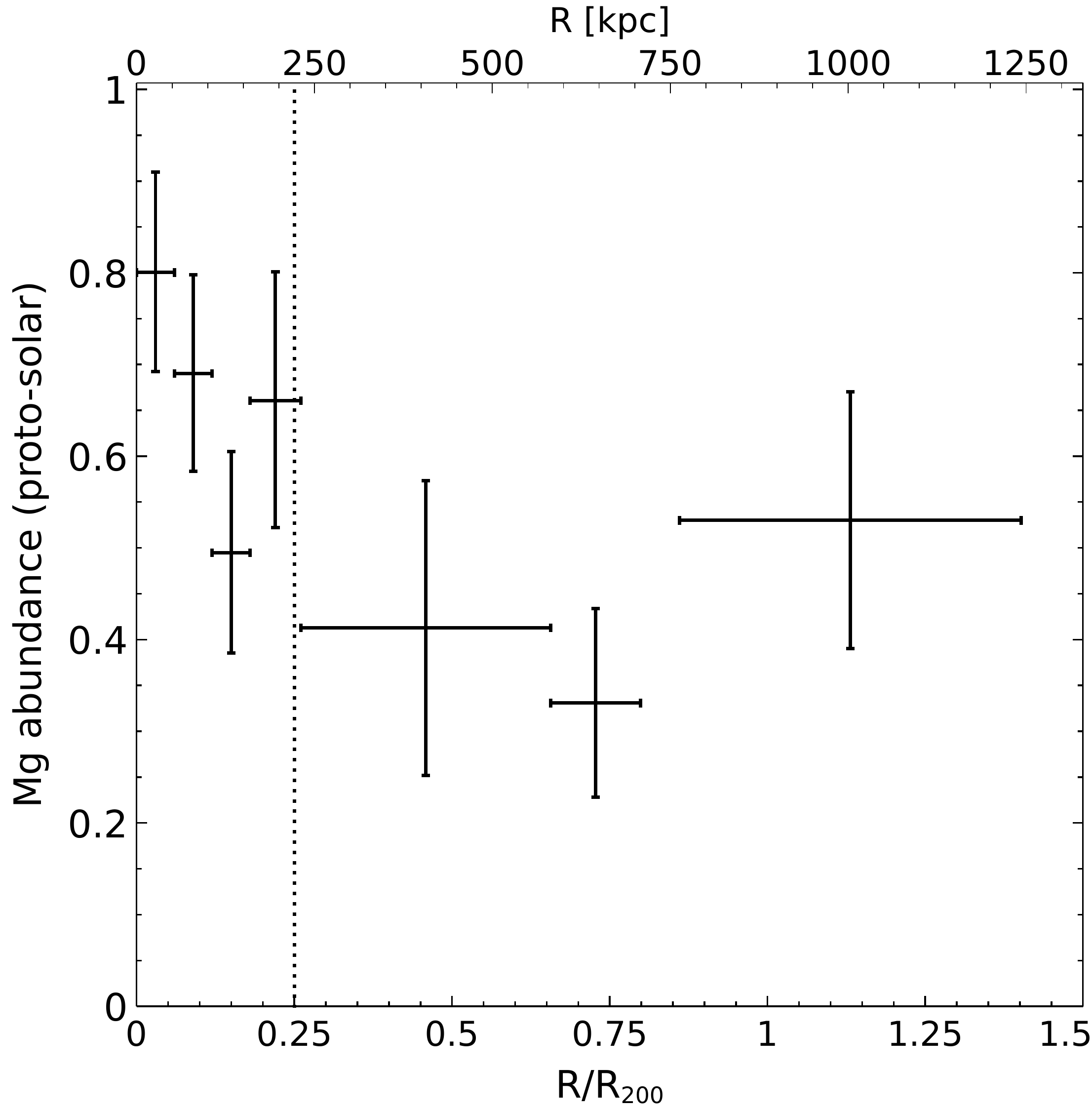}\\
\hspace{-50pt} \includegraphics[width=0.35\textwidth]{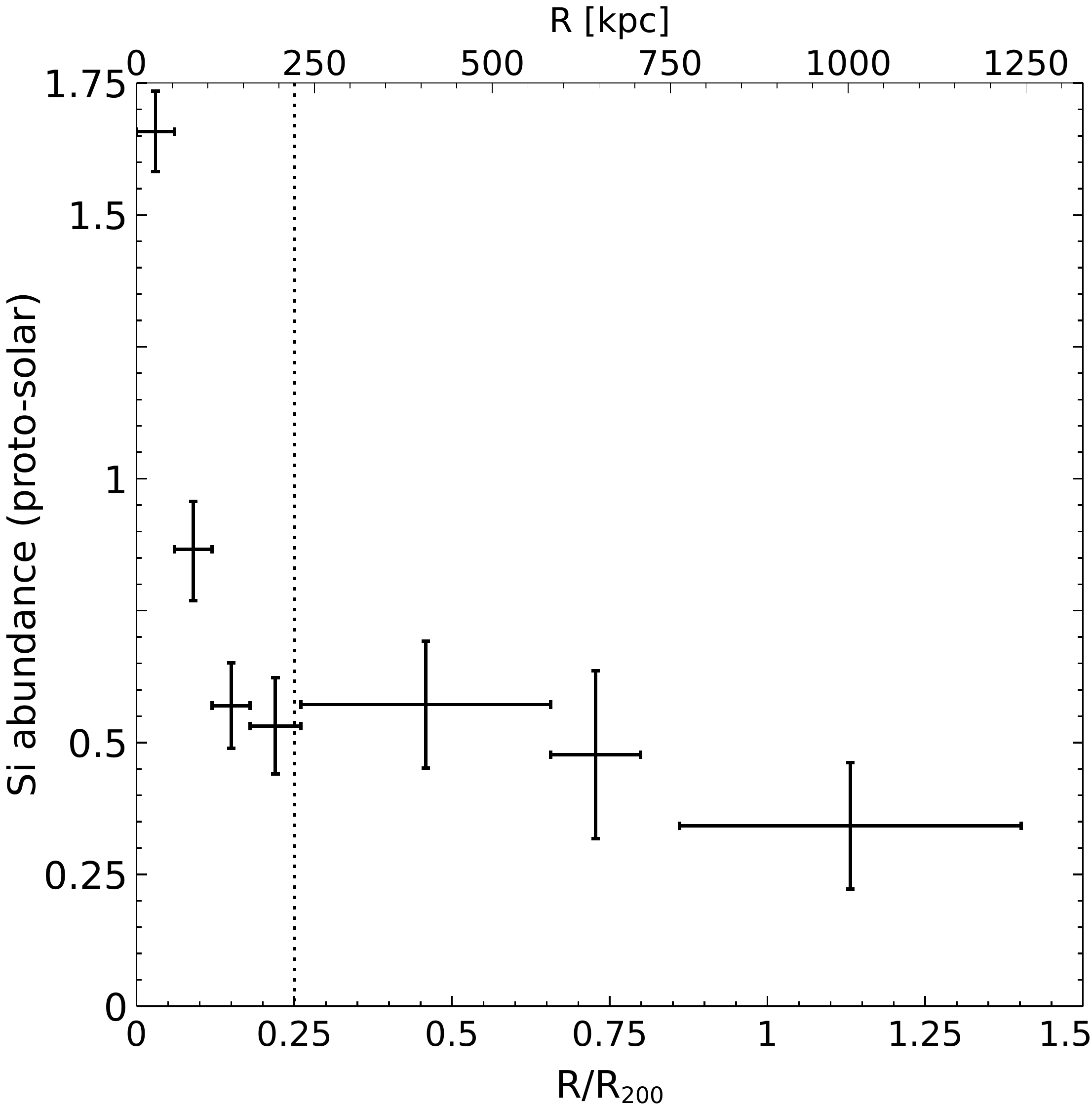} & \hspace{-130pt} \includegraphics[width=0.35\textwidth]{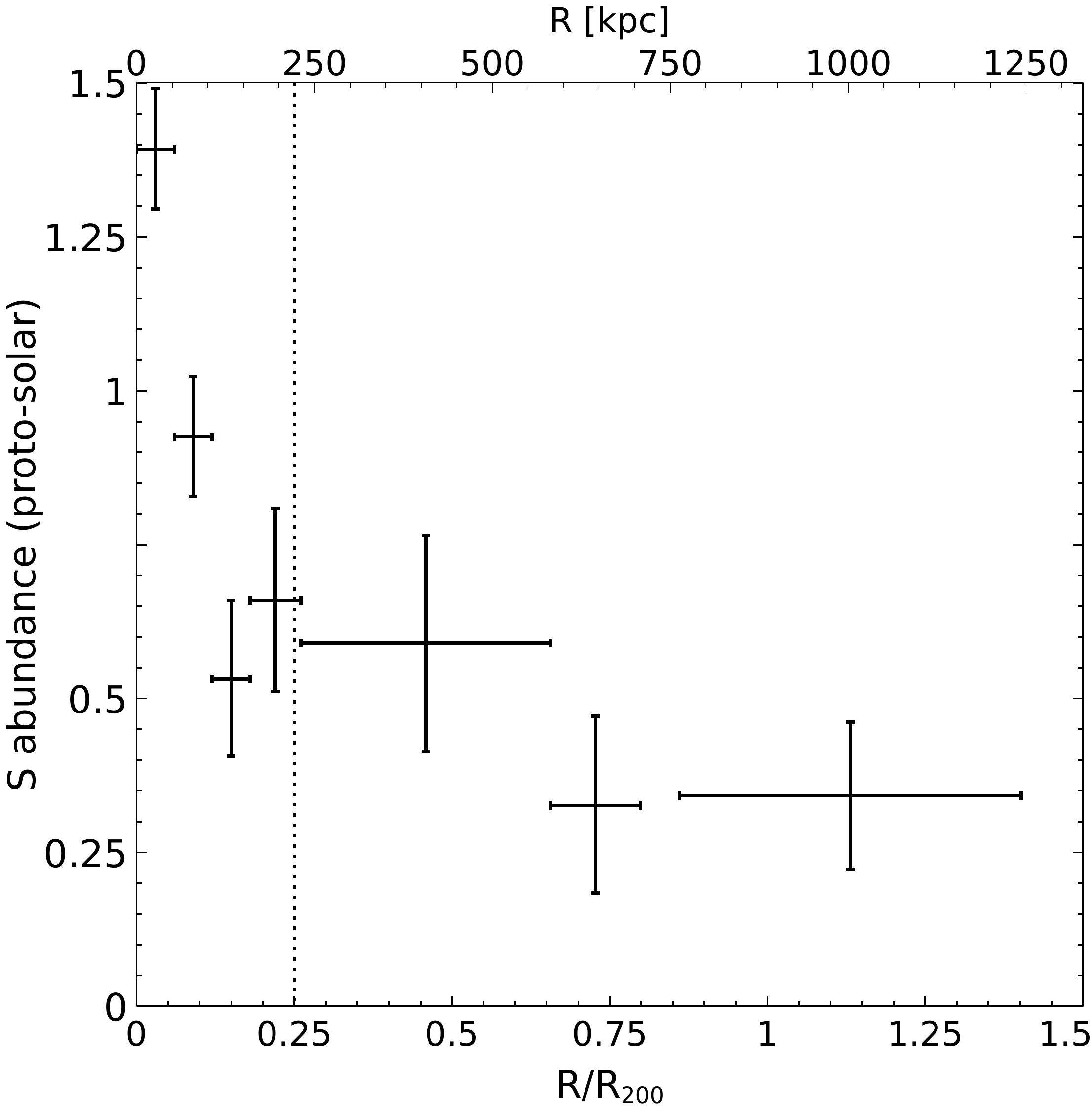} & \hspace{-100pt} \includegraphics[width=0.35\textwidth]{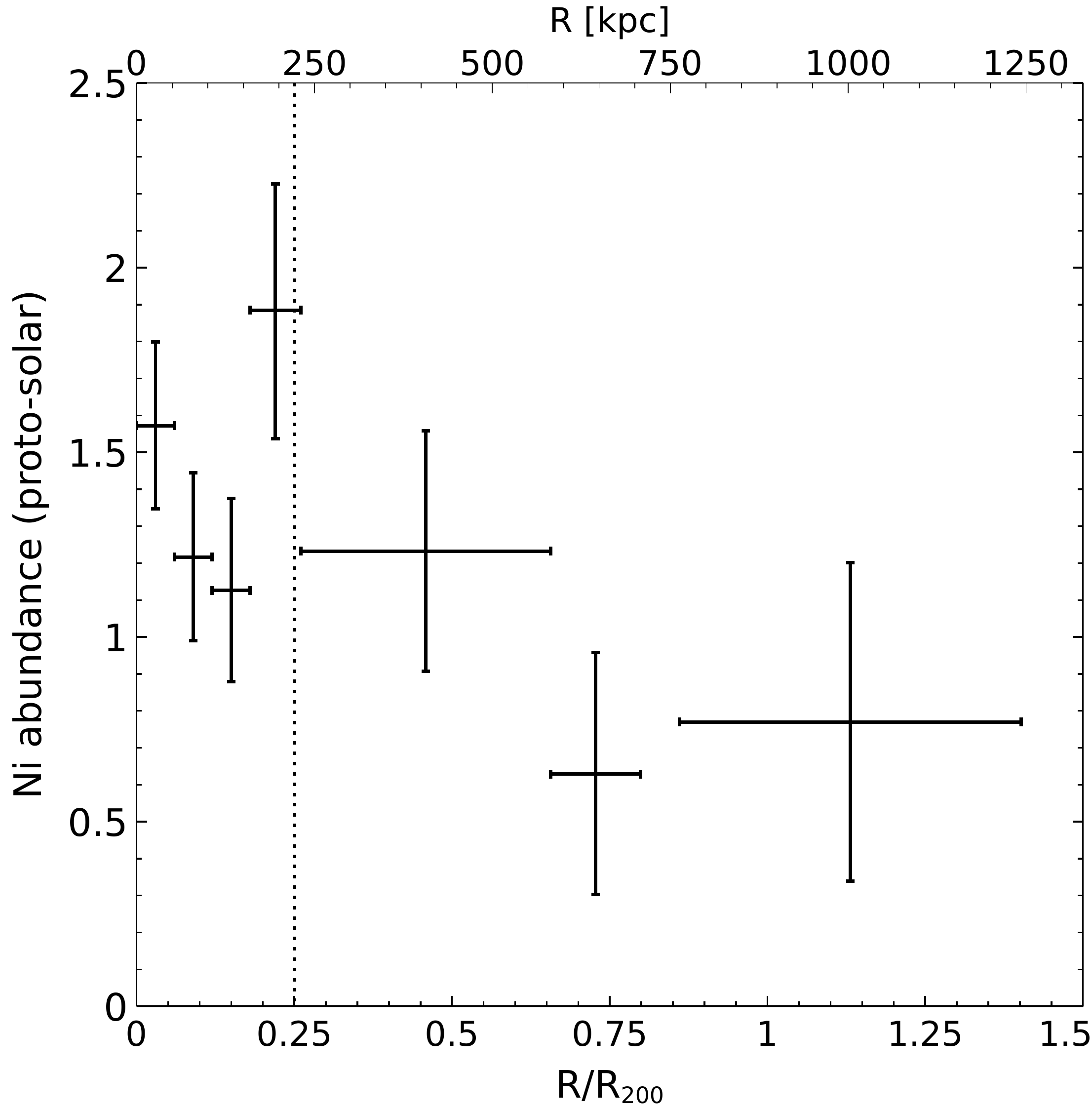} \\
\end{tabular}
\caption{ Top left: NXB subtracted 
and exposure corrected mosaic 
$\suzaku$ image of MWK4 in the 0.5-2 keV energy band.
Magenta circle indicates R$_{200}$. 
Green annuli represent  
the regions used for spectra extraction.
Top right: mosaic $\chandra$ image of 
MKW4 in the 0.5-7.0 keV 
energy band. The resolved point 
sources are marked in green elliptical regions. 
Middle and Bottom: radial profiles of different elements
from the group center to the outskirts.
Vertical dotted line indicate 0.25R$_{200}$.}
\label{fig:image_mkw4}
\end{figure*}

\renewcommand{\thefigure}{A2}
\begin{figure*}
\begin{tabular}{ccc}
%\hspace{-113pt}
& \hspace{-150pt} \includegraphics[width=0.9\textwidth]{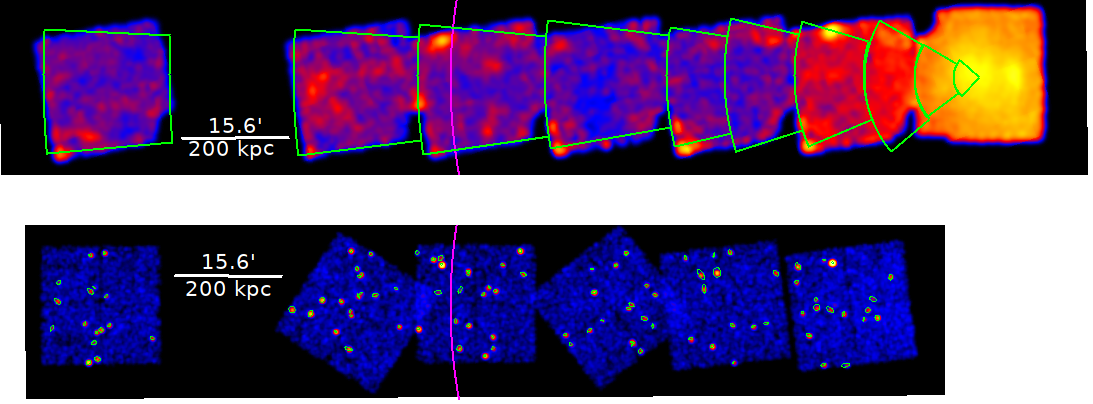} & \\
\hspace{-15pt} \includegraphics[width=0.35\textwidth]{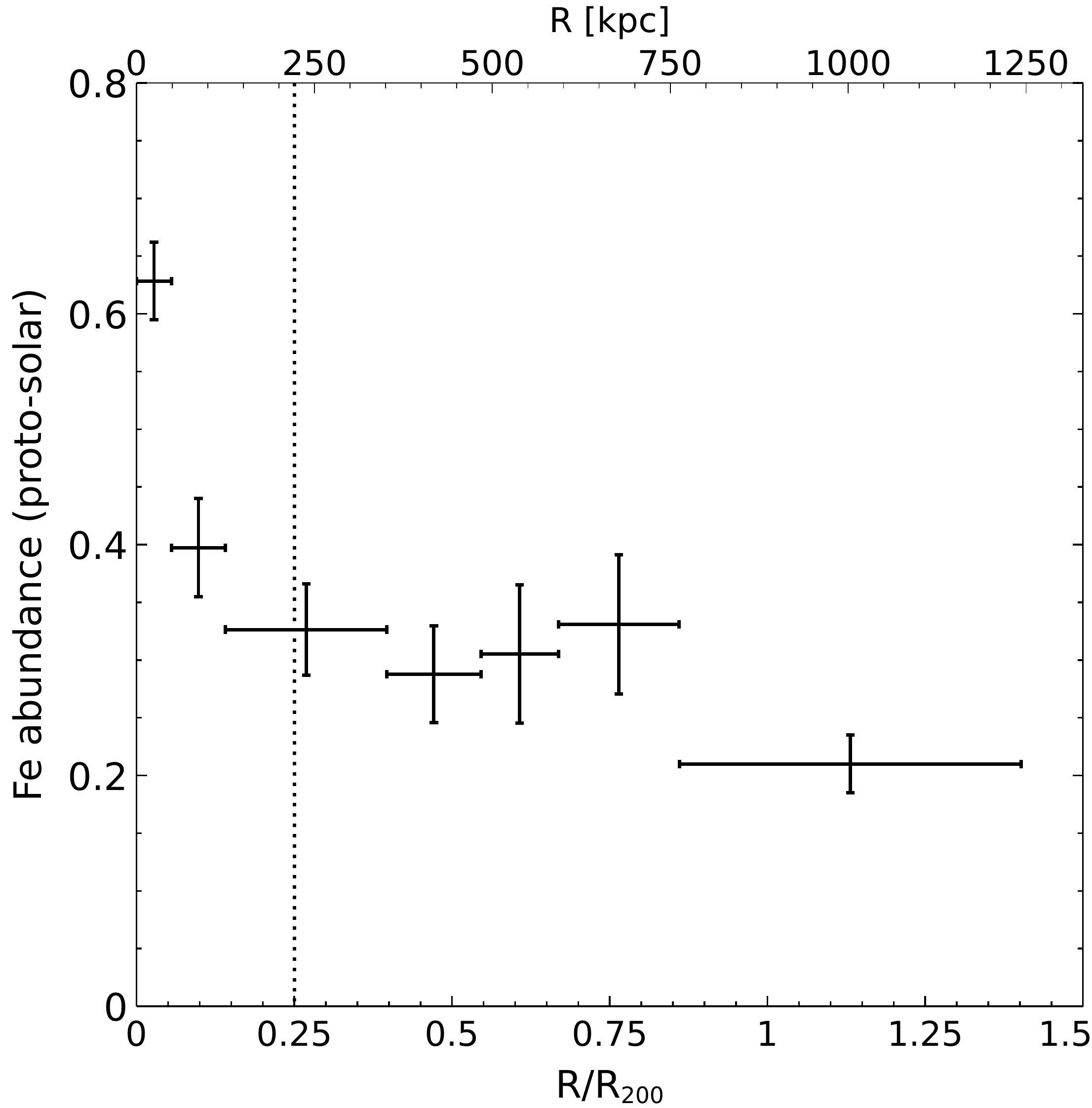} & \hspace{-150pt} \includegraphics[width=0.35\textwidth]{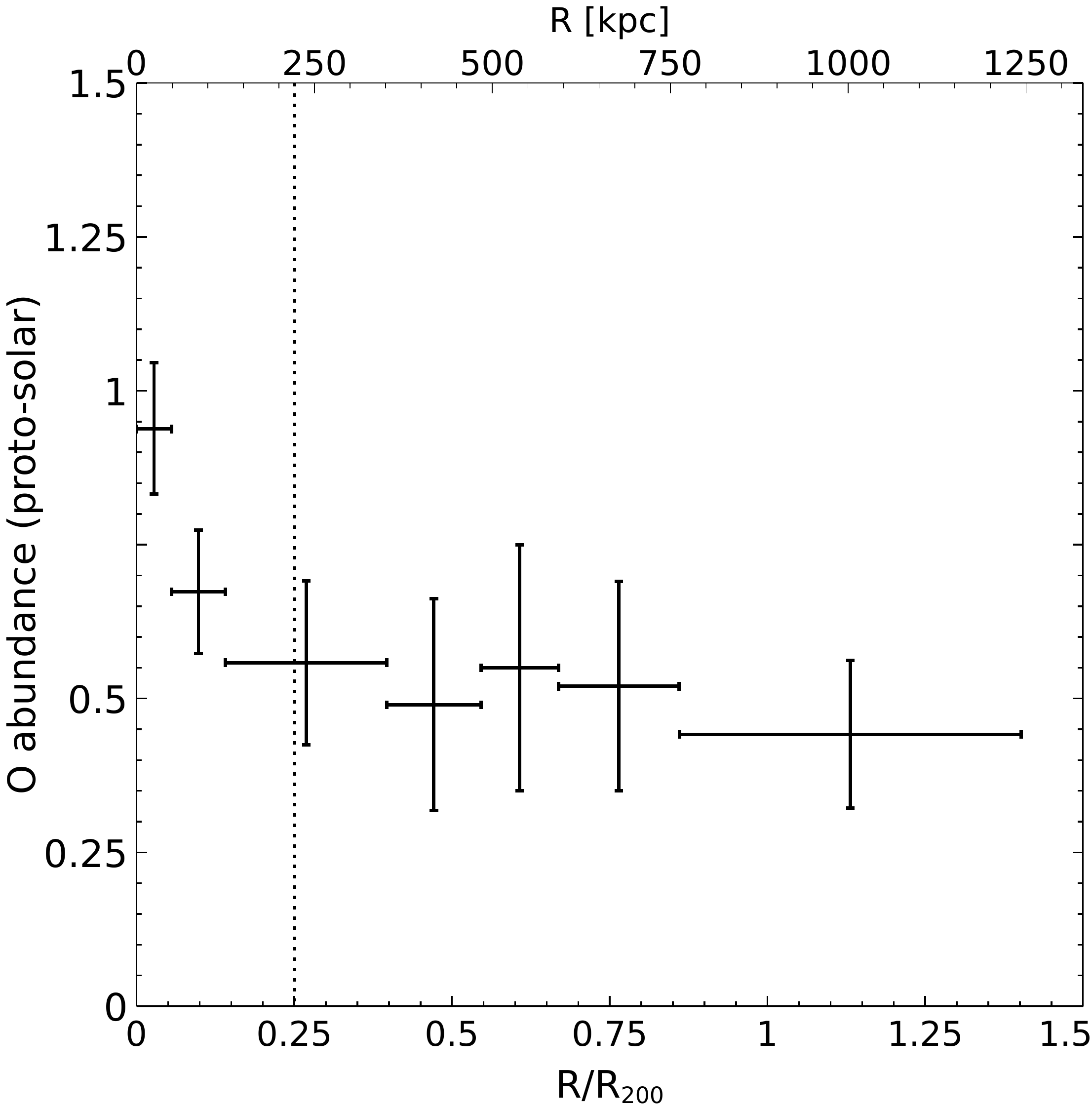} & \hspace{-150pt} \includegraphics[width=0.35\textwidth]{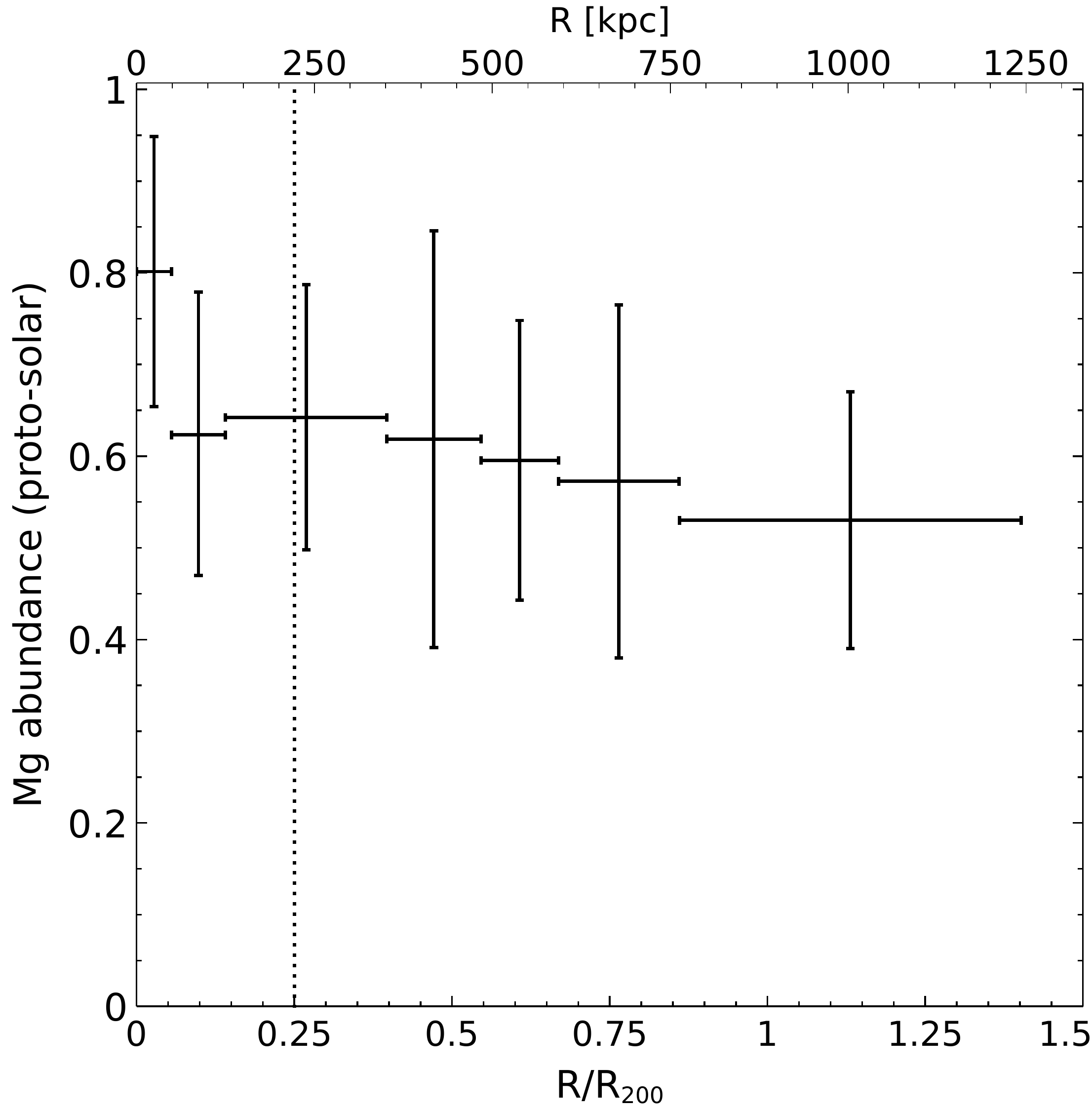}\\
\hspace{-15pt} \includegraphics[width=0.35\textwidth]{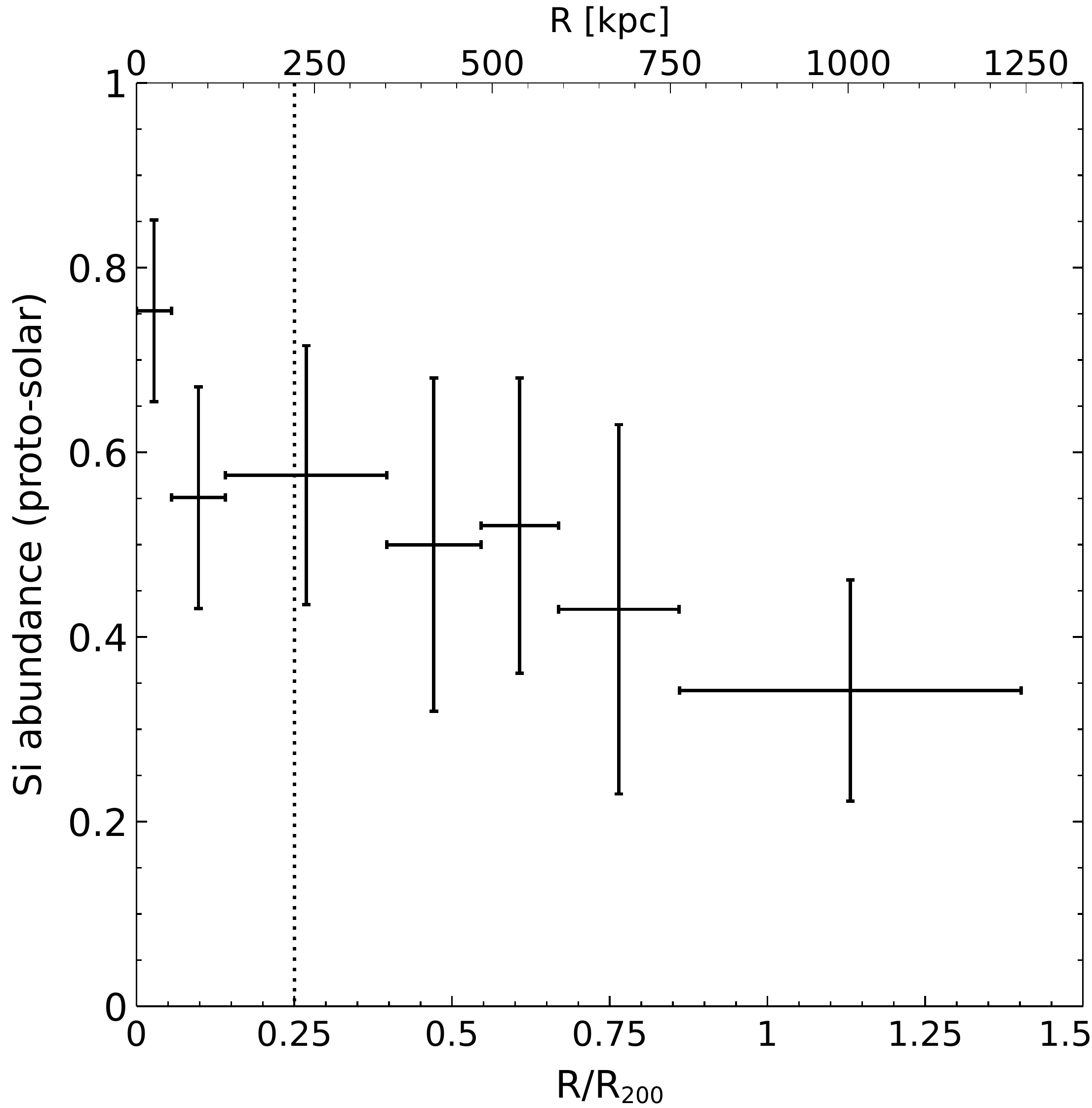} & \hspace{-150pt} \includegraphics[width=0.35\textwidth]{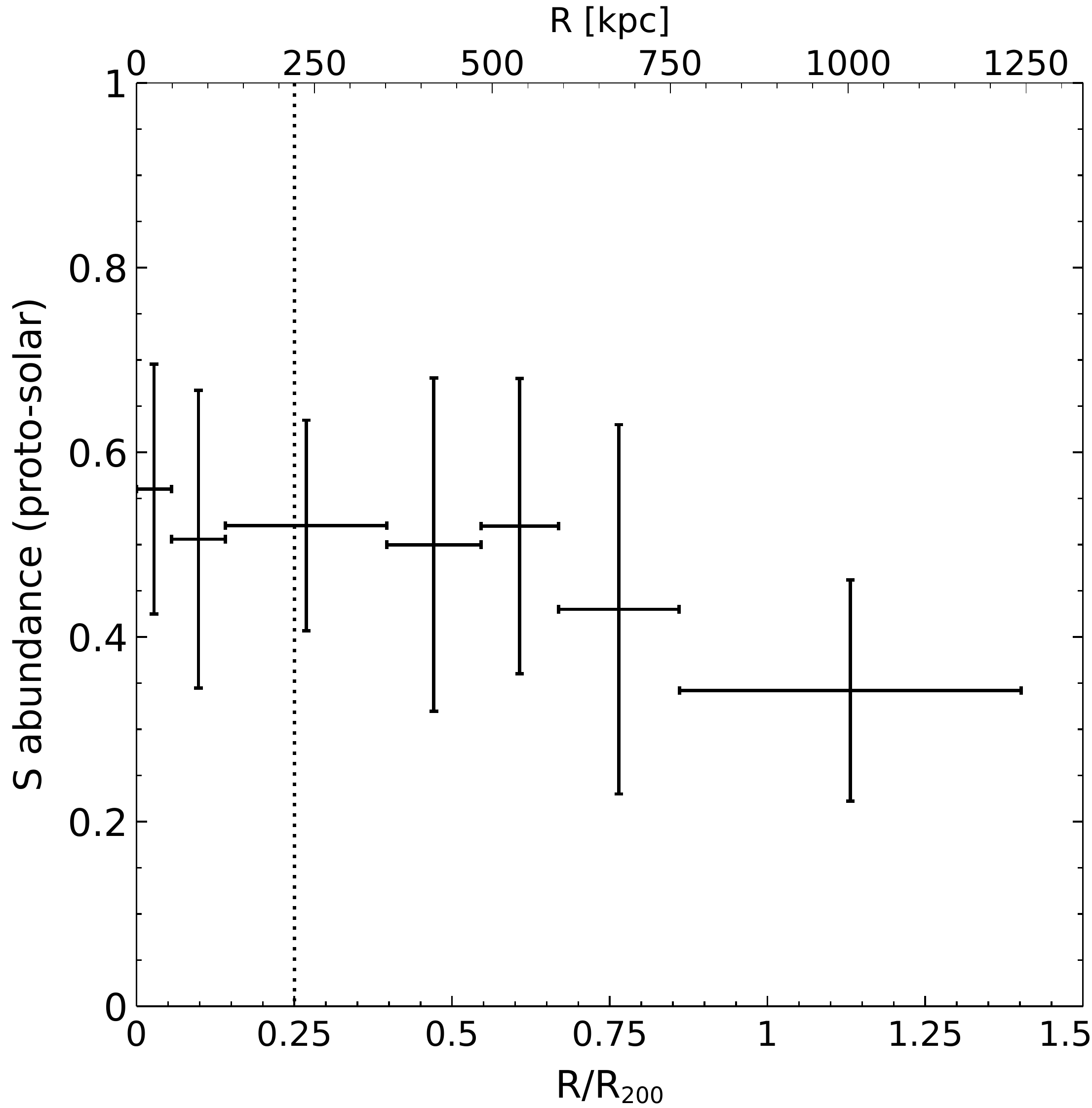} & \hspace{-150pt} \includegraphics[width=0.35\textwidth]{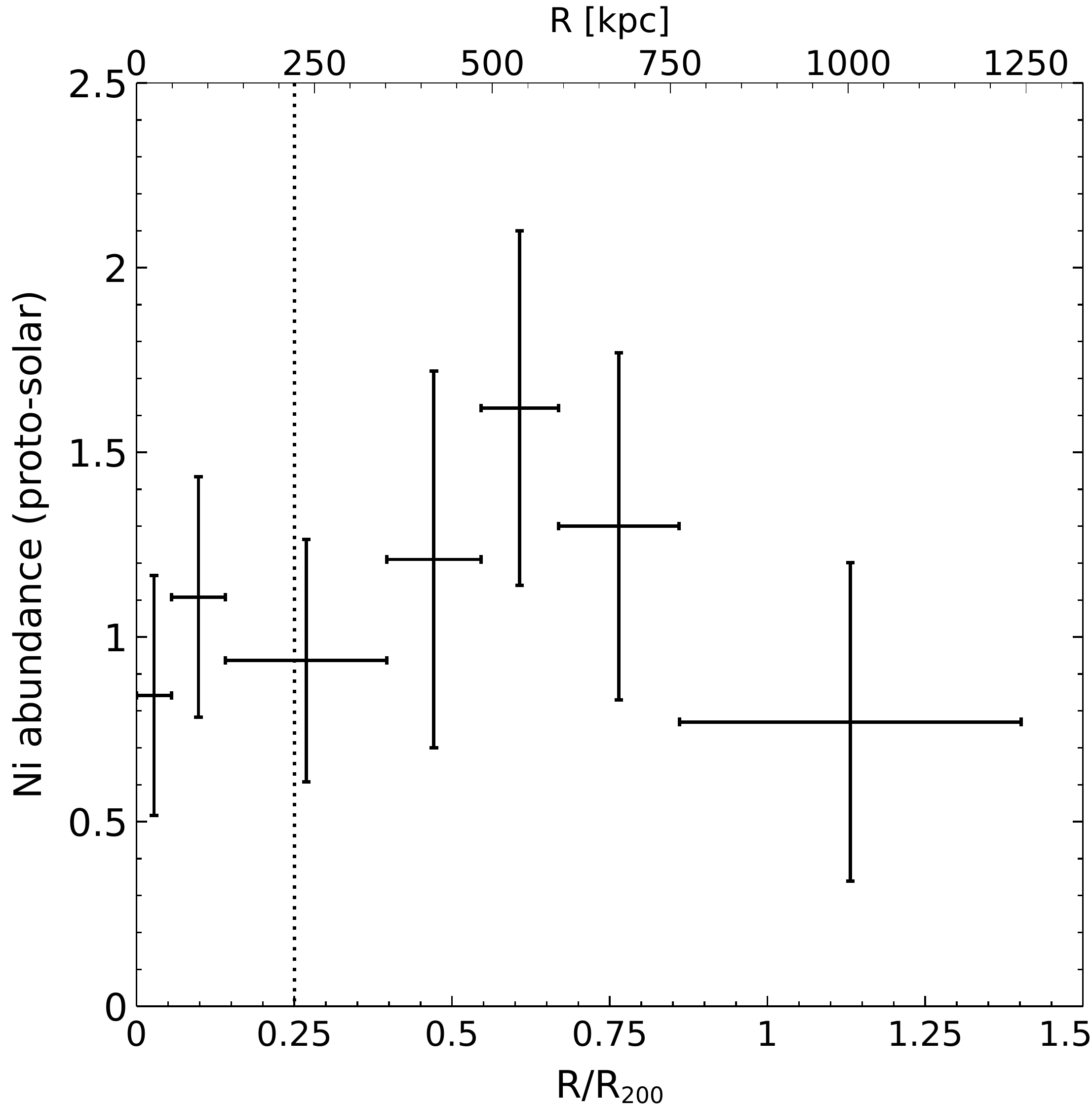} \\
\end{tabular}
\caption{ Antlia; same as Figure \ref{fig:image_mkw4}. 
}
\label{fig:image_antlia}
\end{figure*}

\renewcommand{\thefigure}{A3}

\begin{figure*}
\begin{tabular}{ccc}
& \hspace{-210pt} \includegraphics[width=1.1\textwidth]{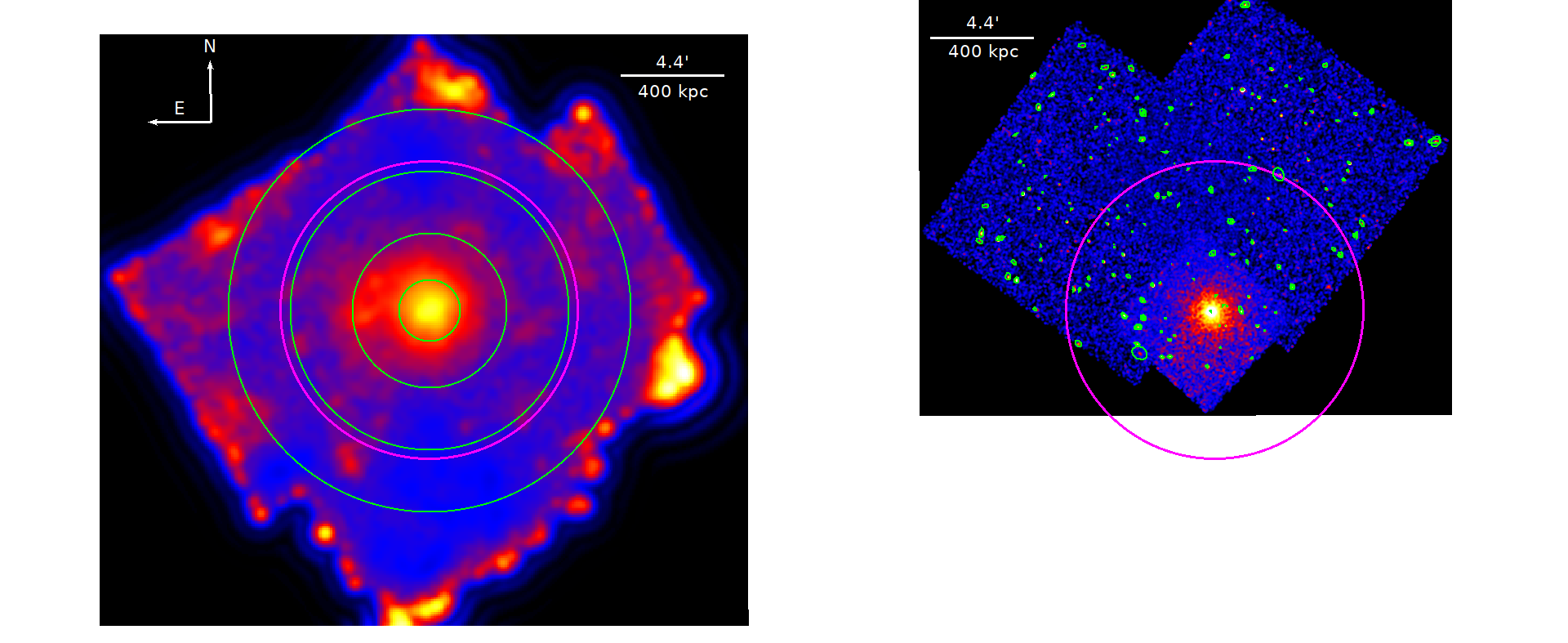} &\\
\hspace{-15pt} \includegraphics[width=0.35\textwidth]{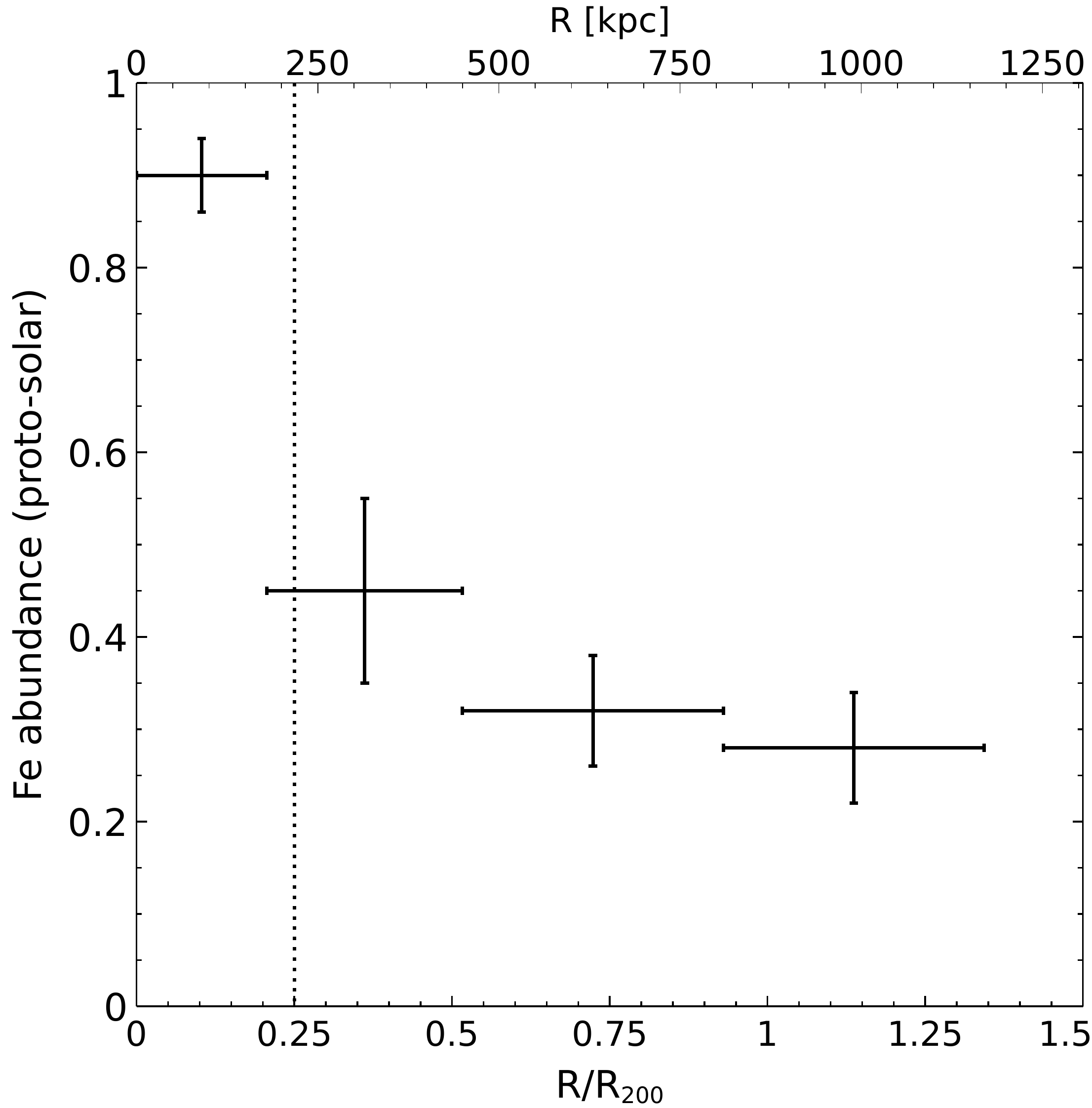} & \hspace{-200pt} \includegraphics[width=0.35\textwidth]{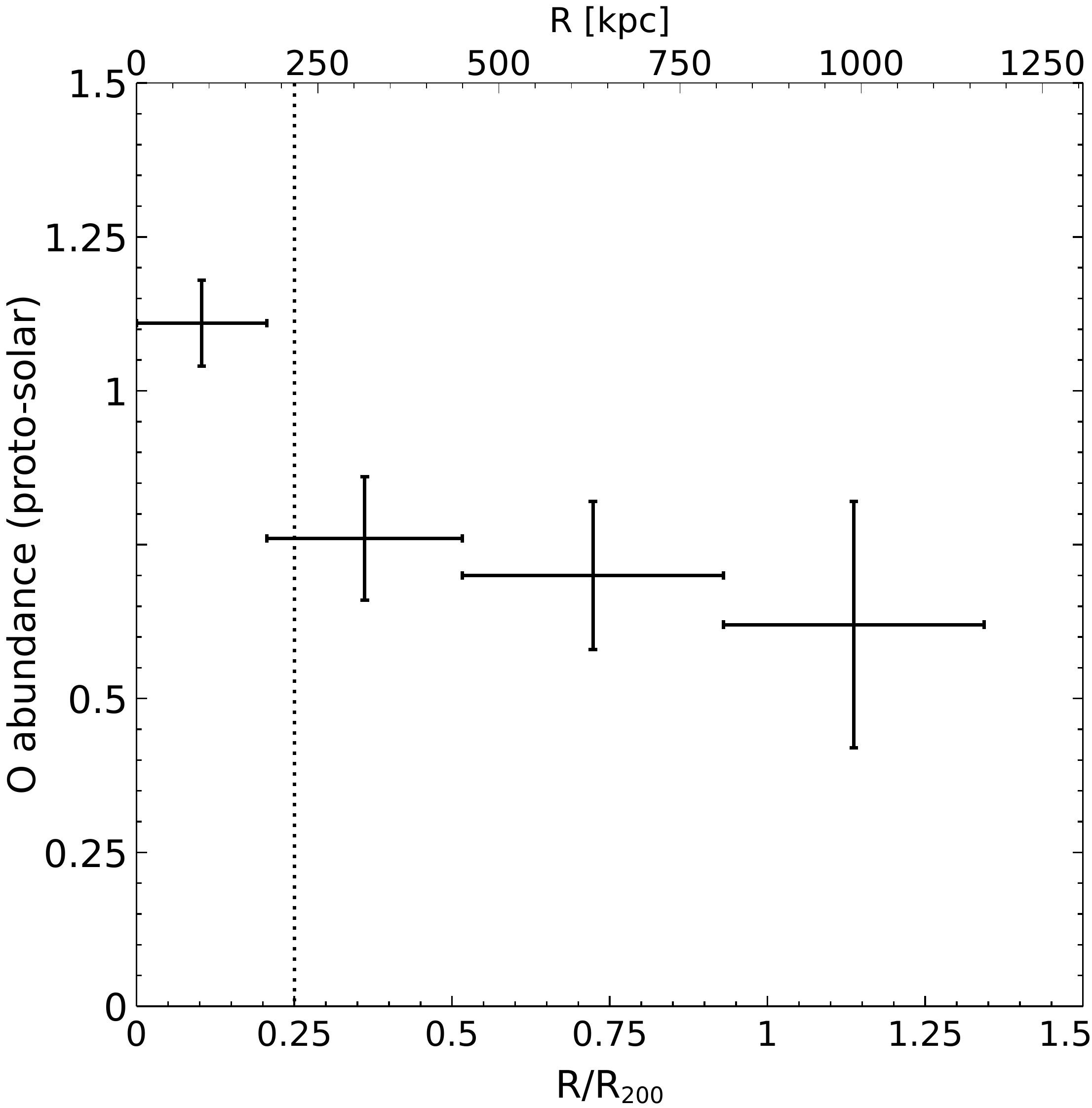} & \hspace{-200pt} \includegraphics[width=0.35\textwidth]{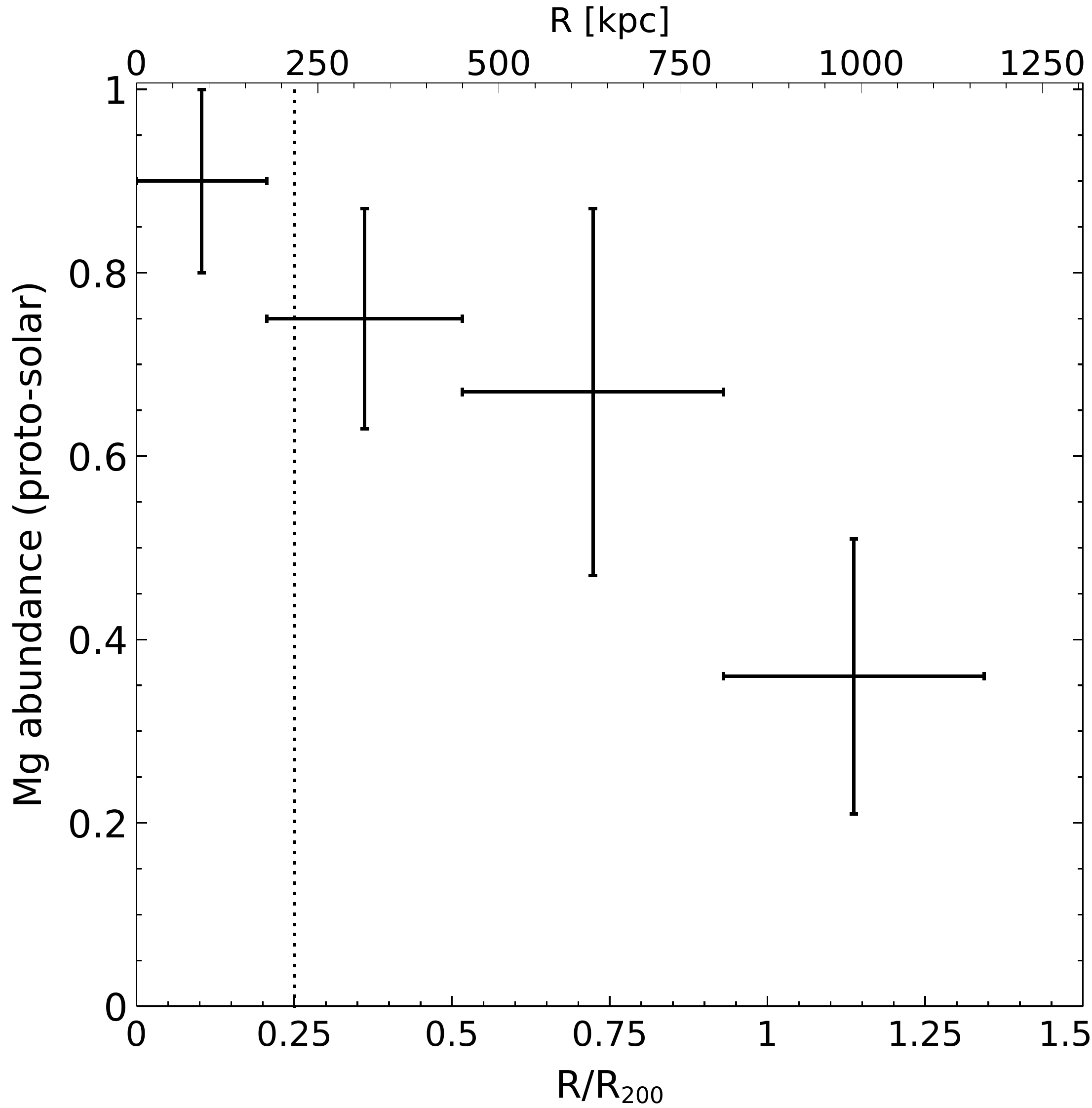}\\
\hspace{-15pt} \includegraphics[width=0.35\textwidth]{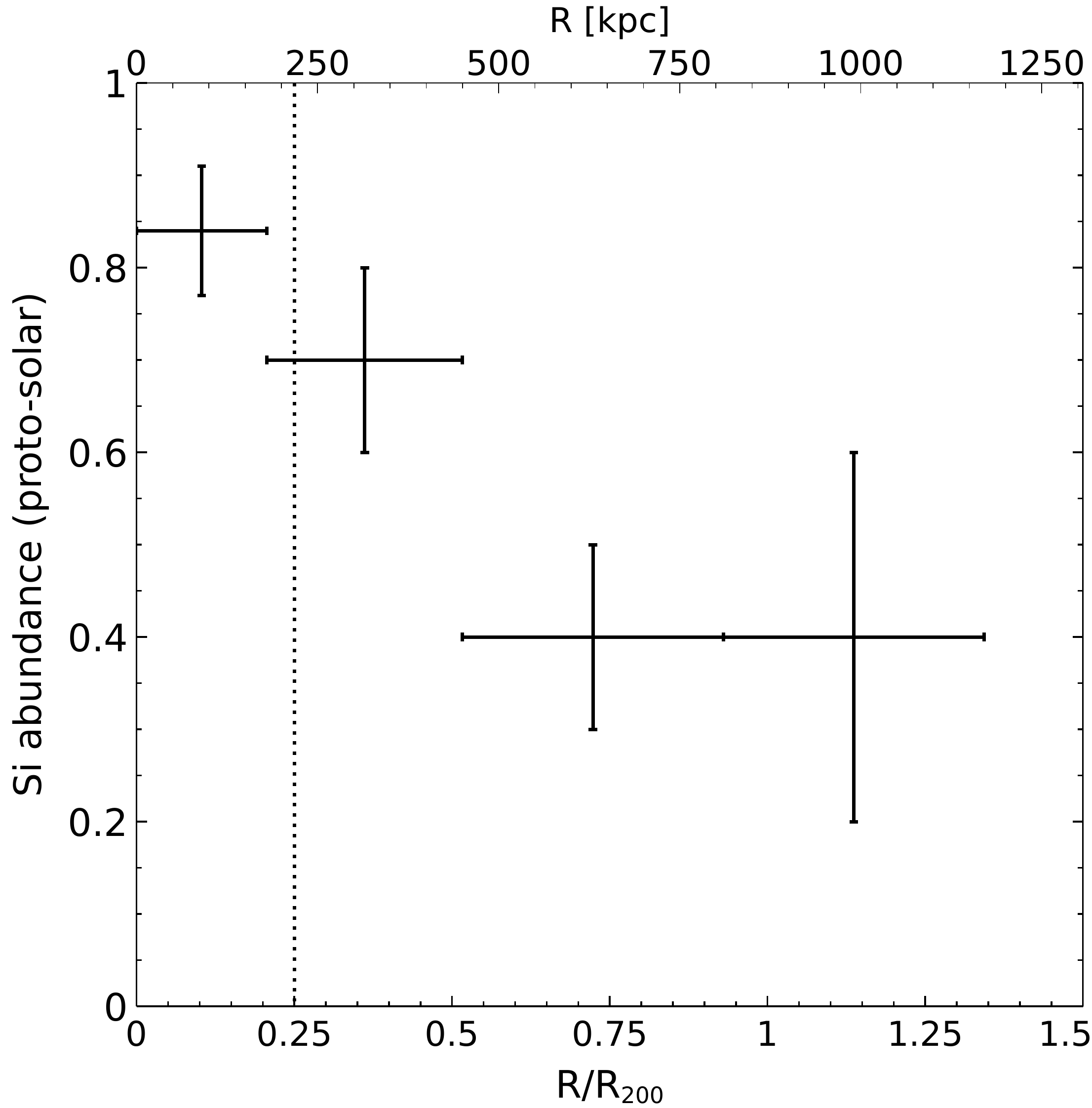} & \hspace{-150pt} \includegraphics[width=0.35\textwidth]{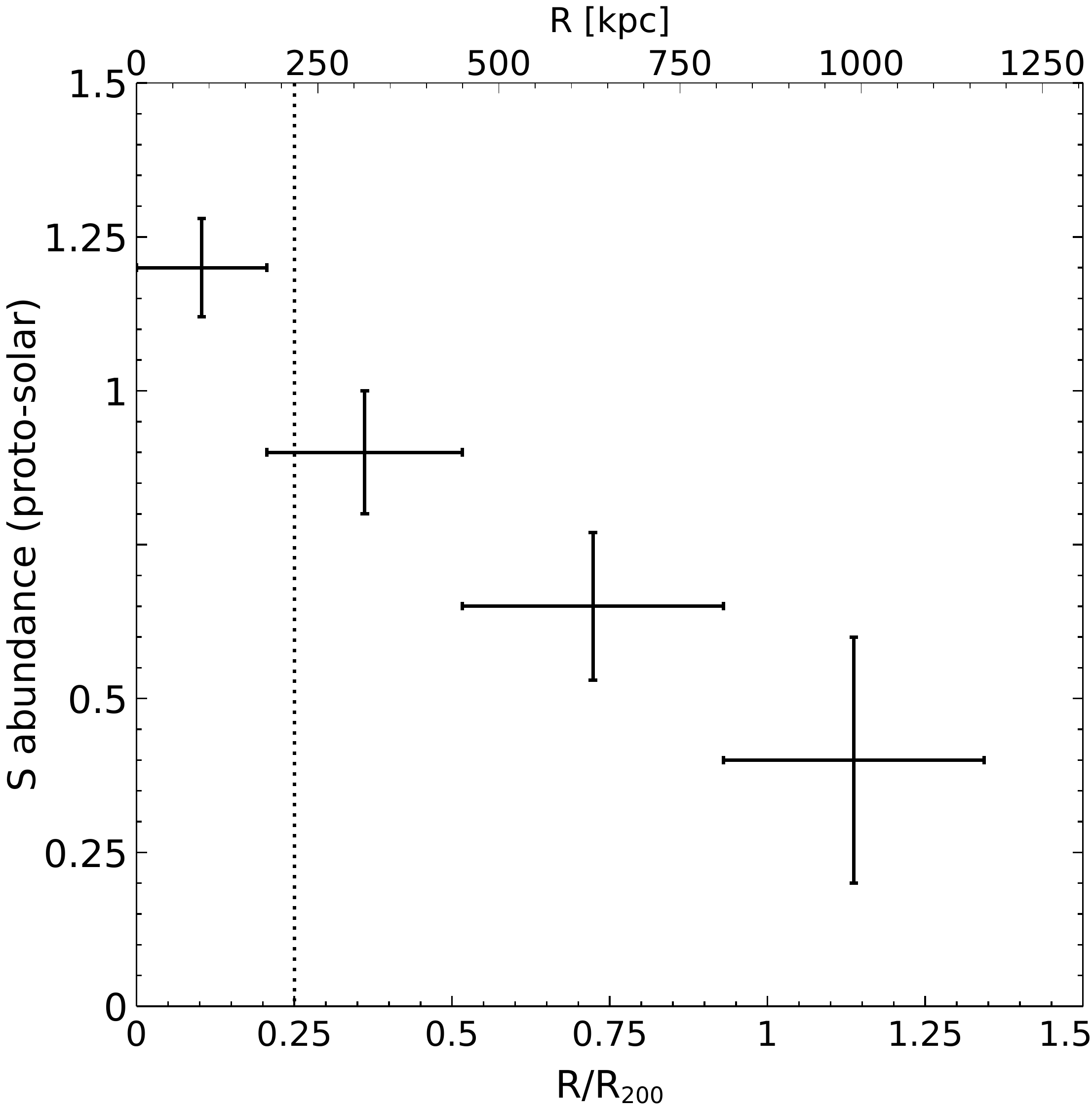}\\
\end{tabular}
\caption{RXJ 1159; same as Figure \ref{fig:image_mkw4}. 
}
\label{fig:image_rxj}
\end{figure*}

\renewcommand{\thefigure}{A4}

\begin{figure*}
\begin{tabular}{ccc}
& \hspace{-110pt} \includegraphics[width=0.75\textwidth]{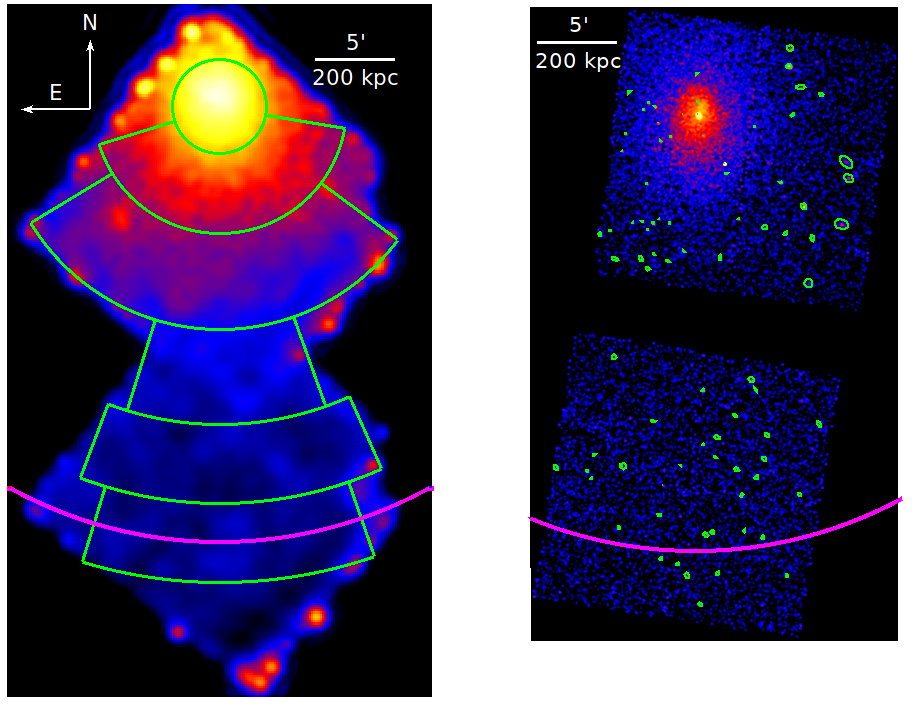}  &\\
\hspace{-15pt} \includegraphics[width=0.35\textwidth]{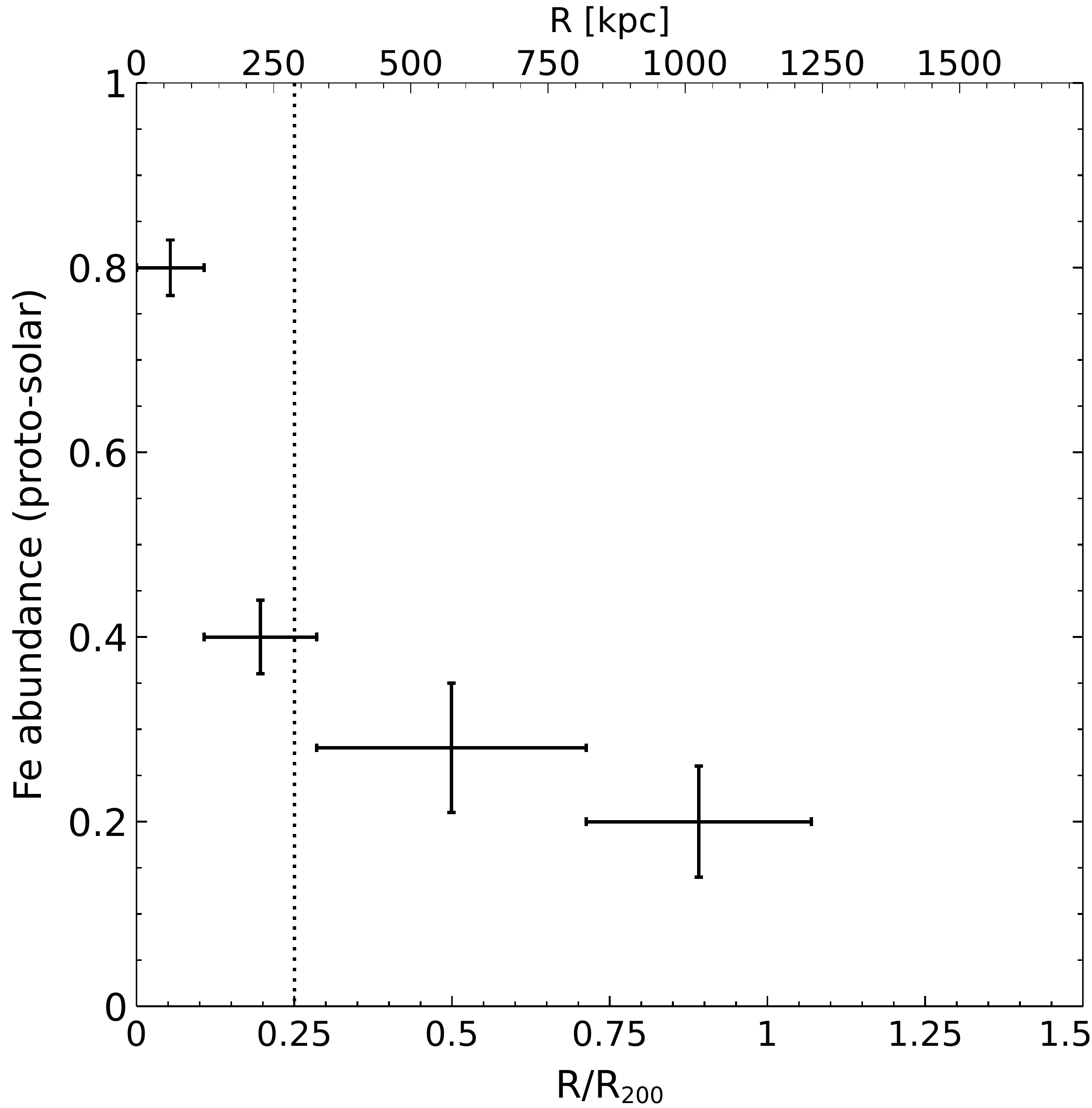} & \hspace{-120pt} \includegraphics[width=0.35\textwidth]{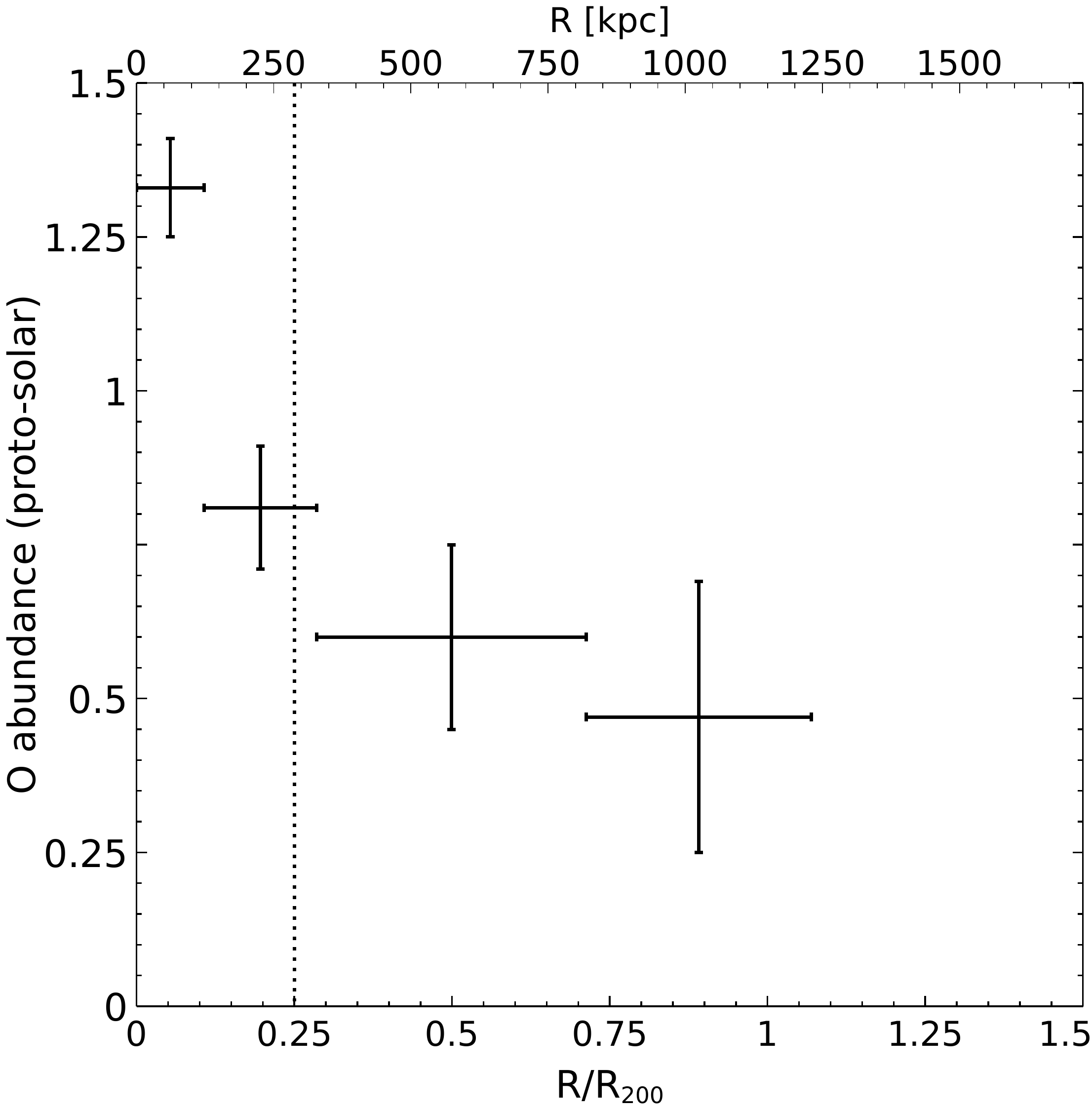} & \hspace{-120pt} \includegraphics[width=0.35\textwidth]{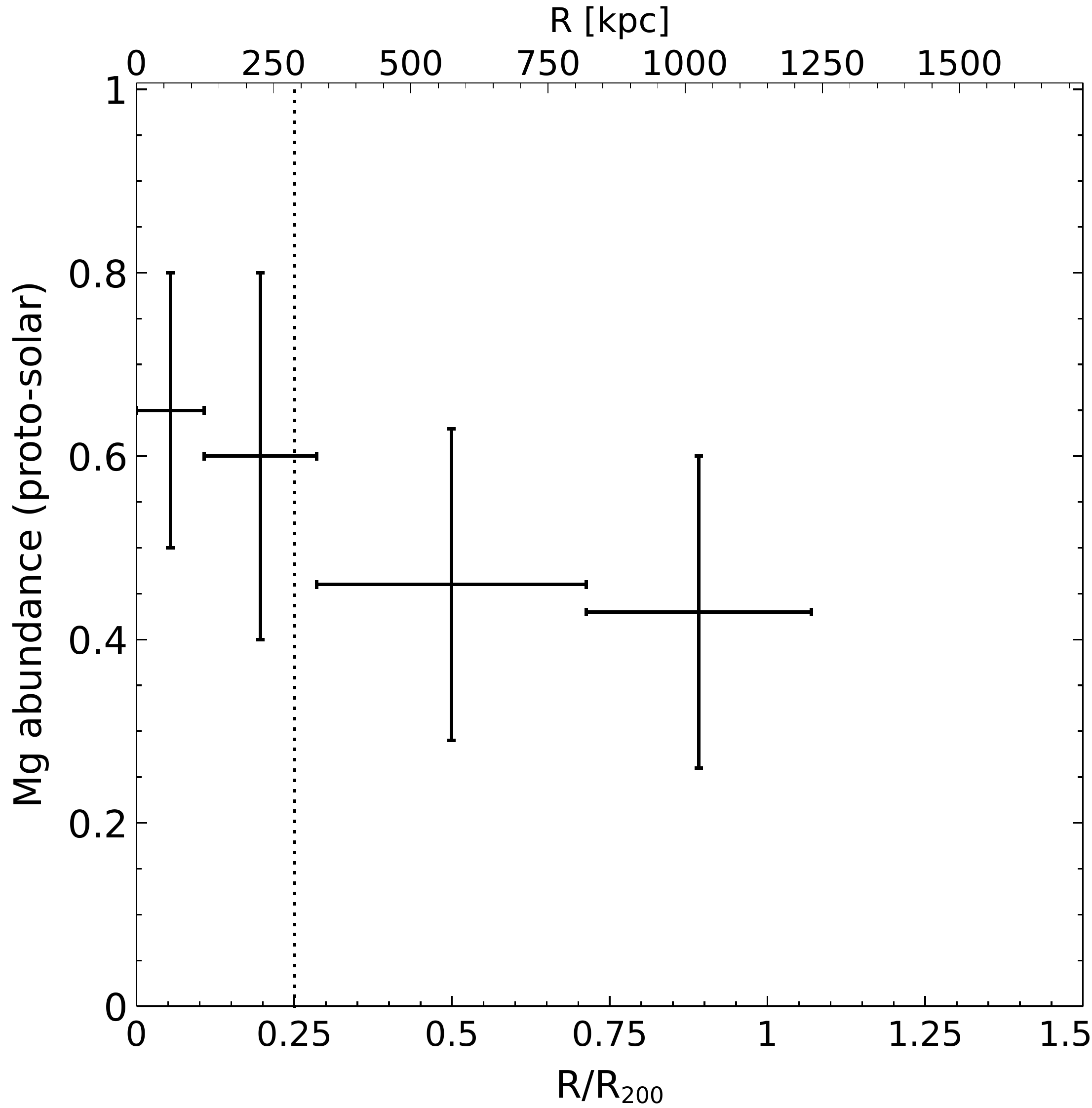}\\
\hspace{-15pt} \includegraphics[width=0.35\textwidth]{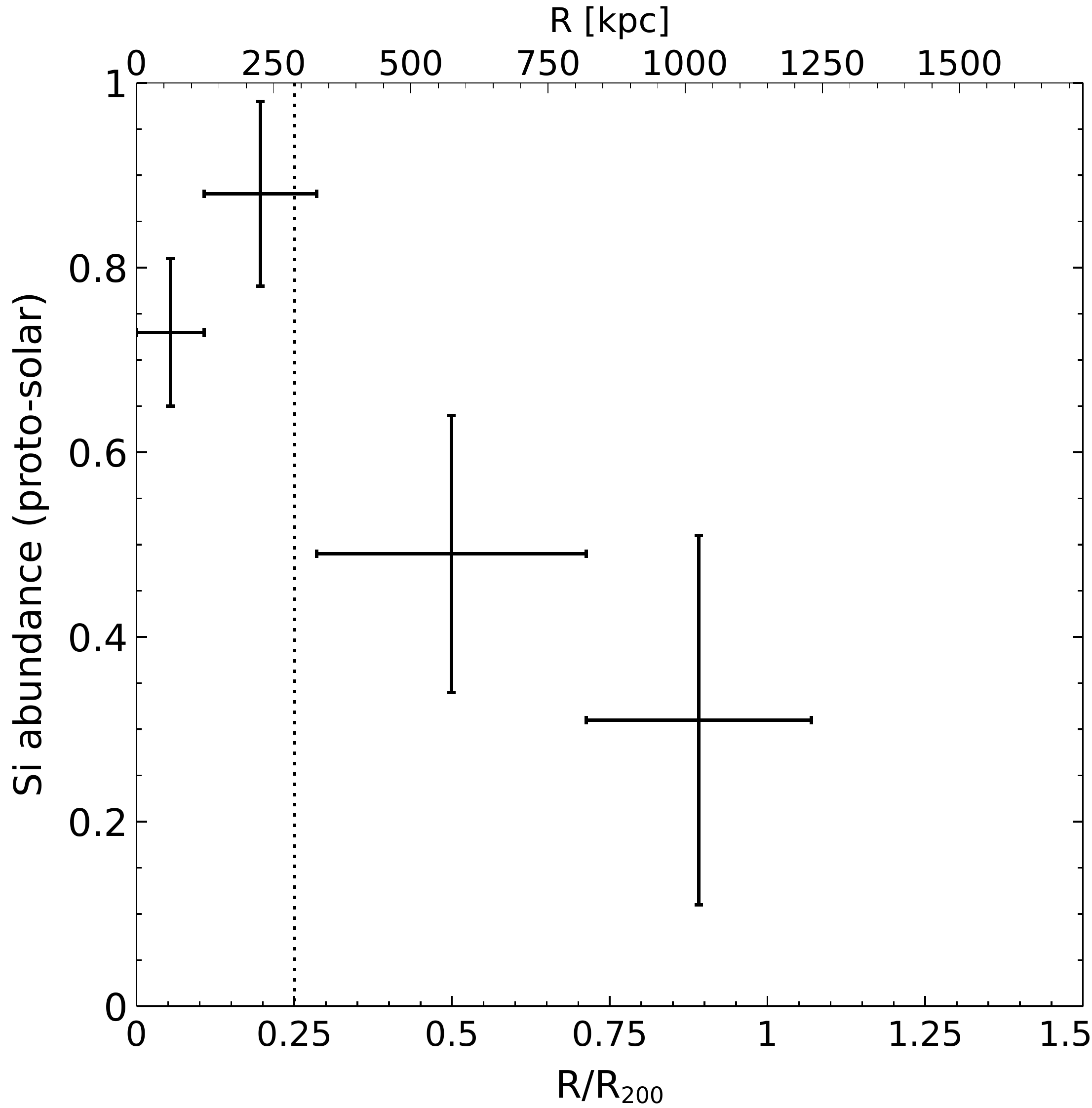} & \hspace{-120pt} \includegraphics[width=0.35\textwidth]{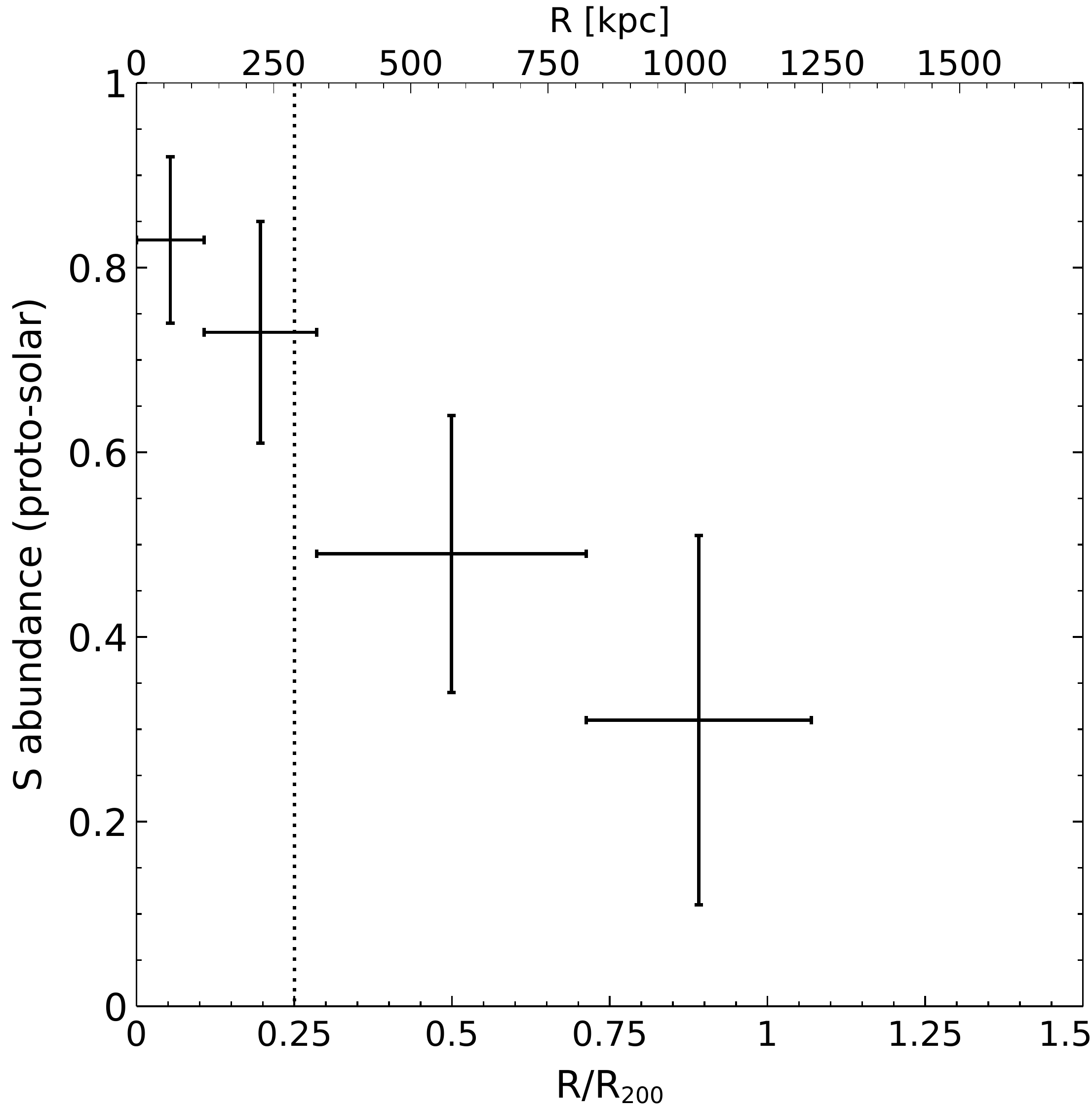}\\
\end{tabular}
\caption{ESO 3060170; same as Figure \ref{fig:image_mkw4}
}
\label{fig:image_eso}
\end{figure*}

\renewcommand{\thefigure}{A5}
\begin{figure*}
\includegraphics[width=0.7\textwidth]{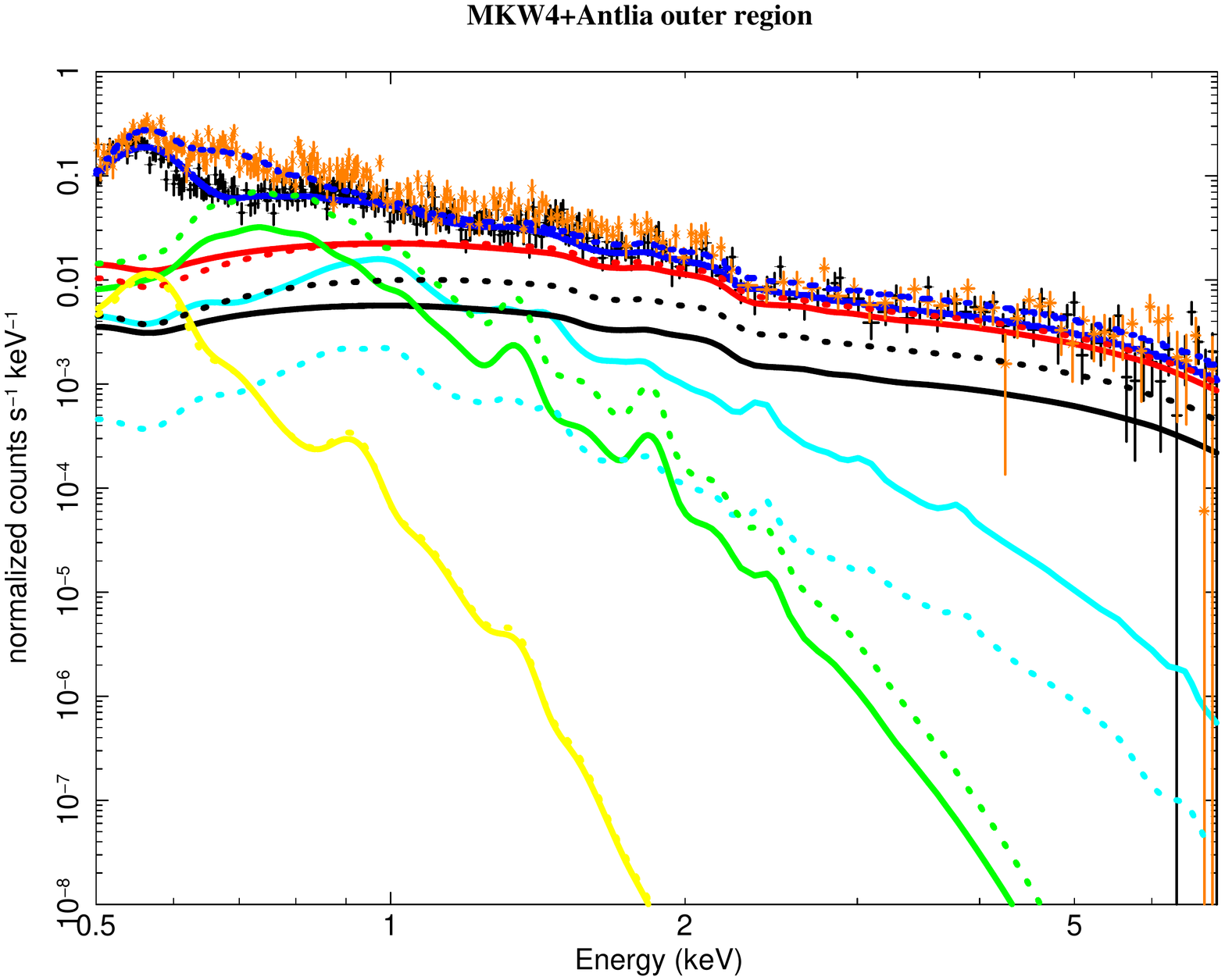}
\begin{tabular}{cc}
\hspace{-20pt} \includegraphics[width=0.6\textwidth]{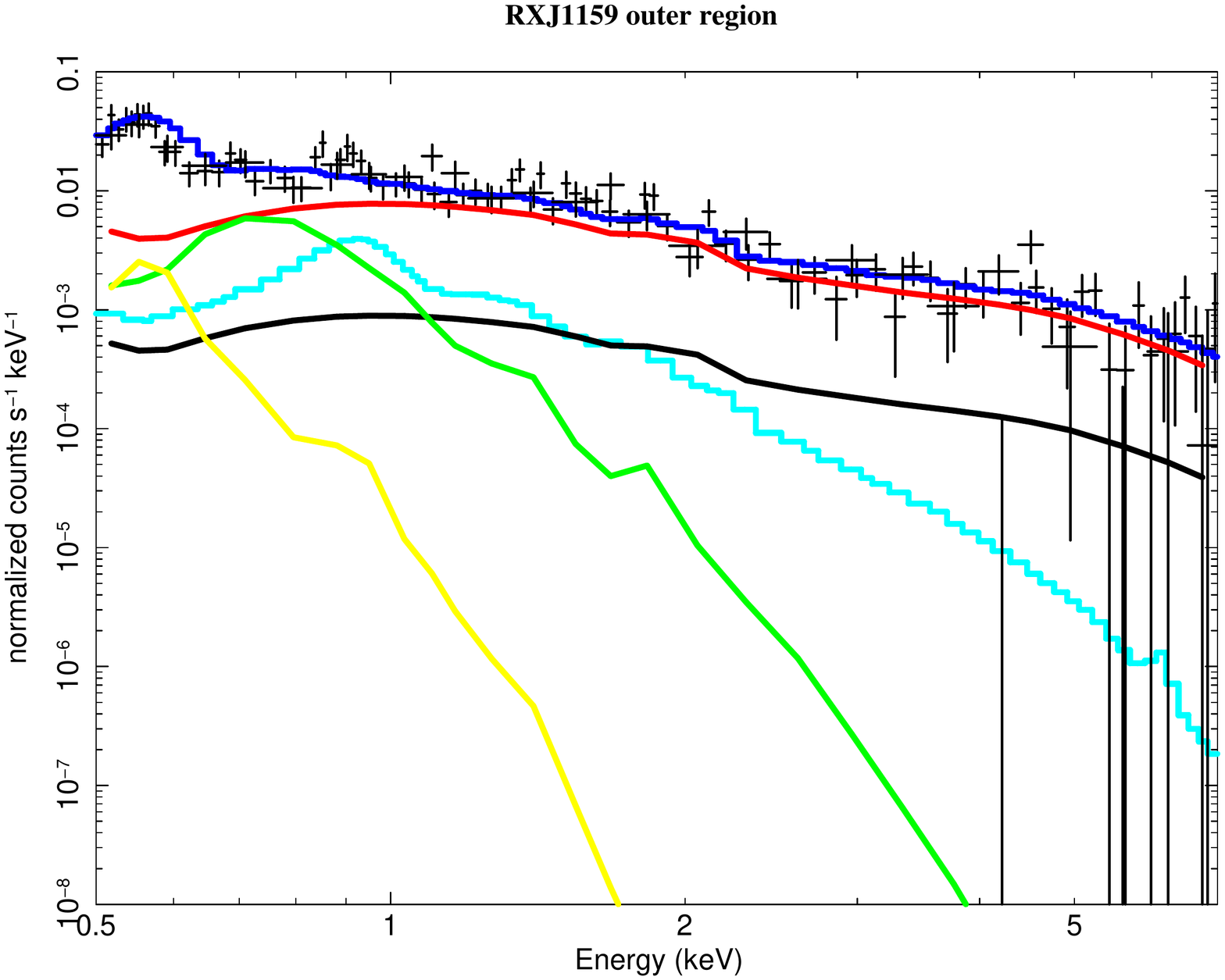} & \hspace{-60pt}
\includegraphics[width=0.6\textwidth]{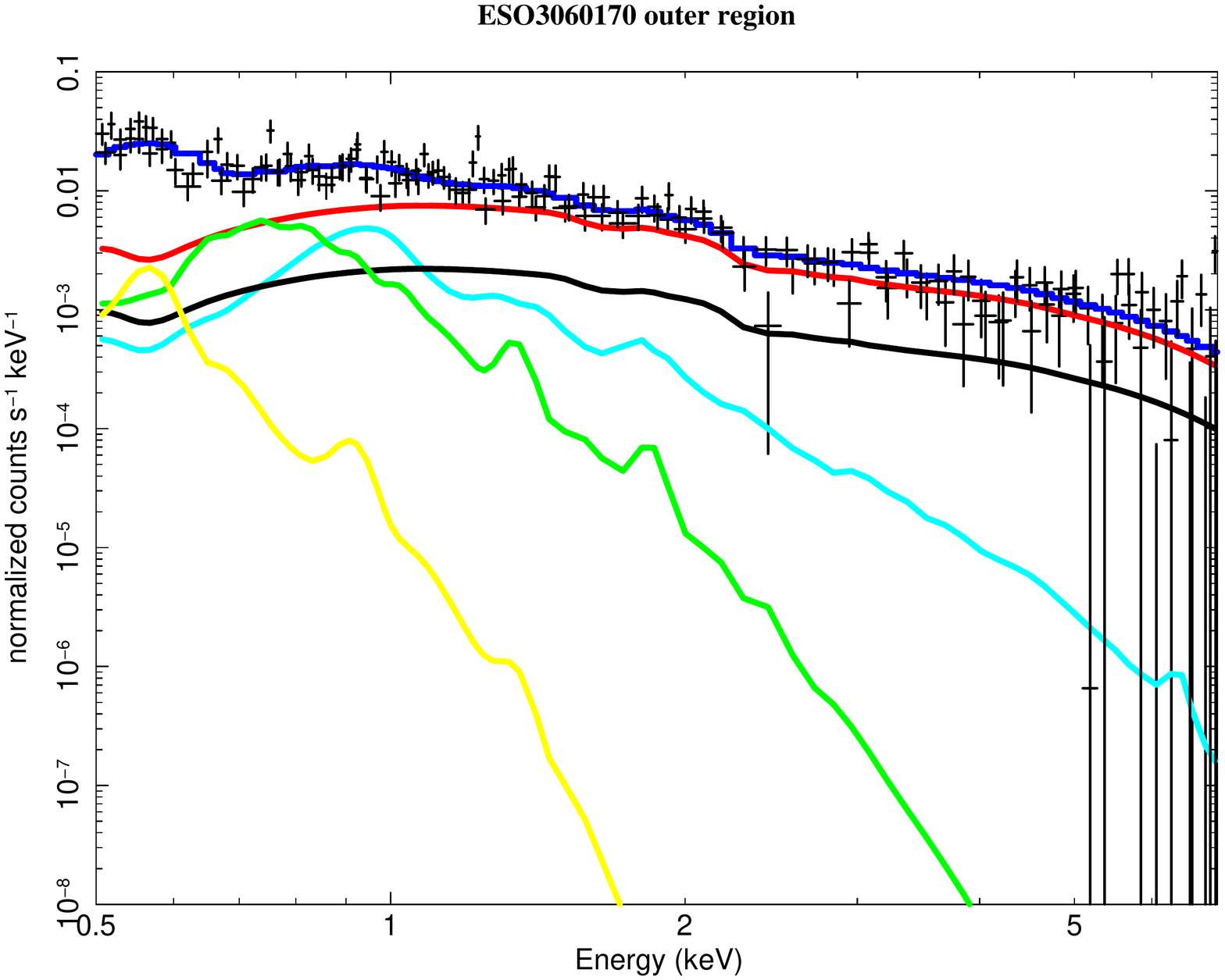}\\
\end{tabular}
\caption{Best-fit results of the spectral analysis
for the outermost regions of four groups.
We show only XIS1 spectra.
Red, black, cyan, green, yellow colored lines
indicate best-fit resolved CXB, unresolved CXB, 
ICM emission, GH, LHB components, respectively.
Blue line shows the best-fit ICM emission and X-ray
background together.
{\it Top panel:} Black data points are for MKW4.
Orange data points are for Antlia.
Solid line represents MKW4 and
dashed line represents Antlia.
{\it Bottom panel:} spectra for RXJ1159 and ESO3060170.}
\label{fig:spectra}
\end{figure*}

%%%%%%%%%%%%%%%%%%%%%%%%%%%%%%%%%%%%%%%%%%%%%%%%%%



%%%%%%%%%%%%%%%%%%%%%%%%%%%%%%%%%%%%%%%

%%%%%%%%%%%%%%%%%%%%%%%%%%%%%%%%%%%%%%%%%%%%%%%%%%

% Don't change these lines
\bsp	% typesetting comment
\label{lastpage}

\end{document}

%% file: macros.tex
\font\manual=manfnt at 7pt \def\dbend{\hbox{\raise0.9ex\hbox{\manual\char127\hspace{0.6em}}}}

\newcounter{INTERNALionstage}

\def\gtsim{\mathrel{\hbox{\rlap{\hbox{\lower4pt\hbox{$\sim$}}}\hbox{$>$}}}}
\def\lesssim{\mathrel{\hbox{\rlap{\hbox{\lower4pt\hbox{$\sim$}}}\hbox{$<$}}}}

\def\ga{{\rm\thinspace gauss}}

\def\la{\mbox{{\rm L}$\alpha$}}

\def\h0{\mbox{{\rm H}$^0$}}
\DeclareMathAlphabet{\vib}{OML}{cmm}{m}{it}
\newcommand*{\satellite}[1]{\textit{#1}}
\newcommand*{\xmm}{\satellite{XMM-Newton}}
\newcommand*{\chandra}{\satellite{Chandra}}
\newcommand*{\suzaku}{\satellite{Suzaku}}
\newcommand*{\asca}{\satellite{ASCA}}